\documentclass[aps,prd,twocolumn,showpacs,superscriptaddress,groupedaddress]{revtex4-2}  
\usepackage[T1]{fontenc}
\usepackage{lmodern} 
\usepackage{graphicx}  
\usepackage{dcolumn}   
\usepackage{bm}        
\usepackage{amssymb}   
\usepackage{amsmath}
\usepackage{bbold}
\usepackage{array}
\usepackage{makecell}
\usepackage{braket}
\usepackage{titlesec} 
\usepackage[colorlinks,citecolor=blue,urlcolor=blue,hypertexnames=true]{hyperref}
\usepackage{subfigure}

\usepackage[usenames,dvipsnames,svgnames,table]{xcolor} 

\renewcommand{\arraystretch}{2}

\newcommand{\be}{\begin{equation}}
\newcommand{\ee}{\end{equation}}
\newcommand{\bea}{\begin{eqnarray}}
\newcommand{\eea}{\end{eqnarray}}
\newcommand{\bml}{\begin{subequations}}
\newcommand{\eml}{\end{subequations}}
\newcommand{\bfig}{\begin{figure}}
\newcommand{\efig}{\end{figure}}

\newcommand{\bfx}{\mbox{\boldmath{$x$}}}

\newcommand{\bmat}{\begin{pmatrix}}
\newcommand{\emat}{\end{pmatrix}}
\usepackage{graphicx,slashed,booktabs,xcolor,multirow,float,
amsfonts,bbold,mathtools,sidecap,tikz,bm,enumitem}

\begin{document}

\definecolor{lime}{HTML}{A6CE39}
\DeclareRobustCommand{\orcidicon}{\hspace{-2.1mm}
\begin{tikzpicture}
\draw[lime,fill=lime] (0,0.0) circle [radius=0.13] node[white] {{\fontfamily{qag}\selectfont \tiny \,ID}}; \draw[white, fill=white] (-0.0525,0.095) circle [radius=0.007]; 
\end{tikzpicture} \hspace{-3.7mm} }
\foreach \x in {A, ..., Z}{\expandafter\xdef\csname orcid\x\endcsname{\noexpand\href{https://orcid.org/\csname orcidauthor\x\endcsname} {\noexpand\orcidicon}}}
\newcommand{\orcidauthorA}{0000-0002-0459-3873}

\widetext


\title{\textcolor{Sepia}{\textbf{\Huge  ${\cal C}$ausality ${\cal C}$onstraint \\on\\ \vspace{0.4cm} ${\cal C}$ircuit ${\cal C}$omplexity from ${\cal COSMOEFT}$}}}

\author{{\large Sayantan Choudhury\orcidA{}${}^{1,2,3}$}}
\email[Corresponding Author]{\\sayantan_ccsp@sgtuniversity.org,\\  sayanphysicsisi@gmail.com}

\author{\large Arghya Mukherjee${}^{3,4}$}
\email{arghya@niser.ac.in}
\author{\large Nilesh Pandey${}^{5}$}
\email{nilesh911999@gmail.com}
\author{\large Abhishek Roy${}^{6}$}
\email{roy.1@iitj.ac.in}

\affiliation{ ${}^{1}$Centre For Cosmology and Science Popularization (CCSP),
SGT University, Gurugram, Delhi- NCR, Haryana- 122505, India,}
\affiliation{${}^{2}$Institute of Physics, Sachivalaya Marg, Bhubaneswar, Odisha - 751005, India.}
\affiliation{${}^{3}$Homi Bhabha National Institute, Training School Complex, Anushakti Nagar, Mumbai - 400085,
	India.}
\affiliation{${}^{4}$National Institute of Science Education and Research, Jatni, Bhubaneswar, Odisha - 752050, India.}
\affiliation{${}^{5}$Department of Applied Physics, Delhi Technological University, Delhi-110042, India.}
\affiliation{${}^{6}$Department of Physics, Indian Institute of Technology Jodhpur,Karwar, Jodhpur - 342037, India.}	

\begin{abstract}
In this article, we investigate the physical implications of the causality constraint via the effective  speed of sound $c_s(\leq 1)$ on Quantum Circuit Complexity (QCC) in the framework of Cosmological Effective Field Theory (COSMOEFT) using the two-mode squeezed quantum states.  This COSMOEFT setup is constructed using the St$\ddot{\text{u}}$ckelberg trick with the help of the lowest dimensional operators,  which are broken under time diffeomorphism.  In this setup, we consider only the contributions from the two derivative terms in the background quasi-de Sitter metric.  Next, we compute the relevant measures of the circuit complexity and their cosmological evolution for different values of $c_s$ by following two different approaches, the  Nielsen’s approach and the Covariance matrix approach.  Using this setup,  we also compute the Von-Neumann and the Rényi entropy,  which finally establishes an underlying  relationship between the entanglement entropy and the circuit complexity. Considering the scale factor and $c_s$ as parameters, our analysis  of the  circuit complexity measures and the entanglement entropy suggests several interesting unexplored features within the window,  $0.024\leq c_s\leq 1$,  which is supported by both causality and cosmological observation.  Finally,  we  comment on the connection between the circuit complexity, the  entanglement entropy and the equilibrium temperature for different $c_s$ values lying within the mentioned window. 
\end{abstract}

\maketitle
\section{\textcolor{Sepia}{\textbf{ \large  Introduction}}}
\label{sec:introduction}

The underlying physical concept of Quantum Circuit Complexity (QCC) is an instrumental tool not only in high energy physics but also in other branches of theoretical physics \cite{Chapman:2021jbh,Chapman:2017rqy,Chapman:2018dem,Chapman:2018lsv,Cano:2018aqi,Barbon:2018mxk,Flory:2018akz,Chapman:2018bqj,Agon:2018zso,Goto:2018iay,Guo_2018,Bernamonti_2019,Goto_2019,Bhattacharyya:2018bbv,Khan:2018rzm,Hackl:2018ptj,Alves:2018qfv,Camargo:2018eof,Camargo:2019isp,Chapman:2018hou,Doroudiani:2019llj,Hashemi:2019aop,Bernamonti_2020,Caceres_2020,Choudhury:2020lja,Brown:2016wib,Couch:2016exn,Chagnet:2021uvi,Koch:2021tvp,Barbon:2015ria,Alishahiha:2015rta,Yang:2016awy,Chapman:2016hwi,Reynolds:2016rvl,Zhao:2017iul,Carmi:2017jqz,Reynolds:2017lwq,Swingle:2017zcd,Fu:2018kcp,Bolognesi:2018ion,Chen:2018mcc,Abt:2018ywl,Hashimoto:2018bmb,Couch:2018phr,Ben-Ami:2016qex,Brown:2017jil,Hashimoto:2017fga,Ali:2019zcj,Chapman:2019clq}. In recent years, this useful tool has  achieved remarkable   popularity  in the domain of quantum field theory as well as  quantum information theory.   Originally this concept was proposed to provide a way to probe the features behind the horizon of black holes by making use of the well-known ``{\it Complexity = Volume} ''  and ``{\it Complexity = Action}'' conjectures  \cite{Susskind:2018pmk,Stanford:2014jda,Susskind:2014rva,Roberts:2014isa,Susskind:2014jwa}.  After these conjectures' success, this concept has been frequently used in the framework of QFT and  AdS/CFT \cite{Jefferson:2017sdb}.  Using both of these conjectures and the underlying concepts, it is possible to study various unexplored features of bulk gravitational physics with the help of quantum information theory. 

 Recently, using both the concept of QCC and the out-of-time ordered correlation (OTOC) functions,\cite{Choudhury:2020yaa,Bhagat:2020pcd, Choudhury:2021tuu,Hashimoto:2017oit,Hashimoto:2020xfr} it has been shown that it is possible  to probe the unknown features of quantum chaos. For example, such studies indicate that it is possible to extract crucial information about the quantum Lyapunov exponent, the scrambling time,  and many more essential quantities to quantify the chaos in a quantum mechanical system. These concepts have further been implemented  in the framework of cosmological islands \cite{Choudhury:2020hil}, the inflationary cosmology \cite{Bhattacharyya:2020rpy,Bhattacharyya:2020kgu}, the  bouncing cosmology \cite{Bhargava:2020fhl}, the  primordial gravitational waves (PGW) \cite{Adhikari:2021ked}, the  black hole gas in arbitrary dimensions \cite{Adhikari:2021pvv},the  supersymmetric quantum field theory \cite{Choudhury:2021qod}, the   quantum neural networks \cite{Choudhury:2020lja} and in many other areas.  The underlying connection between the quantum entanglement in wormholes and the emergent gravitational space-time in terms of the spooky action at a distance has been an active area of interest in recent times.  In this context, the quantum entanglement entropy is interpreted as the minimal area of an {\it Einstein-Rosen bridge} (ER) in the context of the physics of a wormhole.  On the other hand,  in the classical picture, the minimal area of ER grows over a longer time span.  But in the dual gravitational picture, the corresponding system approaches thermal equilibrium very fast,  compared to the ER growth time scale.  To express this phenomenon in a  simple language, in Ref. \cite{Susskind:2018pmk}, the author introduced the concept of quantum circuit complexity.  In the context of black hole physics,  it was further conjectured that the growth rate of quantum circuit complexity ($C$) with respect to time scale ($t$) is equal to the product of entropy ($S$) and the equilibrium temperature ($T$) of the black hole i.e.  \be \frac{d C}{dt}=TS, \ee  which is  consistent with the corresponding physical picture.  Here it is important to note that,  in terms of the quantum circuits, this relationship corresponds to $K\sim S$, where the number of qubits in SU($2^{K}$) gates (where the Hilbert space is $2^{K}$ dimensional),   turns out to be the black hole entropy in this context. Recently in Ref. \cite{Eisert:2021mjg},  the authors have pointed out an underlying connection between the quantum circuit complexity and the quantum entanglement in two different contexts.  It is observed that if  the quantum entanglement entropy grows linearly with the corresponding time scale then the quantum circuit complexity also follows the same behaviour.  However,  it is important to study the underlying connection between the complexity growth and the entropy growth separately in various physical systems. This is  because the previously mentioned relationship only holds for the  black hole systems and is not at all expected to be universal.  For this reason, finding a novel relationship or any  deviation from the known  relationship is itself interesting,  which is the main objective of the present work. To study the quantum circuit complexity and the quantum entanglement, here we  employ the framework of Cosmological Effective Field Theory (COSMOEFT). It should be mentioned that apart from the cosmological perspective pursued in the present work,  the concept of EFT  is extremely useful in many other areas of theoretical physics including the condensed matter physics,  the particle physics, the hydrodynamics and so on (see for example Refs.  \cite{Pich:1998xt,Burgess:2007pt,Shankar:1996vk,Donoghue:1995cz,Donoghue:2012zc,Cheung:2007st,Weinberg:2008hq,Agarwal:2012mq,Ozsoy:2017mqc,Ozsoy:2015rna,Burgess:2017ytm,Baumann:2014nda,Baumann:2009ds,Choudhury:2016wlj,Choudhury:2018glz,Choudhury:2013iaa,Choudhury:2014sua,Delacretaz:2016nhw,Delacretaz:2015edn,LopezNacir:2011kk,Naskar:2017ekm,Senatore:2010wk,Baumann:2015nta,Dubovsky:2011sj,Crossley:2015evo,Adhikari:2021ckk,Choudhury:2017glj,Choudhury:2015hvr,Choudhury:2017cos,Aoki:2021wew,Choudhury:2014sxa} for more details). In general, as we go higher/lower and higher/lower in energy,  an EFT framework is expected to be valid only up to a specific  cut-off scale,  commonly known as the UV/IR cut-off scale.  For the high-energy EFT prescriptions, the upper limit of this scale is fixed at the quantum gravity cut-off scale,  which is the Planck scale $M_p(\sim 10^{19} ~{\rm GeV})$. On the other hand, for the low energy EFT, it is  a free parameter.  Depending on which low energy phenomenon is described by the EFT, one should be able to fix this parameter. In this work we  restrict ourselves to the high-energy EFT framework which considers all the operators  allowed by the underlying symmetry. \textcolor{black}{It should be noted that there are two possible approaches towards the development of the  EFT, the top-down approach and the bottom-up approach. A brief review of the two approaches has been included  in the appendix, (\underline{section \ref{appen1}}) and the usefulness of the bottom-up approach in the present context  is discussed}. Let us now concentrate on the more specific details of our methodology.

    To obtain the expressions for the quantum circuit complexity and the entanglement entropy, we use the squeezed quantized modes that originate from the scalar part of the cosmological metric perturbation, commonly identified as the comoving curvature perturbation.  For this purpose, we use the unitary gauge which is useful to apply the St$\ddot{\text{u}}$ckelberg trick~\cite{Ruegg:2003ps,Grosse-Knetter:1992tbp} which further generates the scalar component of the cosmological perturbation or the Goldstone modes by breaking the time diffeomorphism symmetry.  Also, it is important to note that,  in the present work, we have restricted the quantum initial state to the well-known Bunch Davies or Euclidean state.  Here the St$\ddot{\text{u}}$ckelberg trick mimics the role of the spontaneous symmetry breaking  in the $SU(N)$ non-abelian gauge theory. Specifically, in this formalism, it appears that the Goldstone mode is destroyed by the quasi de Sitter background metric.  To formulate this COSMOEFT setup the following prime components are necessary: 
\begin{itemize}
	\item First of all,  to construct the EFT action, the polynomial powers of the temporal perturbed component of the metric ($\delta g^{00}$) are required which is identified as \be \delta g^{00}= \left(g^{00}+1\right). \ee Here $g^{00}$ represents the time component of the background metric.
	\item Further,  to construct the EFT action, the polynomial powers of the temporal perturbed component of the extrinsic curvature at constant time slice ($\delta K_{\mu\nu}$) is required which is identified as \be \delta K_{\mu\nu}=\left(K_{\mu\nu}-a^2Hh_{\mu\nu}\right),\ee 
	where $K_{\mu\nu}$ is the extrinsic curvature at constant time slice,  $a$ is the previously mentioned scale factor in the quasi de Sitter space-time,  $H$ is the corresponding Hubble parameter and $h_{\mu\nu}$ corresponds to  the spin-2 transverse, traceless graviton modes (gravitational waves).
\end{itemize} 
The key idea of constructing an EFT action using St$\ddot{\text{u}}$ckelberg trick can  also be extended to incorporate several scalar fields.  However,  in  this work, we have restricted our discussion to single canonical or non-canonical scalar field  in COSMOEFT.

With this COSMOEFT setup, we study  the cosmological evolution of the quantum circuit complexity and the entanglement entropy  for different causal values of the  speed of sound parameter $c_s\leq 1$.  This framework helps us to analyse both the canonical and a class of non-canonical single field models with the help of a single EFT action. Most importantly it will help us to constrain a wide class of cosmological single scalar field models in the EFT setup.  In  technical terms, we will compute two very important measures of the quantum circuit complexity and the entanglement entropy,  which are the von Neumann and the Rényi entropies in a  specific situation where the interacting part of the Hamiltonian of the EFT setup is parametrized by the two-mode squeezed states after performing the quantization. 
It turns out that the two important  measures of the quantum circuit complexity can be fully parametrized in terms of two parameters in the mentioned two-mode squeezed state formalism,  which are identified as the squeezing amplitude and the squeezing angle.  On the other hand,  the entanglement entropy is only parametrized via the squeezing amplitude. Both of these parameters evolve in the cosmological time scale and their explicit behaviour  can be computed in the  Heisenberg picture for the given quasi-de Sitter background metric which we use to develop the present COSMOEFT setup.  By performing rigorous calculation it will turn out that in Heisenberg quantum picture, the squeezing amplitude and the angle both satisfy two coupled linear differential equations where the cosmological evolution either can be explained through the time scale or in terms of the scale factor appearing in the expression for the quasi de Sitter metric. The coupled differential equations will be solved numerically with a preferred set of  initial conditions on the squeezing amplitude and the angle.  For $c_s=1$ case, it turns out that  this mentioned procedure is quite simple.  On the other hand,  when $c_s<1$, we will have non-trivial contributions of  $c_s$  in the differential equations of the squeezing parameters.  In this case, one has to provide  a specific form of the sound speed parameter which will be expressed in terms of the parameters appearing in the COSMOEFT action.  This procedure is more complicated and involved compared to the previously mentioned case and has significant implications in  a large class of non-canonical scalar models which can be constrained with the help of the EFT approach. 

Moreover, it is interesting to know, between the quantum circuit complexity and the entanglement entropy, (1) which  one is capturing more information in the quantum information theoretic sense?  (2) which one gives the best measure to describe the underlying COSMOEFT setup? It is also interesting to see how the quantum circuit complexity and the entanglement entropy are related to each other within the present context. To address such issues, we shall attempt a parametric analysis of the rate of change of the quantum circuit complexity with respect to the scale factor and the entanglement entropy, where, the effective  speed of sound $c_s$ will be treated as a parameter to explicitly observe the outcome in the causal region $c_s\leq 1$ of the EFT.  Such an  analysis is useful to comment on the equilibrium thermodynamic features of the system under consideration and is also  essential to quantify the deviation from the previous results obtained from the evaporating black holes.  As mentioned before, in the  evaporating black hole context, we have a simple analytic  relationship between the growth rate of the quantum circuit complexity and the entanglement entropy.  Hence it is interesting to know what would be the possible connection between these two quantities within the context of COSMOEFT.  For this reason, a significant portion of the present article has been devoted to this issue.
\\ \\
The key highlights  of the present work are given as follows:

\begin{itemize}
	\item In this work, the prime focus is to study the physical implication of the causality on quantum circuit complexity and the entanglement entropy within the context of COSMOEFT.
	
	\item For this purpose, the COSMOEFT setup is constructed by following the model-independent bottom-up approach where we use St$\ddot{\text{u}}$ckelberg trick which further generates the scalar component of the cosmological perturbation or the Goldstone modes by breaking the time diffeomorphism symmetry.    
	
	\item Considering  different values of the  speed of sound parameter $c_s$  within the preferred range $0.024\leq c_s \leq 1$,  the quantum circuit complexity is studied using two well-known methods,  which are the Covariance matrix method and Nielsen's wave function method. 
	
	\item  The dependence of  the entanglement (von Neumann) entropy  on (1) the scale factor in the quasi-de Sitter space and (2) the speed of sound  $c_s$ has been analysed.

	\item An underlying connection between the rate of change of circuit complexity and the entanglement entropy has been established by treating the effective  speed of sound $c_s$ as a parameter within this COSMOEFT setup.  Also, we comment on the equilibrium thermodynamic behaviour of the system under consideration.

\end{itemize}

The organization of the paper is as follows:

\begin{itemize}

	\item  In \underline{section \ref{v3a1}},  we start our discussion by providing the analytical asymptotic solution for the mode functions generated from the scalar perturbation,  which mimics the role of Goldstone modes in COSMOEFT.  To fix the solution we have chosen Bunch Davies or Euclidean vacuum state as the initial choice of the quantum vacuum state.
	
\item  In \underline{section \ref{v3a2}},  we further provide the details of the quantization of the Hamiltonian of the COSMOEFT in terms of two-mode squeezed states.  In this formalism, we have shown that the interacting part of the quantized Hamiltonian is parametrized in terms of squeezing amplitude and squeezing angle,  where each of them evolves either with respect to the evolutionary time scale or the scale factor.
	
	\item In \underline{section \ref{sec:CircuitComplexity}},   we provide a short review on the subject of quantum circuit complexity,  which we believe will be very helpful for the better understanding purpose of general readers.  Here we provide sufficient details regarding the fundamental definition,  appropriate measures, and various properties along with the well-known computational methods of the circuit complexity developed by Nielsen and collaborators.  
	
	\item In \underline{section \ref{sec:twomode}},  we discuss the computational details of the quantum circuit complexity using the previously mentioned two-mode squeezed state formalism within the framework of COSMOEFT.  Here we have shown that the quantum circuit complexity can be quantified using the well-known Covariance Matrix method and Nielsen's method of wave functions.

	\item In \underline{section \ref{sec:ee}},  we provide the computational details of the entanglement entropy of two-mode squeezed states. 
In this section, we compute Rényi-entropy,  and von-Neumann entropy in this connection in terms of squeezed amplitude parameter. 

	\item In \underline{section \ref{comp}},  we discuss an underlying connection between quantum circuit complexity and entanglement entropy which will be helpful to know which one is the best measure within the framework of COSMOEFT.
	
	\item In \underline{section \ref{sec:numerical}},  we mention and analyse all the numerical outcomes from the features of quantum circuit complexity and entanglement entropy within the framework of COSMOEFT.  In this section, we have studied the behaviour of quantum circuit complexity and entanglement entropy (1) with respect to the scale factor in quasi de Sitter space for the fixed values of the sound speed $c_s$ and (2) with respect to the sound speed $c_s$ for the fixed value of the scale factor at the early time scale of the evolution of our universe.  Also, we have studied the parametric behaviour of the rate of change of the quantum circuit complexity with respect to the scale factor with the entanglement entropy in this context by considering the effective sound speed $c_s$ as a parameter within the region $0.024\leq c_s\leq 1$,  which is both supported by cosmological observation and causality.  This combined analysis helps us to extract various known physical implications of these quantum information theoretic measures for $c_s\leq 1$,  using which one can further able to constraint both single scalar field models as well as a class of non-canonical scalar field models within the framework of COSMOEFT.  Additionally,  this analysis will be going to help us to determine a  relationship between all of these quantum information theoretic measures.  Most importantly,  it will determine which is the best measure in this context. 
	
\item In \underline{section \ref{method}},  we provide a comparison between the outcomes obtained from the computation of quantum circuit complexity using the Covariance Matrix method and Nielsen's method of wave functions.  This comparison will help us to decide from which method we are getting full information from the underlying COSMOEFT setup.
	
	\item Finally in \underline{section \ref{sec:Conclusions}}, we  summarize  the main results obtained in this work and conclude with a brief discussion on the future prospects.
	    \item {\textcolor{black}{In appendix 1, \underline{section \ref{appen1}} we have given the two possible approches towards the construction of EFT.}}
			\item {\textcolor{black}{In appendix 2, we also provide an overview of the model independent bottom up approach of the generic COSMOEFT framework in \underline{section \ref{sec:E2}}.}}

	\item {\textcolor{black}{Further, in appendix 3, \underline{section \ref{v2b}} we have discussed about St$\ddot{\text{u}}$ckelberg trick which helps further to construct an EFT in terms of Goldstone modes. This will further help to develop a consistent COSMOEFT setup for the present analysis.}}

	\end{itemize}

 	\section{\textcolor{Sepia}{\textbf{ \large Mode Equation and Solution for Scalar~Perturbation\label{v3a1}}}}
We start our analysis by simplifying the form of the Goldstone action which can be expressed in terms of the curvature perturbation variable $\zeta(t,{\bf x})$ as:
  \bea S_{\zeta}=\int d^{4}x ~a^3~\left(\frac{M^2_p\epsilon}{c^2_s}\right)\left[\dot{\zeta}^2
   -c^2_s \frac{1}{a^2}(\partial_{i}\zeta)^2
   \right].~~~~~~~\eea
where $\epsilon$ corresponds to the slow-roll parameter and the essential details of obtaining the action is relegated to the appendix (see  equation (\ref{GM}) of section \ref{v2b4} for the final expression). To serve this purpose we introduce a field redefinition:
		      \be  v(\tau,{\bf x})=z~\zeta(\tau,{\bf x})~M_p~~{\rm where}~~z =\frac{a\sqrt{2\epsilon}}{{c}_{s}}\ee
which helps us to recast the action for the curvature perturbation in terms of new variables as:
\bea S=\int d^{3}x~ d\tau ~\left[v^{'2}
		      		   -{c}^2_s \frac{1}{a^2}(\partial_{i}v)^2
		      		  +\frac{1}{z}\frac{d^2z}{d\tau^2}v^2
		      		   \right],~~~~~~~~~~~~\eea
		      		   Here $\tau$ is the conformal time which can be expressed in terms of physical time $t$ as, 
		      		   $\tau=\int \frac{dt}{a(t)}.$
		      		{\textcolor{black}{ The conformal time described here is negative and lying within $-\infty<\tau<0$.}  In terms of conformal time the background metric of quasi-de Sitter space-time can be recast as:
		\bea ds^2=a^2(\tau)\left(-d\tau^2+d{\bf x}^2\right),\eea     
		where the scale factor in terms of conformal time in quasi-de Sitter space can be expressed as: 		 
		      		   \be\label{sf}
		      		    \displaystyle a(\tau) = -\frac{1}{H\tau}\left(1+\epsilon\right)~~~~ \ee

		      		   Now to compute the mode function for the scalar perturbations, it is useful to  map the whole problem in Fourier space by using the  following ansatz for the Fourier transformation:
		      		   \be v(\tau,{\bfx})=\int \frac{d^3k}{(2\pi)^3}~v_{\bf k}(\tau)~\exp(i{\bf k}.{\bf x})\ee
		      	 After expressing the action in Fourier space and varying the action, the equation of motion for scalar perturbation can be written as:
		      		   \be v^{''}_{\bf k}(\tau)+\left({c}^2_{S}k^2-\frac{1}{z}\frac{d^2z}{d\tau^2}\right)v_{\bf k}(\tau)=0.\label{9eqmod}\ee
In the quasi de Sitter space, one can write:
		      		   \be\label{qds}
		      		   \displaystyle \frac{1}{z}\frac{d^2z}{d\tau^2} =  \frac{\left(\nu^2-\frac{1}{4}\right)}{\tau^2}~~~~ \ee
where the parameter $\nu$ can be expressed as:
		      \be\label{nu}
		      		      		   \displaystyle \nu =  \left(\frac{3}{2}+3\epsilon-\eta+\frac{s}{2}\right)~~~~ \ee		   		    
		      		   		      		   where $\epsilon$, $\eta$ and $s$ are the  slow-roll parameters which are defined as:
		      		   		      		   \bea \label{SLP}\epsilon &=& -\frac{\dot{H}}{H^2},~~~~
		      		   		      		   \eta = 2\epsilon-\frac{\dot{\epsilon}}{2H\epsilon},~~~~
		      		   		      		   s=\frac{\dot{c_{s}}}{Hc_{s}}.\eea

		      		The general analytical solution for $v_{\bf k}(\tau)$ can be expressed as:
		      		\be\label{GEN}
		      		 \displaystyle v_{\bf k}(\tau) = \sqrt{-\tau}\left[C_1  H^{(1)}_{\nu} \left(-k{c}_{S}\tau\right) 
		      		+ C_2 H^{(2)}_{\nu} \left(-k{c}_{S}\tau\right)\right]~~\ee
\textcolor{black}{where $H^{(1)}_{\nu} \left(-k{c}_{S}\tau\right)$ and $H^{(2)}_{\nu} \left(-k{c}_{S}\tau\right)$ are the Hankel functions of first and second kind with order $\nu$ respectively. 	Also $C_{1}$ and $C_{2}$ are the arbitrary integration constants which depend on the choice of the initial quantum vacuum.}   In this work, we use the well-known Bunch Davies or the Euclidean vacuum state as the initial state which is fixed by  $C_{1}=1, C_{2}= 0$. Considering this, the general solution for the scalar mode function can be simplified as:
		      		\bea \label{SMF}
		      		 \displaystyle v_{\bf k}(\tau) = \sqrt{-\tau}~  H^{(1)}_{\nu} \left(-k{c}_{S}\tau\right).\eea 
	Now instead of considering  the general  solution, it is useful to inspect certain  limiting cases which correspond to  cosmologically relevant regions namely:
	\begin{enumerate}
	\item \textcolor{Sepia}{\bf The Superhorizon region:}~~ $k{c}_{s}\tau\ll-1$,
	
	\item \textcolor{Sepia}{\bf The Horizon crossing:}~~~~~~~ $k{c}_{s}\tau= -1$,
	
	\item \textcolor{Sepia}{\bf The Subhorizon region:}~~~~~ $k{c}_{s}\tau\gg-1$.
\end{enumerate}		 


\textcolor{black}{For a deeper  understanding of these limiting cases, let us  consider the modes in Eq \eqref{9eqmod}.  These modes for the scalar perturbation can be related to its wavelength by the following eq:
\begin{equation}
\lambda_{\text {phys }}=\frac{2 \pi}{k} a
\end{equation}
and}

\textcolor{black}{\begin{equation}
\lambda_{p h y s} \gg H^{-1} \quad \Leftrightarrow \quad \frac{2 \pi}{k} a \gg-a {c}_{s}\tau \quad \Leftrightarrow \quad|k {c}_{s}\tau| \ll 1,
\end{equation}}

\textcolor{black}{When the physical wavelength $\lambda_{p h y s}$ of the modes is stretched so much such that it will become larger than the Hubble horizon $H^{-1}$,  it is called the super horizon region which corresponds to $k{c}_{s}\tau\ll-1$, whereas, for $k{c}_{s}\tau\gg-1$, where the wavelength of modes are inside the horizon we clasify  it as the subhorizon region. Therefore, the region, where the wavelength of the modes is comparable to the Hubble horizon, is described as the Horizon crossing region for $k{c}_{s}\tau=-1$.}

\textcolor{black}{ Using asymptotic limit for the Hankel function which is given by:}

\textcolor{black}{\begin{align}
\lim _{-k {c}_{S}\tau \rightarrow \infty} H_{\nu}^{(1)}=\pm \sqrt{\frac{2}{\pi}}& \frac{1}{\sqrt{-k {c}_{S}\tau}}\times \\&\exp \left(\mp i\left\{k{c}_{S} \tau+\frac{\pi}{2}\left(\nu+\frac{1}{2}\right)\right\}\right)\nonumber
\end{align}}

\textcolor{black}{\begin{equation}
\lim _{-k {c}_{S}\tau \rightarrow 0} H_{\nu}^{(1)}=\pm \frac{i}{\pi} \Gamma\left(\nu\right)\left(\frac{-k {c}_{S}\tau}{2}\right)^{-\nu}
\end{equation}}
\textcolor{black}{In Eq \eqref{SMF} we write expression for the solution to the mode function for Bunch Davies state and in the sub horizon limit $-k {c}_{S}\tau \rightarrow \infty~~(-k {c}_{S}\tau \gg 1)$ as:}
\textcolor{black}{\begin{equation}
\lim _{-k {c}_{S}\tau \rightarrow \infty} v_{\mathbf{k}}(\tau)=\sqrt{\frac{2}{\pi k}}\left[\exp \left(-i\left\{k {c}_{S}\tau+\frac{\pi}{2}\left(\nu+\frac{1}{2}\right)\right\}\right)\right]
\end{equation}}

\textcolor{black}{and for the superhorizon region $-k {c}_{S}\tau \rightarrow 0~~(-k {c}_{S}\tau \ll 1)$ the expression for the mode function for the scalar modes reduces to:}
\textcolor{black}{\begin{equation}
\lim _{-k {c}_{S}\tau \rightarrow 0} v_{\mathbf{k}}(\tau)=\sqrt{\frac{2}{k}} \frac{i}{\pi} \Gamma\left(\nu\right)\left(\frac{-k {c}_{S}\tau}{2}\right)^{\frac{1}{2}-\nu}
\end{equation}}


\textcolor{black}{At the end, we combine these two asymptotic solutions, which capture the information from cosmologically relevant regions and the simplified asymptotic solution of the mode function is expressed as:}

\begin{widetext}
\bea\label{yu2zxzx}
 \displaystyle v_{\bf k} (\tau) &=&
   2^{\nu-\frac{3}{2}}\frac{1}{i\tau}\frac{1}{\sqrt{2}\left(k{c}_{s}\right)^{\frac{3}{2}}}
	(-k
					{c}_{s}\tau)^{\frac{3}{2}-\nu}\left|\frac{\Gamma(\nu)}{\Gamma\left(\frac{3}{2}\right)}\right| 
	                    \left(1+ik{c}_{s}\tau\right)
\exp\left(- i\left\{k{c}_{s}\tau+\frac{\pi}{2}\left(\nu+\frac{1}{2}\right)\right\}\right). 
\eea 
         \end{widetext}
\section{\textcolor{Sepia}{\textbf{ \large Quantized Hamiltonian and two-mode squeezed states\label{v3a2}}}}
Using the previously derived scalar mode function from the EFT setup one can easily compute the classical Hamiltonian function which can be promoted as an operator in the Heisenberg picture.  Consequently, the quantized version of the Hamiltonian can be written as \cite{Choudhury:2020hil,Bhargava:2020fhl}:
\bea\label{ham}
&&{H}(\tau)=\frac{1}{2}\int d^3{{{\bf k}}}\biggl[\Omega_{{\bf k}}(\tau)\left(\hat{c}^{\dagger}_{{{\bf k}}}\hat{c}_{{\bf k}}+\hat{c}^{\dagger}_{-{{\bf k}}}\hat{c}_{-{{\bf k}}}+1\right)\nonumber\\
&&~~~+i\lambda_{{\bf k}}(\tau)\Biggl(e^{-2i\phi_{{\bf k}}(\tau)}\hat{c}_{{\bf k}}\hat{c}_{-{{\bf k}}}-e^{2i\phi_{{\bf k}}(\tau)}\hat{c}^{\dagger}_{{\bf k}}\hat{c}^{\dagger}_{-{{\bf k}}}\Biggr)\biggr],
\eea
where,  $(\hat{c}^{\dagger}_{{{\bf k}}},\hat{c}^{\dagger}_{-{{\bf k}}})$ and $(\hat{c}_{{\bf k}},\hat{c}_{-{{\bf k}}})$ are the creation and annihilation operators for the two momentum modes having momenta ${\bf k}$ and $-{\bf k}$.  In this quantized Hamiltonian first three terms are the outcome of the quantization from the free part of the EFT and the rest of the contributions are appearing from the quantization of the interacting part of the EFT action.  Most importantly,  this interaction part in the present context is parametrized in terms of two time-evolving parameters,  these are the squeezing amplitude and the squeezing angle.   Particularly  the quantized part of the interaction Hamiltonian is written in terms of two-mode squeezed states contributions after performing the quantization.  Here we introduce two new symbols $\Omega_{\bf k}(\tau)$ and $\lambda_{\bf k}(\tau)$ which are defined as:
\bea
\Omega_{{\bf k}}(\tau):&=&2^{2(\nu-2)}~k~\left|\frac{\Gamma(\nu)}{\Gamma\left(\frac{3}{2}\right)}\right|^{2}\left(1+c^{2}_s\right), \\ 
\lambda_{{\bf k}}(\tau):&=&\sqrt{\left\{\frac{\Omega_{{{\bf k}}}(\tau)}{2}\biggl(\frac{1-c^{2}_s}{1+c^{2}_s}\biggr)\right\}^{2}+\biggl(\frac{z'(\tau)}{z(\tau)}\biggr)^{2}}.  
\eea
and one can write:
\bea \frac{z'(\tau)}{z(\tau)}&=&\frac{a'(\tau)}{a(\tau)}+\frac{1}{2} \frac{\epsilon'(\tau)}{\epsilon(\tau)}-\frac{c'_s}{c_s}.\eea
As the quantized Hamiltonian is in our hands,  we can now determine the expression for the unitary time evolution operator from these Goldstone modes in the COSMOEFT framework.

\textcolor{black}{The structure of the Hamiltonian in Eq \eqref{ham} is similar to the Hamiltonian given in \cite{Albrecht:1992kf} except the differences are in the expression for $\Omega_{{\bf k}}(\tau)$, $\lambda_{{\bf k}}(\tau)$ and $\frac{z'(\tau)}{z(\tau)}$. Now here we have used a similar factorization for the unitary time evolution operator. One can notice that the term in Hamiltonian \eqref{ham} with $\Omega_{{\bf k}}(\tau)$ represents the free part of the Hamiltonian and the interacting piece is the squeezing part with coupling strength $\lambda_{{\bf k}}(\tau)$. The evolution operator can be represented as a product of squeezing and a rotation operator.}

\bea
U(\tau,\tau_0) = \hat{\mathcal{S}}(r_{{\bf k}}(\tau,\tau_0) ,\phi_{{\bf k}}(\tau))\hat{\mathcal{R}}(\theta_{{\bf k}}(\tau)),
\eea
where the above factorization is only possible if we specifically use the single field two-mode squeezed state formalism in the present context.  However, there exists other possibilities  using which one can  determine the structure of the unitary time evolution operator.  For more details on this issue see refs.  \cite{Bhargava:2020fhl,Choudhury:2020hil,Adhikari:2021pvv,Adhikari:2021ked}.  
Here we note that $\hat{\mathcal{R}}(\theta_{{\bf k}}(\tau))$ is the rotation operator which can be expressed as:
\bea
\label{eq:rotation}
\hat{\mathcal{R}}(\theta_{{\bf k}}(\tau) ) = \exp\left( -i\theta_{{\bf k}}(\tau)\big( \hat{c}_{{\bf k}}\hat{c}_{{\bf k}}^{\dagger} + \hat{c}_{-{{\bf k}}}^{\dagger}\hat{c}_{-{{\bf k}}} \big) \right),
\eea
and $ \hat{\mathcal{S}}(r_{{\bf k}}(\tau,\tau_0) ,\phi_{{\bf k}}(\tau))$ is the squeezing operator,  which is defined as:
\begin{eqnarray}
\label{eq:Squeezed}
\nonumber
\hat{\mathcal{S}}(r_{{\bf k}}(\tau) ,\phi_{{\bf k}}(\tau))&=& \exp\bigg(r_{{\bf k}}(\tau) \big[ e^{-i \phi_{{\bf k}}(\tau)}\hat{c}_{{\bf k}}\hat{c}_{-{{\bf k}}}  \\ &&~~~~~~~~~~~~~~~~~~- e^{i \phi_{{\bf k}}(\tau)}\hat{c}_{-{{\bf k}}}^{\dagger}\hat{c}_{{\bf k}}^{\dagger} \big]\bigg).~~~~~
\end{eqnarray}
Here the squeezing amplitude and squeezing angle are characterized by 
$r_{{\bf k}}(\tau)$ and $\phi_{{\bf k}}(\tau)$,  which evolve with respect to the evolutionary time scale. It should be mentioned here that the rotation operator   $\hat{\mathcal{R}}(\theta_{{\bf k}}(\tau))$ leads to an unwanted phase factor which is not at all important for the computation of the quantum circuit complexity and the quantum entanglement entropy.  For this reason, we  ignore such irrelevant contribution in the present computation. 
 
 At this point, we provide  a few more information required to obtain the analytical expressions of the quantum circuit complexity and the quantum entanglement entropy functions  from various relevant measures. One of the prime input information is about the construction of the initial quantum state.  In this description, the ground state is characterized by the free part of the quantized Hamiltonian and for the computation of the quantum circuit complexity, we fix this as the initial quantum mechanical state.  Using this initial quantum state one can  construct the two-mode squeezed state  which is considered as the final target state for the computation.  
 
 With this specific choice of the  initial and the target state, one can now compute the expressions for the quantum circuit complexity and quantum entanglement entropy in terms of the squeezing amplitude $r_{{\bf k}}(\tau)$ and squeezing angle $\phi_{{\bf k}}(\tau)$.
Now, our next job is to determine the time evolution of the squeezing amplitude $r_{{\bf k}}(\tau)$ and squeezing angle $\phi_{{\bf k}}(\tau)$ which will  determine the quantum circuit complexity and the quantum entanglement entropy in the context of COSMOEFT.  This can be done by knowing the time evolutionary behaviour of the time-dependent squeezed operator $\hat{\mathcal{S}}(r_{{\bf k}}(\tau) ,\phi_{{\bf k}}(\tau))$.  Technically this is described by the Schr\"{o}dinger equation within the present framework.
which gives rise to the following two coupled differential equations of the squeezing amplitude $r_{{\bf k}}(\tau)$ and the squeezing angle $\phi_{{\bf k}}(\tau)$:
\bea
\label{eq:evolution1a}
&&\frac{dr_{{\bf k}}(\tau)}{d\tau} =\lambda_{{\bf k}}(\tau)\cos\left(2\left(\tilde{\phi}_{{\bf k}}(\tau)-\phi_{{\bf k}}(\tau)\right)\right), \\
\label{eq:evolution1b}
&&\frac{d\phi_{\bf k}(\tau)}{d\tau} =\Omega_{{{\bf k}}}(\tau)\nonumber\\
&&+\lambda_{{{\bf k}}}(\tau)\coth\left(2r_{{{\bf k}}}(\tau)\right)\sin\left(2\left(\tilde{\phi}_{{\bf k}}(\tau)-\phi_{{\bf k}}(\tau)\right)\right),~~~~
\eea
where $\tilde{\phi}_{{\bf k}}(\tau)$ is given by the following expressions:
\bea \tilde{\phi}_{{\bf k}}(\tau)&=&-\frac{\pi}{2}+\frac{1}{2}~{\rm tan}^{-1}\left[\frac{\Omega_{{{\bf k}}}(\tau)}{2}\biggl(\frac{z'(\tau)}{z(\tau)}\biggr)\biggl(\frac{1-c^{2}_s}{1+c^{2}_s}\biggr)\right].~~~~~~\eea
Expressions for $\lambda_{{\bf k}}(\tau)$ and $\Omega_{{{\bf k}}}(\tau)$ are already mentioned before in this section.  Here we fix the boundary condition at the late time scale $\tau=\tau_0$ which fixes the squeezing amplitude and the squeezing angle at $r_{\bf k}(\tau_0)=1$ and $\phi_{\bf k}(\tau_0)=1$. With this boundary condition, one can obtain the time evolution of squeezing amplitude $r_{{\bf k}}(\tau)$ and the squeezing angle $\phi_{{\bf k}}(\tau)$ at any scale.   

 However, solving these above-mentioned equations becomes much more simpler when they are expressed in terms of the scale factor rather than the conformal time.   For this reason, we use:
\bea \label{schange}  \frac{d}{d\tau}=a^{'}\frac{d}{da}.\eea
It is easy to recast these coupled differential equations in terms of the scale factor as:
\begin{widetext}
\bea
\label{eq:evolution2a}
\frac{dr_{{\bf k}}(a)}{da} &=&\frac{\lambda_{{\bf k}}(a)}{a^{'}}\cos\left(2\left(\tilde{\phi}_{{\bf k}}(a)-\phi_{{\bf k}}(a)\right)\right), \\
\label{eq:evolution2b}
\frac{d\phi_{\bf k}(a)}{da} &=&\frac{\Omega_{{{\bf k}}}(a)}{a^{'}}+\frac{\lambda_{{{\bf k}}}(a)}{a^{'}}\coth\left(2r_{{{\bf k}}}(a)\right)\sin\left(2\left(\tilde{\phi}_{{\bf k}}(a)-\phi_{{\bf k}}(a)\right)\right),
\eea
\end{widetext}
The boundary condition is fixed  at the late time scale $\tau=\tau_0$ where $a(\tau_0)=a_0=1$ which fixes the squeezing amplitude and the squeezing angle at $r_{\bf k}(a_0=1)=1$ and $\phi_{\bf k}(a_0=1)=1$. Thus, considering this boundary conditions, one can  obtain the evolved squeezing amplitude $r_{{\bf k}}(a)$ and squeezing angle $\phi_{{\bf k}}(a)$ at any value of the scale factor. 

\textcolor{black}{It is instructive to mention the reason behind  such a particular choice of the boundary condition at late time scale $\tau=\tau_0$ which represents the present day universe at $t=13.8$ billion years. The  scale factor and the values of the squeezing amplitude and phase, that are chosen at $\tau=\tau_0$, are in accordance with the observed data from the supernova explosion and the Hubble space telescope observations. We have not fixed the boundary condition at early time scale $\tau_{in}=-\infty$ or $t=0$ which represents the universe at the time of the big bang due to the uncertainty of the type of the universe at that time (it could be inflationary, cycling, bouncing or of some other kind).}

\section{\textcolor{Sepia}{\textbf{ \large Notes on Quantum Circuit Complexity}}}
\label{sec:CircuitComplexity}
In this section,  we give an overview of the topic of quantum circuit complexity which is becoming a  popular area of research in theoretical physics.  The prime objective of quantum information processing science is to build the quantum circuits for search algorithms or Shor's factoring \cite{shor1, Shor:1994jg, nielsen_chuang_2010},  which can be able to perform various unitary operations.  Similarly,  in the framework of computer science, the same job is performed by complexity \cite{cscomplexity1, cscomplexity2},  which is technically defined as the minimum number of gates required to perform a computation.  This is completely a classical definition till now.  In the quantum mechanical context, this is interpreted as the minimum number of quantum gates required to perform a unitary operation  \cite{Barenco:1995na} and it  provides the estimation of the quantum circuit complexity \cite{Aaronson:2016vto, DiVincenzo_2000}.  However, the task of implementing these unitary operations in logic gates is not  simple and a deeper understanding is required to perform such operations in quantum circuits.

There are many useful approaches people have been following for the last few years to estimate the quantum circuit complexity.  One of the path-breaking approaches is to geometrically quantify  the quantum circuit complexity measure \cite{nielsen2005geometric,Nielsen2,nielsen2006geometric}.  The key idea is to find out the optimal quantum circuit which mimics the role of solving problems of determining geodesics in the context of the Riemannian geometry.  In this approach, the Riemannian metric is defined in the space of $n$ number of qubits.  The corresponding distance measure ${\cal D}(I,U)$,  which is measuring the distance between the identity $I$ and the target unitary  quantum operation $U$, is exactly interpreted as the quantum circuit complexity measure in the present context of the discussion.  In this terminology,  minimizing this distance measure ${\cal D}(I,U)$ is physically interpreted as computing the geodesic length in the Riemannian geometry spanned by $n$ number of qubits. With the systematic rigorous development of the problem, it is now expected that this distance  ${\cal D}(I,U)$ can be regarded as a good measure of the quantum circuit complexity in this context.  For this underlying reason, one can reliably use various computational methods based on the Riemannian geometry to study the unexplored features of the quantum circuit complexity.  The methods which are frequently used in this computation, are the Levi-Civita connection, the geodesic, and the curvature, .  

\textcolor{black}{To demonstrate this concept, let us first consider a unitary operator $U$ which transforms a reference state $\ket{\Psi_{R}}$ to the desired target state $\ket{\Psi_{T}}$ via the following operation:} 
\bea
\label{eq:unitary}
\ket{\Psi_{T}} = U \ket{\Psi_{R}}
\eea
\textcolor{black}{This unitary operation plays a pivotal role in the context of quantum computation where it is interpreted as an ordered operation of elementary or universal unitary gates $U_i \forall i=1,2,\cdots,d$ represented by:}
\bea U = U_1U_2\cdots U_d,\eea
where $d$ is the depth of the quantum circuit. \textcolor{black}{The Eq (\ref{31equ}) represents a distance measure between the desired state and the achieved state which is obtained through the unitary transformation applied to the initial reference state. It tells us how closely the unitary transformation $U$ in Eq (\ref{eq:unitary}) has transformed the initial reference state $\ket{\Psi_{R}}$ to the desired target state $\ket{\Psi_{T}}$. There is also a parameter $\Delta$ which is called tolerance \cite{Jefferson:2017sdb,Bhattacharyya:2018bbv} and the transformation $U$ is successful if the two states are sufficiently close to each other such that they satisfy the given relation:}
\bea\label{31equ}
||\ket{\Psi_{T}} -  U \ket{\Psi_{R}}||^2 \leq \Delta
\eea 
The inclusion of such tolerance is perfectly consistent with realistic applications where it is extremely difficult to perform an exact unitary operation which can be interpreted in terms of the ordered discrete set of unitary $U_i \forall i=1,2,\cdots,d$ quantum operations.  It is important to mention here that,  in the literature, there exists an infinite number of possibilities using which one can be able to get the required target quantum state $\ket{\Psi_{T}}$ from the reference state $\ket{\Psi_{R}}$.  In all of these formalisms, depth of the optimal circuit is measured by the quantum circuit complexity.

In the context of the computation of quantum circuit complexity measure using the geometric technique in refs.  \cite{nielsen2005geometric,Nielsen2,nielsen2006geometric}, the authors have used the principles of Hamiltonian control.  This approach is commonly known as Nielsen's geometric approach towards the circuit complexity.  This idea is further used in various contexts of quantum mechanical systems to explore the unknown features of the complexity measures.  In this geometric technique, people don't measure the number of discrete gates to perform the previously mentioned unitary operations.  In Nielsen's geometric approach the unitary operator $U$ is path ordered and can be expressed in terms of the time-dependent Hamiltonian $H(\tau)$ as:
\bea
\label{eq:control}
U &=& \overleftarrow{\mathcal{P}} \text{ exp}\left[ -i\int_{0}^{1}d\tau H(\tau) \right], \eea
where the Hamiltonian operator can be written in terms of hermitian operators $\mathcal{O}_I$ and control functions $Y^I(\tau)$ as:
\be H(\tau) = \sum_{I} Y^I(\tau)\mathcal{O}_I
\ee
Here it is important to note that,  the path ordering $\mathcal{P}$ operation is the geometric version of the time ordering operation.  It physically signifies that the quantum circuit is made up of non-commuting operators which are acting from the right side hand to the left hand side.  In this description, $Y^I(\tau)$ is interpreted as a particular gate which is added at a time $s$.

These possible paths finally construct the form of the unitary operations which are expressed as:
\bea
\label{eq:trajectory}
U = \overleftarrow{\mathcal{P}} \text{ exp}\left[ -i\int_{0}^{\tau}d\tau' H(\tau') \right]. 
\eea
Here the boundary conditions  are fixed at:
\begin{enumerate}
\item At time scale $\tau = 0$ the unitary operator is fixed at $U(\tau = 0)= \mathrm{1}$.
\item  At time scale $\tau = 1$ the unitary operator is fixed at $U(\tau = 1)= U$.
\end{enumerate}
Also, the hermitian operators $\mathcal{O}_I$ and the control functions $Y^I(\tau)$ satisfy the following equation:
{\textcolor{black}{
\bea
\label{eq:v}
\sum_{I} Y^I(\tau)\mathcal{O}_I = \partial _{\tau} U(\tau)U^{-1}(\tau),
\eea
}}
which is representing a time-dependent Schrödinger equation describing path-ordered unitary time evolution. 

Instead of finding out the infinite possible ways to construct the unitary operations, here we will restrict ourselves only to the procedures which give the optimal answer.  For this reason, we consider a cost function ${\cal F}(U, \vec{Y}(\tau))$ which is a local functional.  For each possible optimal path, the cost is given by the following expression:
{\textcolor{black}{
\bea
\label{eq:geod}
\mathcal{D}(U(\tau)) = \int_{0}^{\tau}d\tau'~{\cal  F}(U(\tau'),\partial_{\tau}{U}(\tau'))
\eea
}}
Nielsen explicitly elaborated in his computation that,  the variational method of minimizing this cost functional is nothing but determining the optimized quantum circuit in the present context. 

Here this cost function ${\cal F}(U, \vec{Y}(\tau))$ should satisfy the following four properties :
\begin{enumerate}
\item  \underline{\textcolor{Sepia}{\bf Continuity}}:\\ 
 The cost functional should be continuous which is described by the symbol ${\cal F} \in \mathcal{C}^0$.  This is quite reasonable from the physics perspective.
\item  \underline{\textcolor{Sepia}{\bf Positivity}}:\\
The technical definition of the cost function ${\cal F}$ demands positivity which can be mathematically stated as:
\bea
{\cal F}(U,v) \geq 0,
\eea
 where equality holds if $v = 0$ which further implies that the reference and target quantum states are identical.
\item   \underline{\textcolor{Sepia}{\bf Positive homogeneity}}:\\
For any positive real number $\lambda$ and any arbitrary vector $v$,  the following condition has to be satisfied:
{\textcolor{black}{ \bea {\cal F} (U,\lambda v) = \lambda~ {\cal F} (U,v). \eea }}
 This condition in technical language is identified as the positive homogeneity criteria.
\item  \underline{\textcolor{Sepia}{\bf Triangle Inequality}}:\\
Last but not the least,  the cost functional ${\cal F} $ satisfies the triangle inequality in the present context:
\bea
{\cal F} (U,v+u) \leq {\cal F} (U,v) + {\cal F} (U,u)~~\forall~(u,v) 
\eea
The equality holds if the tangent vectors $v$ and $u$ are along the same ray which is coming out from the origin. 
\end{enumerate}
If the continuity condition is extended to ${\cal F}  \in C^\infty$, then  the corresponding manifold is known as {\it Finsler Manifold}.  In this connection, it is important to note that Nielsen's geometric approach is based on computing the expression for geodesic in Finsler geometry and this calculation of the geodesic length is interpreted as the quantum circuit complexity.  

In literature, there exists different choices for the cost functions ${\cal F} (U,v)$  using which one can estimate quantum circuit complexity. These choices depend on how one defines the elementary gates to compute complexity measures from the corresponding setup.  Some of the possibilities are quoted below which will be useful for this purpose:
\bea
  \label{eq:cost}
{\cal F}_1(U,Y) &=& \sum_I |Y^I|, \\ {\cal F}_2(U,Y) &=& \sqrt{\sum_I |Y^I|^2 },  \\{\cal F}_m(U,Y) &=& \sum_I m_I|Y^I|, \\
  {\cal F}_n(U,Y) &=& \sqrt{\sum_I n_I|Y^I|^2 }. 
\eea
A few remarks regarding the  definitions of the cost functions are as follows:
\begin{itemize}
\item Here the cost function ${\cal F}_1$ is the linear one which measures the counting of individual gates in the quantum circuit.  

\item Here ${\cal F}_2$ is the quadratic cost functional which is physically interpreted as measuring the proper distance in the corresponding manifold under consideration. 

\item Here the cost functions ${\cal F}_{m}$ (for linear measure) and ${\cal F}_{n}$ (for quadratic measure) are defined in presence of additional penalty factors $m_I$ and $n_I$,  which are used to physically represent the fact that only certain direction is  preferred over all other possibilities in this context.  These possibilities are very useful to describe the physical situations where the non-local qubits are discarded from consideration and the elementary gates are only coupled to the local  qubits.  

\end{itemize}

One can further introduce a more general class of homogeneous and inhomogeneous family of cost functions,  which are given by the following expressions:
\bea
    \label{eq:k_cost}
    {\cal F}_\beta(U,Y) &=& \sum_I |Y^I|^\beta \\
    {\cal F}_{\frac{1}{\beta}}(U,Y) &=& \sum_I |Y^I|^{\frac{1}{\beta}}
\eea
where the parameter $\beta \geq 1$ physically signifies the degree of homogeneity.  Here it is important to point out that  $ {\cal F}_\beta$ was first introduced in the context of AdS/CFT to match the results obtained from the well-known "Complexity = Action" and "Complexity = Volume" conjectures,  which have been mentioned before.  

\section{\textcolor{Sepia}{\textbf{ \large Quantum Circuit Complexity from two mode squeezed states}}} \label{sec:twomode}
The framework of the two-mode squeezed quantum state is the simplest realization of the generalized multi-mode entangled quantum states using which one can study various quantum mechanical and information-theoretic features,  such as the entanglement entropy, the  quantum circuit complexity, the  quantum discord, the  logarithmic negativity and many more.  In this article, we  only focus on two aspects out of all of these possibilities,  which are the entanglement entropy and the quantum circuit complexity.  Here and in the next section, we discuss how, using such useful two-mode squeezed state formalism, one can quantify these two measures within the framework of COSMOEFT.   See refs.  \cite{gerry_knight_2004} for more details on two-mode squeezed states.

We have already defined the two-mode squeezed state operator in the previous half of the paper, to remind ourselves again let us write down this expression again in this section which will be useful for further discussion.  In this context, the two-mode squeezing operator is described as:
\begin{eqnarray}
\label{eq:Squeezed1}
\hat{\mathcal{S}}(r_{{\bf k}},\phi_{{\bf k}})&=& \exp\bigg(r_{{\bf k}} \big[ e^{-i \phi_{{\bf k}}}\hat{c}_{{\bf k}}\hat{c}_{-{{\bf k}}} - e^{i \phi_{{\bf k}}}\hat{c}_{-{{\bf k}}}^{\dagger}\hat{c}_{{\bf k}}^{\dagger} \big]\bigg).~~~~~
\end{eqnarray}
where $r_{{\bf k}}$ and $\phi_{{\bf k}}$ are the squeezing parameters which represent the corresponding amplitude and angle.  Here it is important to note that,  these parameters are lying within the window,  $0 \leq r_{{\bf k}} < \infty$ and $0 \leq \phi_{{\bf k}} \leq 2\pi$.  In our computation, the two-mode squeezed state is considered to be the target quantum state and this can be constructed by operating the squeezing operator $\hat{\mathcal{S}}(r_{{\bf k}},\phi_{{\bf k}})$ on two-mode initial vacuum state:
\begin{eqnarray}
\label{eq:Squeezed1}
  \ket{\Psi_{\text{sq}}}_{{\bf k},-{\bf k}}&=&\hat{\mathcal{S}}(r_{{\bf k}},\phi_{{\bf k}})  \ket{0_{{\bf k}},0_{-{\bf k}}}.~~~~~
\end{eqnarray}
Here it is important to note that,  the initial vacuum state structure is given by:
\bea \ket{0_{{\bf k}},0_{-{\bf k}}}:=\ket{0}_{{\bf k}}\ket{0}_{-{\bf k}}.\eea

In terms of the two-mode occupation number states,  one can further express  the target two-mode squeezed state which is given by the following expression:
\bea
\label{eq:stateX}
    \ket{\Psi_{\text{sq}}}_{{\bf k},-{\bf k}} = \frac{1}{\text{cosh}r_{\bf k}} \sum_{n= 0}^\infty (-1)^n e^{in\phi_{{\bf k}}}(\text{tanh}r_{\bf k})^n \ket{n_{\bf k},n_{-{\bf k}}}.~~~~~~~
\eea
Using this target two-mode squeezed quantum states we  can now  compute the expressions for the quantum circuit complexity measure and the entanglement entropy.  Using these initial reference and target quantum states one needs to construct a wave function which is Gaussian in nature.

To serve this purpose let us first write down the position and the momentum variables by the following expressions:
\bea
 \hat{q}_{{{\bf k}}} &=& \frac{1}{\sqrt{2\Omega_{{\bf k}}}} \big (\hat{c}^\dagger_{{{\bf k}}} + \hat{c}_{{{\bf k}}} \big)\\
 \hat{p}_{{{\bf k}}} &=&i \sqrt{\frac{\Omega_{{\bf k}}}{2} }\big (\hat{c}^\dagger_{{{\bf k}}} - \hat{c}_{{{\bf k}}} \big)
\eea
where they satisfy the following commutation condition:
\bea [\hat{q}_{{{\bf k}}},\hat{p}_{{{\bf k'}}}] = i\delta^3({{\bf k}}-{{\bf k'}}). \eea
Consequently,  in the position space, the reference and target states can be expressed as:
\bea
\label{eq:ref}
\nonumber
   \Psi_R(q_{{\bf k}},q_{-{\bf k}}) &=& \langle q_{{\bf k}}, q_{-{\bf k}}  \ket{0_{{\bf k}},0_{-{\bf k}}}\\
    &=& \left( \frac{\Omega_{\bf k}}{\pi} \right)^{\frac{1}{4}}\exp\left( -\frac{\Omega_{\bf k}}{2}\big(q_{{\bf k}}^2+ q_{-{\bf k}}^2 \big)\right),\\
    \label{eq:target}
    \Psi_T(q_{{\bf k}},q_{-{\bf k}})&=&\langle q_{{\bf k}}, q_{-{\bf k}}\ket{\Psi_{\text{sq}}}_{{\bf k},-{\bf k}}  \\
    &=& \frac{ e^{A\big(q_{{\bf k}}^2+ q_{-{\bf k}}^2 \big) - Bq_{{\bf k}}q_{-{\bf k}}}}{\text{cosh}r_{\bf k}\sqrt{\pi}\sqrt{1-e^{-4i\phi_{\bf k}}\text{tanh}^2r_{\bf k}}}      
\eea
where the coefficients $A$ and $B$  in terms of the squeezing amplitude $r_{\bf k}$ and the squeezing angle $\phi_{\bf k}$ are given by the following expressions:
\bea
    A &=& \frac{\Omega_{\bf k}}{2} \bigg(\frac{e^{-4i\phi_{\bf k}}\text{tanh}^2r_{\bf k}+1}{e^{-4i\phi_{\bf k}}\text{tanh}^2r_{\bf k}-1}\bigg), \\
    B &=& 2\Omega_{\bf k} \bigg(\frac{e^{-2i\phi_{\bf k}} \text{tanh}r_{\bf k}}{e^{-4i\phi_{\bf k}}\text{tanh}^2r_{\bf k}-1}\bigg).
\eea
We further introduce a few more relevant quantities which are written in terms of $A$,  $B$ and $\Omega_{\bf k}$ as:
\bea
\label{eq:o}
    \Sigma _{\bf k} &=& \left(-2A +B\right),\\
    \label{eq:p}
    \Sigma _{-{\bf k}} &=& -\left(2A +B\right),\\
     \label{eq:q}
    \omega _{\bf k}&=& \omega_{-{\bf k}}= \frac{\Omega_{\bf k}}{2}
\eea
In the following subsections and next section, we will discuss the covariance matrix method and the Nielsen's method to compute the quantum circuit complexity and the entanglement entropy using the two-mode squeezed state formalism developed in this section.   

\subsection{\textcolor{Sepia}{\textbf{ \large  Complexity via Covariance matrix method}}}
Here we discuss  the covariance matrix method  to quantify the quantum circuit complexity measure.  It will turn out that the final expression of circuit complexity computed in this method solely depends on the squeezing amplitude $r_{\bf k}$ and is completely independent of the squeezing angle $\phi_{\bf k}$.

As we have already seen  that in the present context, the initial reference and target states,  represented by Eq \eqref{eq:ref} and Eq \eqref{eq:target} are Gaussian wave functions,  one can easily use the covariance matrix method for the computation of the quantum circuit complexity measure.  The corresponding covariance matrices for the initial reference state and the target state are  given by :
\bea
    G_k^{s=0}&=&\displaystyle 
\begin{pmatrix}
\displaystyle \frac{1}{\Omega_{\bf k}} & 0 & 0 & 0 \\
0 & \displaystyle\Omega_{\bf k} & 0 & 0\\
0 & 0 &\displaystyle \frac{1}{\Omega_{\bf k}} & 0 \\
0 & 0 & 0 & \displaystyle\Omega_{\bf k}
\end{pmatrix}
\\
    G_k^{s=1} &=& 
\begin{pmatrix}
\displaystyle\frac{1}{\text{Re}(\Sigma _{{\bf k}})} & \displaystyle-\frac{\text{Im}(\Sigma _{{\bf k}})}{\text{Re}(\Sigma _{{\bf k}})} & 0 & 0 \\
\displaystyle-\frac{\text{Im}(\Sigma _{{\bf k}})}{\text{Re}(\Sigma _{{\bf k}})} & \displaystyle \frac{|\Sigma_{\vec{k}}|^2}{\text{Re}(\Sigma _{{\bf k}})} & 0 & 0 \\
0 & 0 & \displaystyle \frac{1}{\text{Re}(\Sigma _{-{\bf k}})} &\displaystyle -\frac{\text{Im}(\Sigma _{-{\bf k}})}{\text{Re}(\Sigma _{-{\bf k}})} \\
0 & 0 & \displaystyle-\frac{\text{Im}(\Sigma _{-{\bf k}})}{\text{Re}(-\Sigma _{{\bf k}})} & \displaystyle\frac{|\Sigma_{-{\bf k}}|^2}{\text{Re}(\Sigma _{-{\bf k}})}
\end{pmatrix}\nonumber\\
&&
\eea
where the symbols $\Sigma _{{\bf k}}$ and $\Sigma _{-{\bf k}}$ are  defined in Eq \eqref{eq:o} and Eq \eqref{eq:p}.  The good part of the computation using the covariance matrix method is that the above-mentioned covariance matrices for the initial reference and the target state capture the full information of the corresponding Gaussian wave functions.  Using this method, the computed quantum circuit complexity measures the logical quantum gates required to transform the reference covariance matrix $G_k^{s=0}$ to target covariance matrix $G_k^{s=1}$.  For  computational simplicity, we now factorize the $4 \times 4$ reference and target covariance matrices into two non-trivial $2 \times 2$ symmetric blocks.  For the initial reference state we have the following two symmetric blocks:
\bea
 G_{k=0}^{s=0} &=& 
\begin{pmatrix}
\displaystyle \frac{1}{\Omega_{\bf k}} & 0 \\
0 & \displaystyle\Omega_{\bf k}\\
\end{pmatrix}, 
\\
 G_{k=1}^{s=0} & =&  
\begin{pmatrix}
\displaystyle\frac{1}{\Omega_{\bf k}} & 0 \\
0 & \displaystyle\Omega_{\bf k} \\
\end{pmatrix}
\eea
On the other hand, for the target state we have:
\bea
G_{k=0}^{s=1} &=&
\begin{pmatrix}
\displaystyle\frac{1}{\text{Re}(\Sigma _{{\bf k}})} & \displaystyle-\frac{\text{Im}(\Sigma _{{\bf k}})}{\text{Re}(\Sigma _{{\bf k}})} \\
\displaystyle-\frac{\text{Im}(\Sigma _{{\bf k}})}{\text{Re}(\Sigma _{{\bf k}})} & \displaystyle\frac{|\Sigma_{\vec{k}}|^2}{\text{Re}(\Sigma _{{\bf k}})}
\end{pmatrix}, 
\\
    G_{k=1}^{s=1} &=&
\begin{pmatrix}
\displaystyle\frac{1}{\text{Re}(\Sigma _{-{\bf k}})} &\displaystyle -\frac{\text{Im}(\Sigma _{-{\bf k}})}{\text{Re}(\Sigma _{-{\bf k}})} \\
\displaystyle-\frac{\text{Im}(\Sigma _{-{\bf k}})}{\text{Re}(\Sigma _{-{\bf k}})} & \displaystyle\frac{|\Sigma_{-{\bf k}}|^2}{\text{Re}(\Sigma _{-{\bf k}})}
\end{pmatrix}, 
\eea
Here the total complexity is given by the sum of the individual complexities of the two blocks,  where we have to perform an additional sum over all $\Omega _{\bf k}$.

Now to make the calculation simpler we perform the following change of basis for each block:
\bea
   && \Tilde{G}^{s=1} = SG^{s=1}S^{-1}, \\
   &&\Tilde{G}^{s=0} = SG^{s=0}S^{-1}.
\eea
where $S$ is a similar matrix which is given by:
\bea
    S =
    \begin{pmatrix}
   \displaystyle \sqrt{\Omega_{\bf k}} & 0\\
    0 & \displaystyle\frac{1}{\sqrt{\Omega_{\bf k}} }
    \end{pmatrix}
\eea
 This further implies: 
\bea
\Tilde{G}^{s=0} &=& \mathbb{1}_{2\times 2},\\
    \Tilde{G}^{s=1} &=& 
    \begin{pmatrix}
\displaystyle\frac{\Omega_{\bf k}}{\text{Re}(\Sigma _{{\bf k}})} & \displaystyle-\frac{\text{Im}(\Sigma _{{\bf k}})}{\text{Re}(\Sigma _{{\bf k}})} \\
 & \\
\displaystyle-\frac{\text{Im}(\Sigma _{{\bf k}})}{\text{Re}(\Sigma _{{\bf k}})} & \displaystyle \frac{|\Sigma_{{\bf k}}|^2}{\Omega_k \text{Re}(\Sigma _{{\bf k}})}
\end{pmatrix}, 
\eea
In this method the corresponding similarity transformation of the wave functions can be represented as:
 \bea
 \Tilde{G}^s = \Tilde{U}(\tau)\Tilde{G}^{s=0}\Tilde{U}^{-1}(\tau).
 \eea
 Further, these similarity transformations are parametrized with the help of logical gates which satisfy the $SL(2,R)$ algebra and the corresponding structure of the similarity matrix is given by:
\begin{widetext}
\bea
    \Tilde{U}(\tau) = 
    \begin{pmatrix}
   \displaystyle \text{cos}\mu(\tau)\text{cosh}\rho(\tau)-\text{sin}\theta(\tau)\text{sinh}\rho(\tau)~~~~~ &~~~~~ \displaystyle-\text{sin}\mu(\tau)\text{cosh}\rho(\tau)+\text{cos}\theta(\tau)\text{sinh}\rho(\tau) \\
  \displaystyle  \text{sin}\mu(\tau)\text{cosh}\rho(\tau)+\text{cos}\theta(\tau)\text{sinh}\rho(\tau) & 
   \displaystyle \text{cos}\mu(\tau)\text{cosh}\rho(\tau)+\text{sin}\theta(\tau)\text{sinh}\rho(\tau)
    \end{pmatrix}
\eea
\end{widetext}
where,  $\mu, \rho, \theta$ are the coordinates which characterize  $SL(2,R)$ group.  Next, we set the following two boundary conditions:
\bea
        \Tilde{G}^{s=1} &=& \Tilde{U}(\tau = 1)\Tilde{G}^{s=0}\Tilde{U}(\tau =1)^{-1} \\
        \Tilde{G}^{s=0} &=& \Tilde{U}(\tau = 0)\Tilde{G}^{s=0}\Tilde{U}(\tau =0)^{-1}
\eea
These boundary conditions applied with the  $SL(2,R)$ parametrized similarity transformations further give:
\begin{widetext}
\bea
\label{eq:bc}
\text{cosh}2\rho(1)&=& \frac{\Omega_{\bf k}^2 + |\Sigma_{\bf k}|^2}{2\Omega_{\bf k} \text{Re}(\Sigma_{\bf k})},~~~
\text{tan}(\theta(1)+\mu(1))= \frac{\Omega_{\bf k}^2 - |\Sigma_{\bf k}|^2}{2\Omega_{\bf k} \text{Im}(\Sigma_{\bf k})},~~~
\rho(0)=0,~~~
\theta(0) + \mu(0) = c.
\eea
\end{widetext}
For the simplification purpose we choose the following constraints:
\begin{enumerate}
\item $\displaystyle \mu(\tau = 1) = \mu(\tau = 0)= 0$.
\item $\displaystyle \theta(\tau = 0) =\theta(\tau = 1) = c = \text{tan}^{-1}\left(\frac{\Omega_{\bf k}^2 - |\Sigma_{\bf k}|^2}{2\Omega_{\bf k} \text{Im}(\Sigma_{\bf k})} \right)$.
\end{enumerate}

Given these constraint conditions, the final form of the  $SL(2,R)$ metric for $\Tilde{U}$ is given by:
\bea
    ds^2 &=&\bigg( d\rho^2 + \text{cosh}2\rho~\text{cosh}^2\rho d\mu^2 +\text{cosh}2\rho~\text{sinh}^2\rho~ d\theta^2 \nonumber
   \\
   &&~~~~~~~~~ ~~~~~~~~~~~~-\text{sinh}^22\rho~d\mu~ d\theta\bigg).
\eea
Now from the boundary conditions stated in Eq \eqref{eq:bc},  we get the following expressions:
\bea
    \rho_{\bf k}(\tau = 1) &=& \frac{1}{2}\text{cosh}^{-1} \left[ \frac{\Omega_{\bf k}^2 + |\Sigma _{{\bf k}}|^2 }{2\Omega_{\bf k}\text{Re}(\Sigma _{{\bf k}})} \right],\\
 \rho_{-{\bf k}}(\tau = 1) &=& \frac{1}{2}\text{cosh}^{-1} \left[ \frac{\Omega_{-{\bf k}}^2 + |\Sigma _{-{\bf k}}|^2 }{2\Omega_{-{\bf k}}\text{Re}(\Sigma _{-{\bf k}})} \right].    
\eea
Here the simple geodesic is represented by a straight line which is given by,  $\rho(\tau)=\rho(\tau=1)\tau$.  
Finally,  to get the expression for the quantum circuit complexity measure using the covariance matrix method, we have to sum  the contributions coming from two momentum modes in squeezed state formalism.
Consequently,  considering the contributions from both linear and quadratic cost functions, the corresponding circuit complexity measures $C_1$ and $C_2$, can be computed as:
\begin{widetext}
\bea
        \label{eq:ccCovariance}
         C_1 &=&\rho_{\bf k}(\tau = 1) +\rho_{-{\bf k}}(\tau = 1)= \frac{1}{2}\Bigg\{\text{cosh}^{-1} \left(\frac{\Omega_{\bf k}^2 + |\Sigma _{{\bf k}}|^2 }{2\Omega_{\bf k}\text{Re}(\Sigma _{{\bf k}})} \right)  +\text{cosh}^{-1} \left( \frac{\Omega_{-{\bf k}}^2 + |\Sigma _{-{\bf k}}|^2 }{2\Omega_{-{\bf k}}\text{Re}(\Sigma _{-{\bf k}})} \right)\Bigg\} ,\\
 C_2&=& \sqrt{\rho_{\bf k}(\tau = 1)^2 +\rho_{-{\bf k}}(\tau = 1)^2} = \frac{1}{2} \sqrt{\left( \text{cosh}^{-1} \left[ \frac{\Omega_{\bf k}^2 + |\Sigma _{{\bf k}}|^2 }{2\Omega_{\bf k}\text{Re}(\Sigma _{{\bf k}})} \right]  \right)^2 +\left( \text{cosh}^{-1} \left[ \frac{\Omega_{-{\bf k}}^2 + |\Sigma _{-{\bf k}}|^2 }{2\Omega_{-{\bf k}}\text{Re}(\Sigma _{-{\bf k}})} \right]  \right)^2 }.
\eea
\end{widetext} 
Further substituting the explicit form of the functions $\Sigma _{{\bf k}}$  and $\Sigma _{-{\bf k}}$, the quantum circuit complexity measure reduces to the following simplified form:
\bea
    \label{eq:ccCovariance1}
    C_1&=& 4~r_{\bf k} \\
    C_2&=& 2\sqrt{2}~r_{\bf k}.
\eea
These results are completely independent of the squeezing angle $\phi_{\bf k}$ and only characterized by $r_{\bf k}$.  These two circuit complexities computed from linear and quadratic cost functions are related to each other by the following expression: 
\bea C_1 = \sqrt{2}~C_2.\eea  
For small squeezing amplitude $r_{\bf k} \rightarrow 0$, $C_1 , C_2 \approx 0$ which implies the reference and target states are the same. 

\subsection{\textcolor{Sepia}{\textbf{\large   Complexity via Nielsen's wave-function method}}}

Nielsen's approach deals with wave functions which finally give rise to expression for the quantum circuit complexity for the two-mode squeezed states,  which we will show,  depend on both the squeezing parameters,  the squeezing amplitude $r_{\bf k}$ and the squeezing angle $\phi_{\bf k}$.  However,  the general philosophical construction of computing the quantum circuit complexity measure of Nielsen's approach is similar to that of the covariance matrix method.  In this approach,  we will directly quantify the quantum circuit complexity measure using the initial reference and the target states using the two-mode squeezed state formalism.

Let us start with the exponent of the target two-mode squeezed states as stated in Eq \eqref{eq:target}, 
\textcolor{black}{ If we use Eq \eqref{eq:niTarget}, then in general the matrix ${\mathcal{W}}$ is not diagonal and is of the following form:
\bea
\label{eq:niTarget}
    \Psi_{\text{T}} = \mathcal{N}\text{exp}\left(-\frac{1}{2}q_a {\mathcal{W}}^{ab}q_b\right)
\eea
$$\mathcal{W}=\left[\begin{array}{cc}
-2 A+B & B \\
-B & -2 A
\end{array}\right].$$
Here the corresponding basis is $\{q_k, {q}_{-k}\}$.}
\textcolor{black}{But after diagonalizing $\mathcal{W}$ in the rotated new basis $\{\tilde{q}_k=\frac{1}{\sqrt{2}}\left(q_k+q_{-k}\right), \tilde{q}_{-k}=\frac{1}{\sqrt{2}}\left(q_k-q_{-k}\right)\}$ with eigen values $-2A\pm B$, we get the matrix $\tilde{\mathcal{W}}$ which is given in Eq\eqref{diago}. Now, the target wavefunction \eqref{eq:target}, in diagonal form can be expressed as: }
\bea
\label{eq:nielTarget}
    \Psi_{\text{T}} = \mathcal{N}\text{exp}\left(-\frac{1}{2}\Tilde{q_a} \Tilde{\mathcal{W}}^{ab}\Tilde{q_b}\right)
\eea
where, $\mathcal{N}$ is the normalization constant of the wave function, which is represented by the denominator in Eq (\ref{eq:target}) and,  the matrix $ \Tilde{\mathcal{W}}$ can be expressed as:
\bea \label{diago}
    \Tilde{\mathcal{W}} &=& 
    \begin{pmatrix}
     \displaystyle  -2A+B & 0\\
    0 &  \displaystyle  -2A -B
    \end{pmatrix} 
   =
    \begin{pmatrix}
    \displaystyle  \Sigma _{{\bf k}} & 0\\
    0 &  \displaystyle  \Sigma _{-{\bf k}} 
    \end{pmatrix}.
\eea
Also,  the initial reference state is represented by the following expression:
\bea
\label{eq:nielReference}
        \Psi_{\text{R}} &=& \mathcal{N}\text{exp}\left( -\frac{\Omega_{\bf k}}{2}\big(q_{{\bf k}}^2+ q_{-{\bf k}}^2 \big)\right).
\eea
In this context the reference and target wave functions of the Gaussian  structure can be generally written as:
\begin{equation}
    \Psi^\tau = \mathcal{N}\text{exp}\left( -\frac{1}{2}\left( v_a. \mathcal{K}_{ab}^\tau.v_b  \right) \right)
\end{equation}
where,  the basis vector $v = (\tilde{q}_{{\bf k}}, \tilde{q}_{-{\bf k}})$ and $\mathcal{K}^\tau$ is a diagonal matrix.  For the target state eq: \eqref{eq:nielTarget} and the reference state \eqref{eq:nielReference}, the structure of this matrix is given by:
\bea
    \mathcal{K}^{\tau=1}  &=& 
    \begin{pmatrix}
   \displaystyle   \Sigma _{{\bf k}} & 0\\
    0 &  \displaystyle  \Sigma _{-{\bf k}} 
    \end{pmatrix}= \mathcal{W},
\\
    \mathcal{K}^{\tau=0} &=&
    \begin{pmatrix}
    \displaystyle   \Omega_{\bf k} & 0\\
    0 &  \displaystyle  \Omega_{-{\bf k}}
    \end{pmatrix}.
\eea
Further,  the unitary transformation on the matrix $\mathcal{K}$ is given by:
\bea
    \mathcal{K}^\tau = \mathcal{U}(\tau).\mathcal{K}^{\tau = 0 }.\mathcal{U}^T(\tau),
\eea
and the corresponding  boundary conditions are given by the following expressions:
\bea
     \mathcal{K}^{\tau = 1} &=& \mathcal{U}(\tau = 1). \mathcal{K}^{\tau = 0 }.\mathcal{U}^T(\tau = 1), \\
      \mathcal{K}^{\tau = 0} &=& \mathcal{U}(\tau = 0). \mathcal{K}^{\tau = 0 }.\mathcal{U}^T(\tau = 0).
\eea
Here the unitary operator $\mathcal{U}$ can be parametrized as in terms of eqn \eqref{eq:control} provided at $\tau = 1$,  the target state is reached.  
In principle the elements of $\mathcal{K}^{\tau = 1}$ and $\mathcal{K}^{\tau = 0}$ matrices are complex and for this reason the logical gates are parametrized by $GL(2,\mathbb{C})$ group.  In this context,  the tangent vectors $Y^I$ as appearing in Eq \eqref{eq:v} are complex and can be expressed in terms of its generators $\mathcal{O}_I$ by the following expression:
\bea
    Y^I = \text{Tr}(\partial _{\tau} U(\tau)U^{-1}(\tau)\mathcal{O}^T_I)
\eea
where the generators satisfy the following constraint:
\bea
\mathcal{O}^T_I=\mathcal{O}^{-1}_I,\quad\text{Tr}(\mathcal{O}_I.\mathcal{O}_J^T) = \delta^{IJ},  \forall I,J = 0,1,2,3\quad
\eea
The corresponding $GL(2,\mathbb{C})$ metric is then given by the following expression:
\bea
ds^2 = G_{IJ}dY^IdY^{*J}. 
\eea
Now for our computation we fix $G_{IJ} = \delta^{IJ}$ which will further fix the penalty factors.  Further,  we fix the off-diagonal elements in $GL(2,\mathbb{C})$ matrix to zero as they increase the distance between the initial reference and the target states. Then in this diagonal basis,  the unitary operator $U(\tau)$ takes the following simplified form:
\bea
    U(\tau) = \text{exp}\left(\sum_{i \in ({\bf k},-{\bf k})}\mu^i(\tau)\mathcal{O}_i^{diag}\right),
\eea
where $\mu^i(\tau)$ represents complex parameters and $\mathcal{O}_i^{diag}$ are the corresponding generators in the diagonal basis having  $i$ number of identities at the position of the diagonal elements.  In the new diagonal basis, the corresponding metric takes the following simplified structure:
\bea
    ds^2 =\sum_{i \in ({\bf k},-{\bf k})} (d\mu^{i,\text{Re}})^2 +\sum_{i \in ({\bf k},-{\bf k})}  (d\mu^{i,\text{Im}})^2,
\eea
where the notation Re and Im signify the real and imaginary part of the complex parameters $\mu$.  Here the geodesic is a straight line,  which is given by:
\bea
    \mu^{i,p}(\tau) = \mu^{i,p}(\tau =1)+\mu^{i,p}(\tau =0)\forall i\in ({\bf k},-{\bf k})\quad
\eea
For $p = \text{Re and Im}$.  Using these boundary conditions we get:
\bea
        \mu^{i,\text{Re}}(\tau = 0) &=& \mu^{i,\text{Im}}(\tau = 0) = 0, \\
        \mu^{i,\text{Re}}(\tau = 1) &=& \frac{1}{2}\text{ln}\left| \frac{\Sigma _{i}}{\omega _{i}}\right|, \\
        \mu^{i,\text{Im}}(\tau = 1) &=& \frac{1}{2}\text{tan}^{-1}\frac{\text{Im}(\Sigma_{i})}{\text{Re}(\Sigma _{i})}.
\eea
where the index $i\in ({\bf k},-{\bf k})$ correspond to the two modes in the squeezed state formalism.

Finally,  the quantum circuit complexity measure for the linear and the quadratic cost functions,  $C_1(\Omega_{\bf k})$ and $C_2(\Omega_{\bf k})$ are given by:

\begin{widetext}
\bea
\nonumber
  C_1&=& \mu^{{\bf k},\text{Re}}(\tau = 1) + \mu^{-{\bf k},\text{Re}}(\tau = 1)+ \mu^{{\bf k},\text{Im}}(\tau = 1) + \mu^{-{\bf k},\text{Im}}(\tau = 1)\\
  &=&  \frac{1}{2} \Bigg\{ \text{ln}\left| \frac{\Sigma _{{\bf k}}}{\omega _{{\bf k}}}\right| +  \text{ln}\left| \frac{\Sigma _{-{\bf k}}}{\omega _{-{\bf k}}}\right| + \text{tan}^{-1}\frac{\text{Im}(\Sigma _{{\bf k}})}{\text{Re}(\Sigma _{{\bf k}})} + \text{tan}^{-1}\frac{\text{Im}(\Sigma_{-{\bf k}})}{\text{Re}(\Sigma _{-{\bf k}})}\Bigg\}, 
 \\
\nonumber
      C_2&=& \sqrt{(\mu^{{\bf k},\text{Re}}(\tau = 1))^2 + (\mu^{-{\bf k},\text{Re}}(\tau = 1))^2+ (\mu^{{\bf k},\text{Im}}(\tau = 1))^2 + (\mu^{-{\bf k},\text{Im}}(\tau = 1)})^2\\
      &=&  \frac{1}{2} \sqrt{ \Bigg(  \text{ln}\left| \frac{\Sigma _{{\bf k}}}{\omega _{{\bf k}})}\right|\Bigg)^2  + \Bigg(\text{ln}\left| \frac{\Sigma _{-{\bf k}}}{\omega _{-{\bf k}})}\right|  \Bigg)^2 
   + \Bigg(\text{tan}^{-1}\frac{\text{Im}(\Sigma _{{\bf k}})}{\text{Re}(\Sigma _{{\bf k}})}  \Bigg)^2 + \Bigg(\text{tan}^{-1}\frac{\text{Im}(\Sigma _{-{\bf k}})}{\text{Re}(\Sigma _{-{\bf k}})}  \Bigg)^2}
\eea
\end{widetext}
Further using Eq \eqref{eq:o},  \eqref{eq:p} and \eqref{eq:q}, the quantum circuit complexity measure can be expressed in terms of the squeezing parameters,  $r_{\bf k}$ and $\phi_{\bf k}$ as:
\begin{widetext}
\begin{eqnarray}
 C_1 &=&  \left|\text{ln}\left| \frac{1+\text{exp}(-2i\phi_{\bf k})\text{tanh}r_{\bf k}}{1-\text{exp}(-2i\phi_{\bf k})\text{tanh}r_{\bf k}}\right|\right| + \left| \text{tanh}^{-1}(\text{sin}(2\phi_{\bf k})\text{sinh}(2r_{\bf k})) \right|, \\
C_2 &=& \frac{1}{\sqrt{2}}\sqrt{\left(\text{ln}\left| \frac{1+\text{exp}(-2i\phi_{\bf k})\text{tanh}r_{\bf k}}{1-\text{exp}(-2i\phi_{\bf k})\text{tanh}r_{\bf k}}\right|\right)^2 + \left( \text{tanh}^{-1}(\text{sin}(2\phi_{\bf k})\text{sinh}(2r_{\bf k})) \right)^2} . 
\end{eqnarray}
\end{widetext}
As a special case, if we set $\phi_{\bf k} \rightarrow -\pi/2$ and consider large squeezing amplitude $r_{\bf k}$,  then we get back the same  relationship between the different measures of the circuit complexities obtained from the linear and the quadratic cost functions  of the covariance matrix method.  This is given by:
\begin{equation}
    C_1 \approx \sqrt{2}C_2 \approx \left| \text{ln} \left( \frac{1-\text{tanh}r_{\bf k}}{1+\text{tanh}r_{\bf k}} \right) \right| \approx r_{\bf k}
\end{equation}

\section{\textcolor{Sepia}{\textbf{ \large Entanglement entropy of two mode squeezed states}}}
\label{sec:ee}
In this section,  our prime objective is to derive the expression of the  entanglement entropy using the previously mentioned two-mode squeezed state formalism.  This will  help us to make a comparison between the quantum circuit complexity measure and the entanglement entropy,  which we will discuss in the next section in detail. Before going to  further computation, here it is important to note   that the two squeezed modes having momenta ${\bf k}$ and $-{\bf k}$ are (1) entangled with each other,  (2) strongly correlated with each other. and (3) they are highly symmetric with each other.  In this description, the target state which is a squeezed state $\ket{\psi_{\text{T}}}_{{\bf k},-{\bf k}}=\ket{\psi_{\text{sq}}}_{{\bf k},-{\bf k}}$ is the eigenket of the operator $\left(\hat{n}_{\bf k}-\hat{n}_{-{\bf k}}\right)$ having zero eigenvalue,  where $\hat{n}_{\bf k} = \hat{c}_{{\bf k}}^\dagger\hat{c}_{-{\bf k}}$ and $\hat{n}_{-{\bf k}}= \hat{c}^\dagger_{-{\bf k}}\hat{c}_{{\bf k}}$ are physically interpreted as the number operators of the two squeezed modes having momenta ${\bf k}$ and $-{\bf k}$ respectively.  In this setup, if we compute the average number density for photons corresponding to each momentum modes, then it  turns out to be the same and it is given by the following expression:
\bea
    \langle\hat{n_{\bf k}}\rangle= \langle\hat{n}_{-{\bf k}}\rangle = \text{sinh}^2r_{\bf k},
\eea
which will tend  trivially to zero when the squeezing amplitude parameter $r_{\bf k}$ is extremely small.  Here the $\langle\cdots\rangle$ operation is performed using the $\ket{\psi_{\text{sq}}}_{{\bf k},-{\bf k}}$ state.

The reduced density operators for the two squeezed modes having momenta ${\bf k}$ and $-{\bf k}$ are given by the following expressions:
\bea
   \hat{\rho}_{\bf k} &=& \sum_{n = 0}^\infty \frac{1}{(\text{cosh }r_{\bf k})^2}(\text{tanh }r_k)^{2n}|n_{\bf k}\rangle\langle n_{\bf k}|,\\
    \hat{\rho}_{-{\bf k}} &=& \sum_{n = 0}^\infty \frac{1}{(\text{cosh }r_{-{\bf k}})^2}(\text{tanh }r_{-{\bf k}})^{2n}|n_{-{\bf k}}\rangle\langle n_{-{\bf k}}|.
    \quad
\eea
Using these reduced density operators for two-mode squeezed states, one can now compute the entanglement entropy which is derived using the von-Neumann and the Rényi entropy measures.  

In general, the von-Neumann entropy measure can be expressed by the following expression in terms of the reduced density operator:
\bea
    S(\hat{\rho}) = -\text{Tr}\left(\hat{\rho}~\text{ln}\hat{\rho}\right)
\eea
For the pure state, the  von-Neumann entropy is zero and for the mixed state, it is non-zero,  in which case it is meaningful to compute this expression for a given quantum mechanical system.  

In the basis where  the reduced density operator is diagonal, the entanglement entropy can be computed as:
\bea
    S(\hat{\rho})& =& -\sum_j \rho_{jj}~ \text{ln}\rho_{jj}.
\eea
In this diagonal basis, the entanglement entropy of two-mode squeezed states using the von-Neumann entropy measure can be written as:
\bea S(\hat{\rho}_{\bf k})&=&-\text{Tr}\left(\hat{\rho_{\bf k}}~\text{ln}\hat{\rho_{\bf k}}\right),\\
S(\hat{\rho}_{-{\bf k}})&=&-\text{Tr}\left(\hat{\rho}_{-{\bf k}}~\text{ln}\hat{\rho}_{-{\bf k}}\right).\eea
After a simplification, we found that the von-Neumann entropy measure from both the squeezed modes turns out to be the same and is given by the following expression:
\bea
\label{eq:opy}
S(\hat{\rho}_{\bf k}) &=& S(\hat{\rho}_{-{\bf k}})\nonumber\\
    &=& - \sum_{n = 0}^\infty \frac{\text{tanh}^{2n}r_{\bf k}}{\text{cosh}^2r_{\bf k}} \text{ln}\bigg(\frac{\text{tanh}^{2n}r_{\bf k}}{\text{cosh}^2r_{\bf k}}\bigg) \nonumber\\
    &=&\left\{\text{ln(cosh}^2r_{\bf k}) \text{cosh}^2r_{\bf k} - \text{ln(sinh}^2r_{\bf k}) \text{sinh}^2r_{\bf k}\right\}.\quad\quad\quad
\eea
This result shows that the von-Neumann entropy measure is completely independent of the squeezing angle $\phi_{\bf k}$ and it is solely determined by the squeezing amplitude $r_{\bf k}$. 

Further one can generalize the von-Neumann entropy to get the expression for the  Rényi entropy from the reduced density operator:
\bea
   \label{eq:renyi}
    S_\alpha(\hat{\rho})  &=& \frac{1}{1-\alpha} \ln \left({\rm Tr}~\hat{\rho}^{\alpha}\right)\eea
    which can be written for the two-mode squeezed states as:
    \bea S_\alpha(\hat{\rho}_{\bf k})&=& \frac{1}{1-\alpha} \ln \left({\rm Tr}~\hat{\rho}^{\alpha}_{{\bf k}}\right),\\
S_\alpha(\hat{\rho}_{-{\bf k}})&=& \frac{1}{1-\alpha} \ln \left({\rm Tr}~\hat{\rho}^{\alpha}_{-{\bf k}}\right).\eea
After a simplification,  we also found that the Rényi entropy measure from both the squeezed modes turns out to be the same and it is given by the following expression:
    \bea S_\alpha(\hat{\rho}_{\bf k})&=&S_\alpha(\hat{\rho}_{-{\bf k}})\nonumber\\
    &=&\frac{2\alpha \text{ ln coshr}_{\bf k} + \text{ln}(1- \text{tanh}^{2\alpha}r_{\bf k})}{\alpha -1}.   
\eea
where the Rényi parameter $\alpha \geq 0$ and $\alpha\rightarrow 1$ limit gives back the expression for the von-Neumann entropy measure.  Again, we can observe that the Rényi entropy measure is solely described by the squeezing amplitude parameter $r_{\bf k}$ just like the von-Neumann entropy measure. 

This discussion can be further extended to explicitly compute the expression for the effective temperature of the photon. The average photon number of the thermal field is given by: 
\bea
    \langle\hat{n}_i\rangle = \frac{1}{\left(\text{exp}(\omega_i/T)-1\right)}=\text{sinh}^2r_{\bf k}~~~\forall~i\in ({\bf k},-{\bf k}).\quad\quad\quad
\eea 
where we set $\hbar=1$ and Boltzmann constant $k_B=1$ in the natural unit system.   
Then, one can compute the effective temperature as:
\bea
    T &=& \omega_i~ \text{ln}\left( \frac{\langle\hat{n}_i\rangle}{ \langle\hat{n}_i\rangle +1 } \right) = \frac{\omega_i}{2\ln({\rm coth}~r_{\bf k})}~~~\forall~i\in ({\bf k},-{\bf k}).\nonumber\\
    &&
\eea
Here the frequency parameter $\omega_i\forall~i\in ({\bf k},-{\bf k})$ is represented by eqn. (\ref{eq:q}).
\section{\textcolor{Sepia}{\textbf{ \large Quantum Circuit Complexity Vs Entanglement\label{comp}}}}
In this section, our prime objective is to make a comparison between the quantum circuit complexity measure and the entanglement entropy measure,  particularly when both of them are described in the two-mode squeezed state formalism.  Recently in ref.  \cite{Eisert:2021mjg} the authors showed that there exists some underlying relationship between the entangling power and the quantum circuit complexity measure. 

Naively, it seems that the quantum circuit complexity measure and the entanglement entropy are of  different physical origins.  But in ref.  \cite{Eisert:2021mjg}, the authors showed that it is possible to use the entanglement entropy measure to bound the circuit complexity for (1) small value of circuit cost and (2) small value of entanglement.  Additionally,  in ref.  \cite{Eisert:2021mjg} there are a few interesting arguments presented which will help us to understand such connections for the system under consideration. Here we list two of them  point-wise:
\begin{itemize}
\item  Quantum gates which are close to the identity gates perform little entanglement from the entangled states. 

\item If the entanglement entropy measure grows linearly with respect to time then the quantum circuit complexity measure also grows linearly.   Particularly this linear growth of quantum entanglement is a generic feature of various quenched many-body quantum systems.
\end{itemize}
The interesting point is that the circuit complexity measure and the entanglement entropy computed from two-mode squeezed states in this work, perfectly goes well with the results obtained in ref.  \cite{Eisert:2021mjg}.  From the discussion in the previous sections, it is quite clear that the expressions for the quantum circuit complexity measure and the entanglement entropy measure from the covariance matrix method are solely determined by the squeezing amplitude parameter $r_{\bf k}$ and are completely independent of the squeezing angle $\phi_{\bf k}$.  On the other hand,  the same results obtained from Nielsen's wave function method are described by both squeezing amplitude parameter $r_{\bf k}$ and squeezing angle $\phi_{\bf k}$.  For this reason, one can use very easily the results obtained from the covariance matrix method to show the underlying connection between circuit complexity measure and the entanglement entropy in the present context of the discussion.

Using the covariance matrix method the circuit complexity measure from linear and quadratic cost functions can be expressed as:
\bea
    \label{eq:ccCovariance1c}
    C_1&=& 4~r_{\bf k} \\
    C_2&=& 2\sqrt{2}~r_{\bf k}.
\eea
\textcolor{black}{The expression for the Rényi entropy measure in the limit where $S_{\infty}(r_{\bf k} \rightarrow 0)$ is given by:}
\bea
 &&S_{\infty}(r_{\bf k} \rightarrow 0):=\lim_{r_{\bf k} \rightarrow 0} S_{\infty}\approx~ r^2_{\bf k}.
\eea
\textcolor{black}{Now comparing all the three results mentioned above, one can easily write the following  relationship for the squeezing amplitude parameter $r_{\bf k}$ in the $(r_{\bf k} \rightarrow 0)$ limit,   which is given by:}
\bea
&& C_1 = \sqrt{2}~C_2=4  \sqrt{S_{\infty}(r_{\bf k} \rightarrow 0)}.
\eea  
\textcolor{black}{This gives the relationship between the quantum circuit complexity measure and the entanglement entropy measure for the small values of the squeezing amplitude parameter $r_{\bf k}$.}


Now, for small squeezing amplitude parameter $r_{\bf k}$ if we take the $\alpha\rightarrow 1$ limit then we get the following expression for the entanglement entropy:
\bea S_{\rm vN}(r_{\bf k} \rightarrow 0):=\lim_{r_{\bf k} \rightarrow 0} S_{\alpha\rightarrow 1}\approx~ r^2_{\bf k}\left(1-2\ln~r_{\bf k}\right).\quad
\eea
Using this expression further we obtain the following  relationship between all of these measures in the present context:
\bea S_{\rm vN}(r_{\bf k} \rightarrow 0)&=&\frac{C^2_1}{16}\bigg[1-2\ln\bigg(\frac{C_1}{4}\bigg)\bigg]\nonumber\\
&=&\frac{C^2_2}{8}\bigg[1-2\ln\bigg(\frac{C_2}{2\sqrt{2}}\bigg)\bigg].\eea
For small squeezing amplitude parameter $r_{\bf k}$ one can write down the following relation:
\bea r_{\bf k}&\approx&\exp\bigg(-\frac{\omega_{\bf k}}{2T}\bigg)\nonumber\\
&=&\exp\bigg(-\frac{\omega_{-{\bf k}}}{2T}\bigg)\nonumber\\
&=&\exp\bigg(-\frac{\Omega_{\bf k}}{4T}\bigg),\quad \quad \quad\eea
where we have:
\bea \omega_{\bf k}=\omega_{-{\bf k}}=\Omega_{\bf k}/2.,\eea 
where the expression for $\Omega_{\bf k}$ is defined before.  Using this expression one can further write down the expressions for different circuit complexity  and the entanglement measures as:
\bea
    \label{eq:new}
    &&C_1= 4~\exp\bigg(-\frac{\Omega_{\bf k}}{4T}\bigg), \\
    &&C_2= 2\sqrt{2}~\exp\bigg(-\frac{\Omega_{\bf k}}{4T}\bigg),\\
    &&S(r_{\bf k} \rightarrow 0)=\exp\bigg(-\frac{\Omega_{\bf k}}{2T}\bigg),\\
 &&S_{\rm vN}(r_{\bf k} \rightarrow 0)=\bigg(1+\frac{\Omega_{\bf k}}{2T}\bigg) \exp\bigg(-\frac{\Omega_{\bf k}}{2T}\bigg).
\eea

Till now, we have used the results of the circuit complexity measure from the covariance matrix method to establish this connecting relation analytically.  
In the next section, we are going to use Nielsen's wave function method to understand such connections numerically.

\section{\textcolor{Sepia}{\textbf{ \large Numerical Results}}}
\label{sec:numerical}
\begin{figure*}[htb!]
	\centering
	\subfigure[\rm $r_{\bf k}$ ~vs~$a$]{
		\includegraphics[width=8cm]{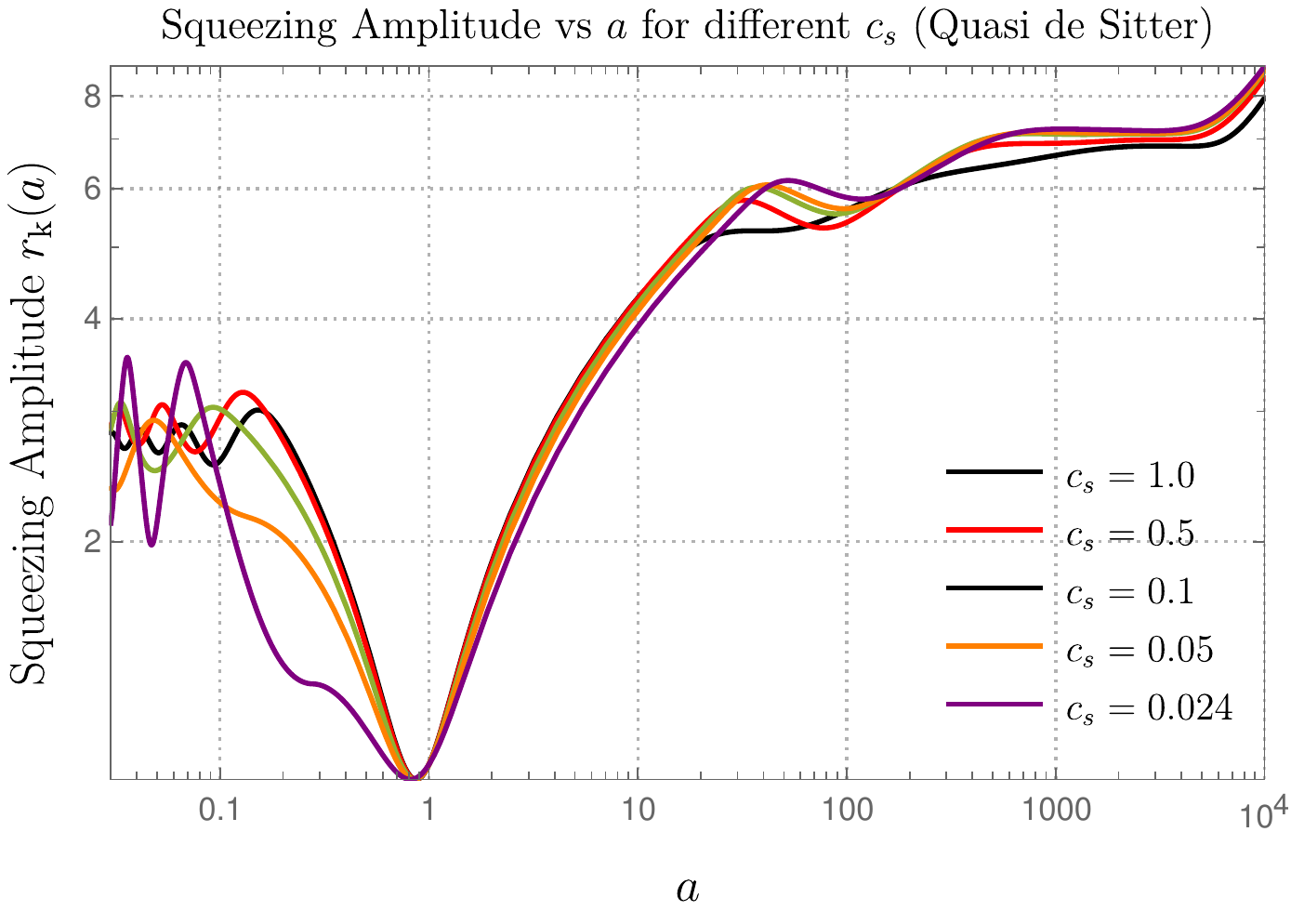}\label{rkphikvsa1}
	}
	\subfigure[\rm  $\phi_{\bf k}$ ~vs~$a$]{
		\includegraphics[width=8.3cm]{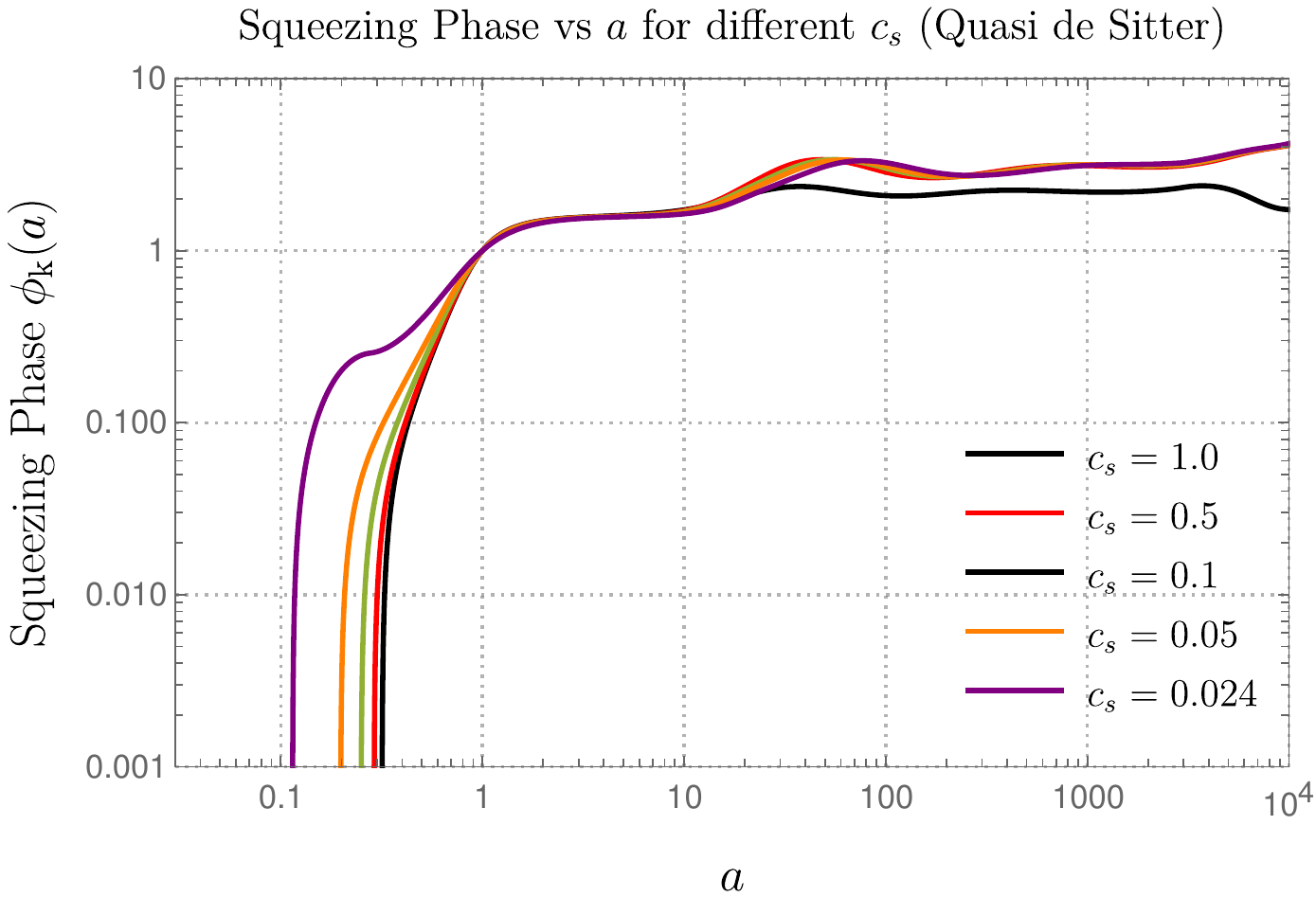}\label{rkphikvsa2}
	}
	\caption{Behaviour of the squeezing amplitude ($r_{\bf k}$) and squeezing angle/phase ($\phi_{\bf k}$) with respect to cosmological scale factor $a$ for the different values of the effective sound speed parameter $c_s$ lying within the window $0.024\leq c_s\leq 1$.  }
	\label{rkphikvsa}
\end{figure*}

\begin{figure*}[htb]
	\centering
	\subfigure[\rm  $r_{\bf k}$ ~vs~$c_s$]{
		\includegraphics[width=8cm]{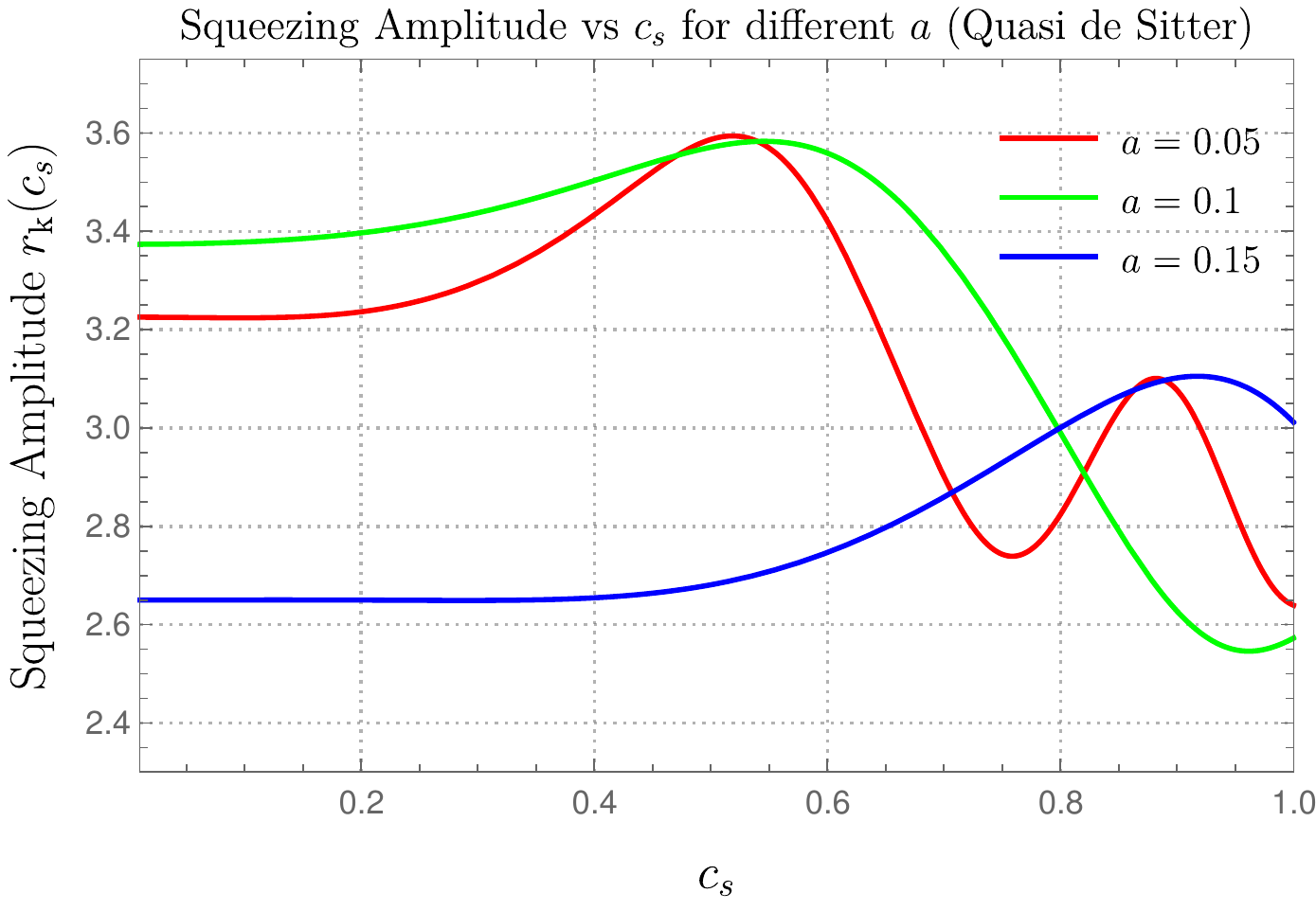}\label{rkphikvscs1}
	}
	\subfigure[\rm $\phi_{\bf k}$ ~vs~$c_s$]{
		\includegraphics[width=8.1cm]{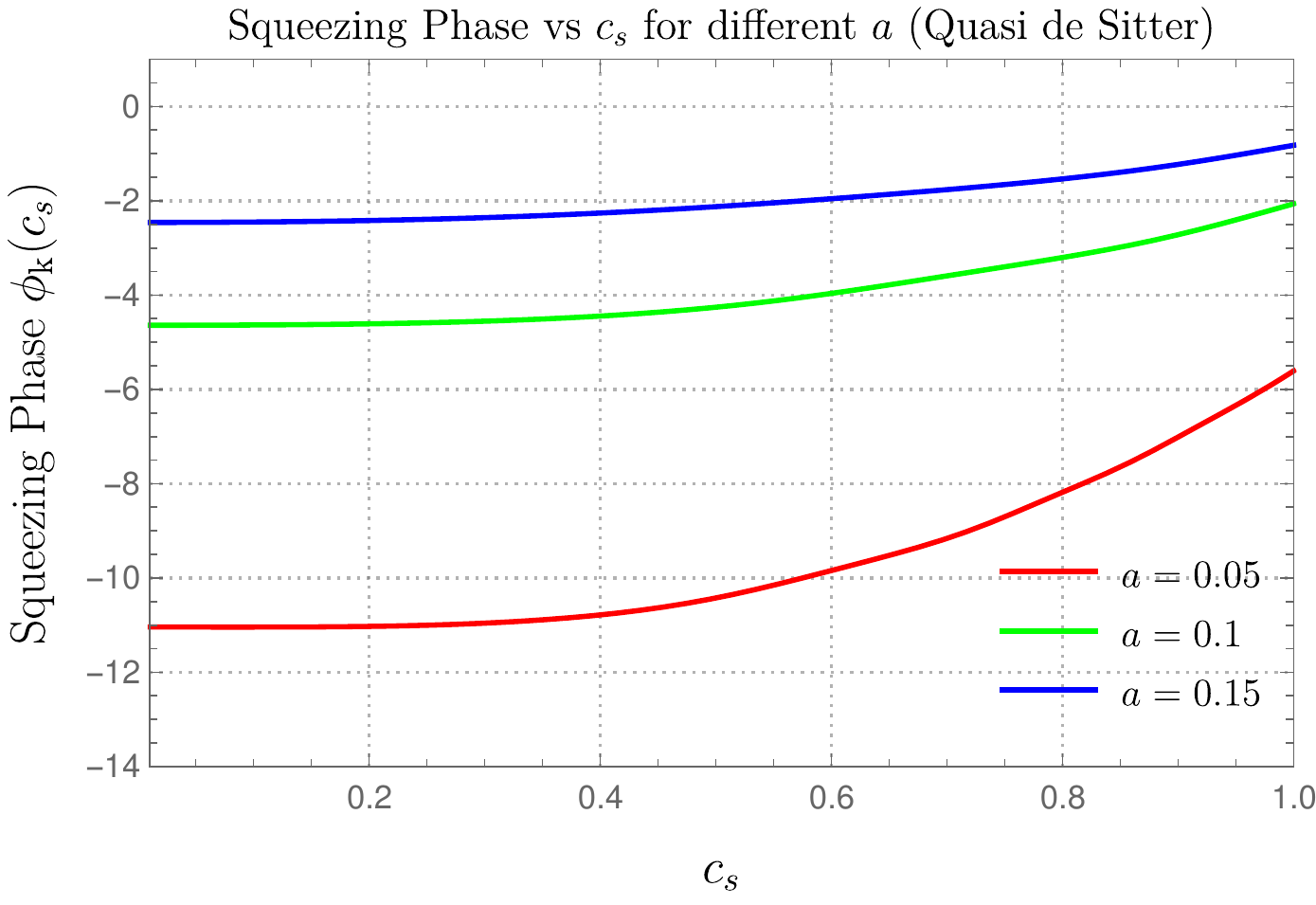}\label{rkphikvscs2}
	}
	\caption{Behaviour of the squeezing amplitude ($r_{\bf k}$) and squeezing angle/phase ($\phi_{\bf k}$) with respect to the effective sound speed parameter $c_s$ for the different values of the cosmological scale factor $a$ fixed at an early time scale of our universe.  }
	\label{rkphikvscs}
\end{figure*}

\begin{figure*}[htb!]
	\centering
	\subfigure[\rm $C_{1}$ ~vs~$a$]{
		\includegraphics[width=7.9cm] {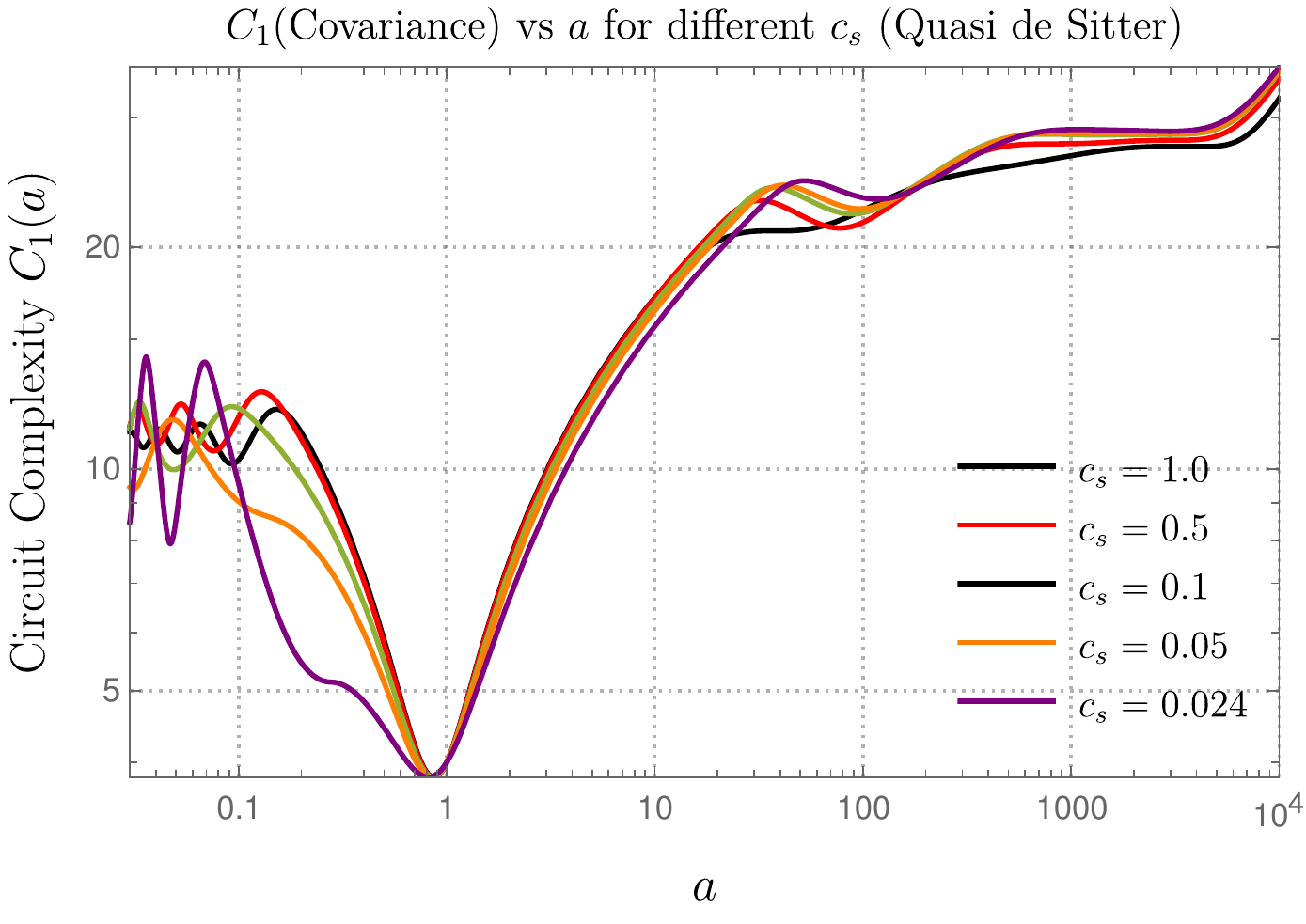}\label{coCvsa1}
	}
	\subfigure[\rm $C_{2}$ ~vs~$a$]{
		\includegraphics[width=8cm] {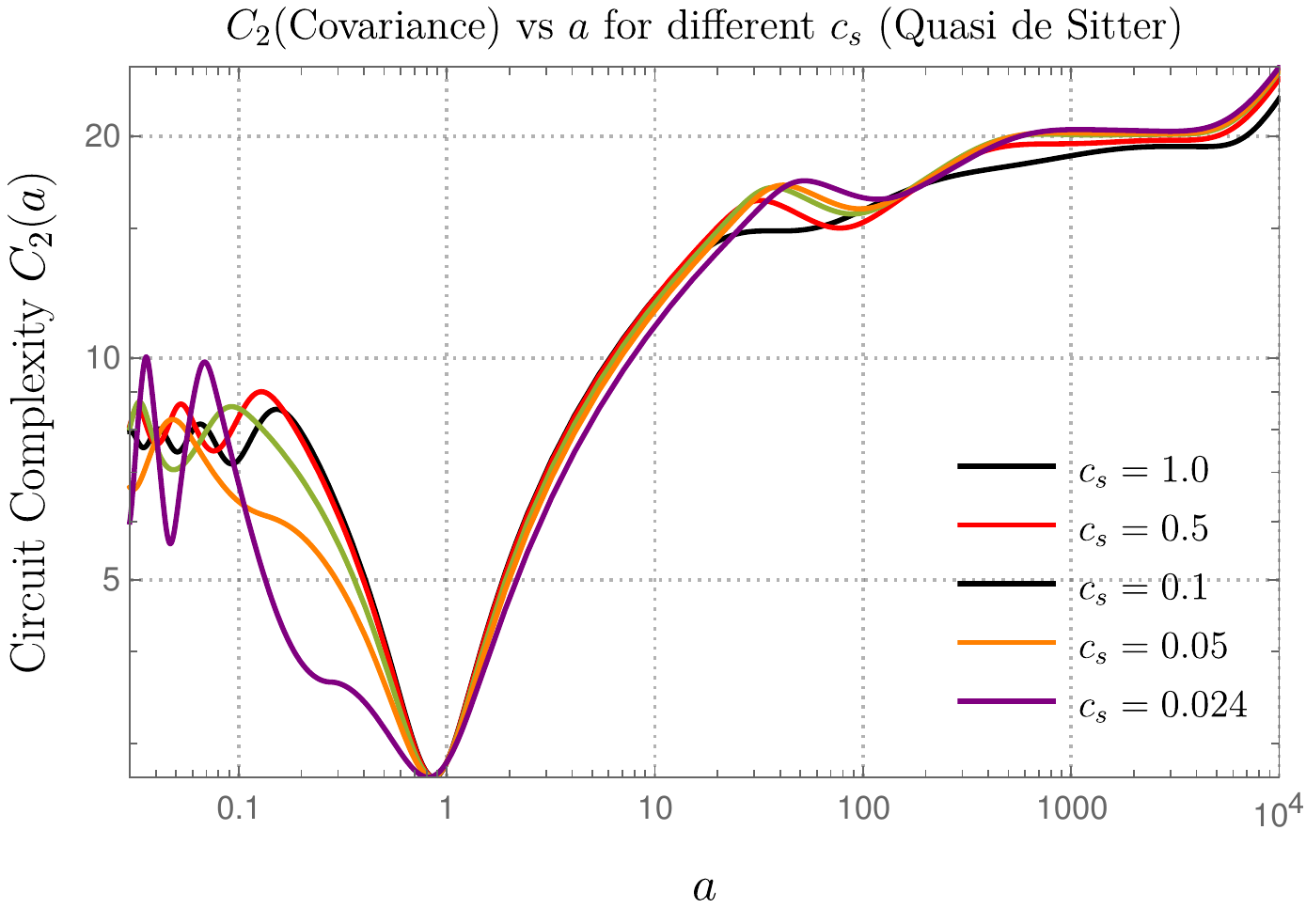}\label{coCvsa2}
	}
	\caption{Behaviour of the circuit complexity from linear cost function ($C_{1}$) and quadratic cost function ($C_{2}$) with respect to cosmological scale factor $a$ for the different values of the effective sound speed parameter $c_s$ lying within the window $0.024\leq c_s\leq 1$.  Here we have used the covariance matrix method.}
	\label{coCvsa}
\end{figure*}

\begin{figure*}[htb!]
	\centering
	\subfigure[\rm $C_{1}$ ~vs~$c_s$]{
		\includegraphics[width=8cm] {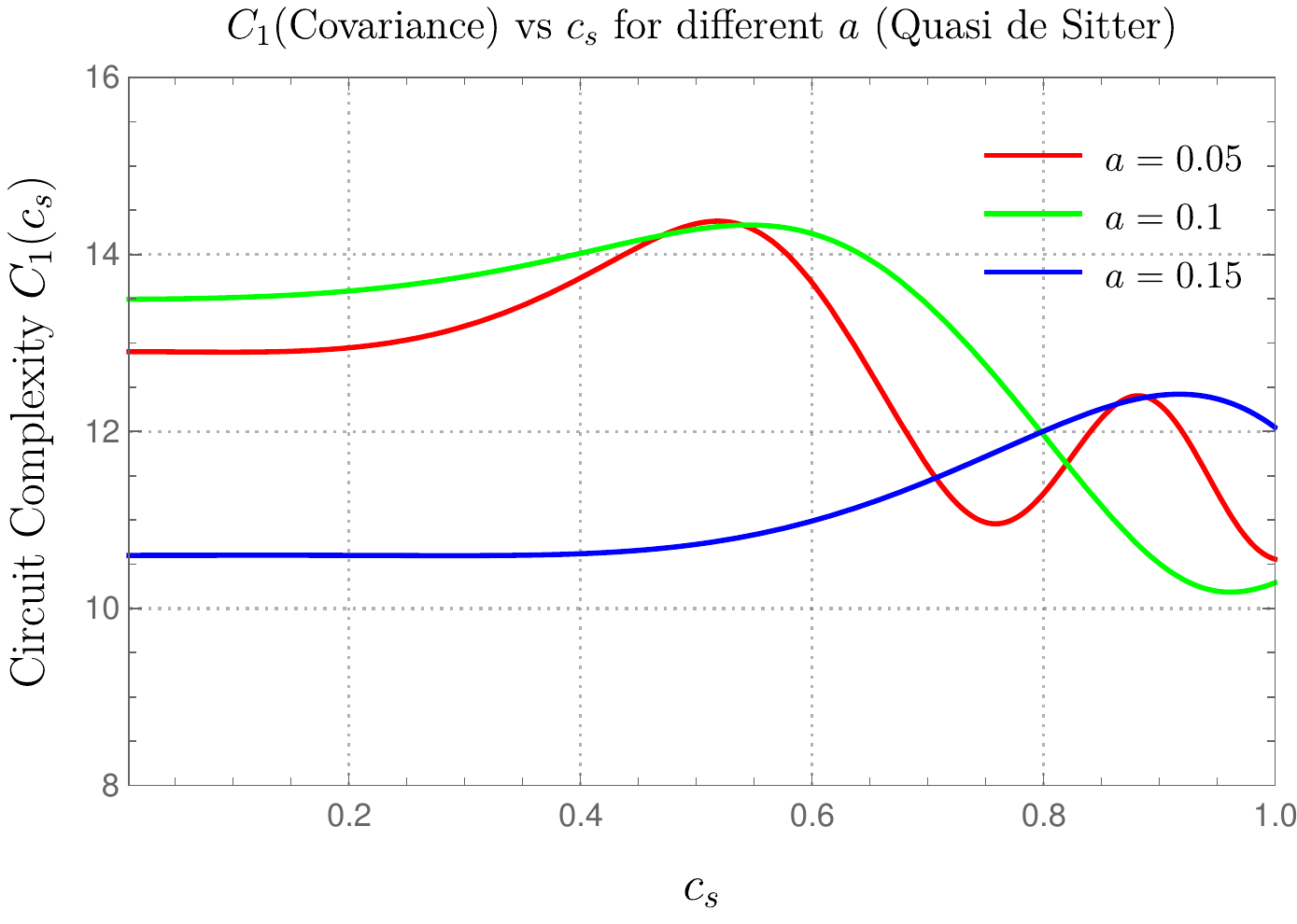}\label{coCvscs1}
	}
	\subfigure[\rm $C_{2}$ ~vs~$c_s$]{
		\includegraphics[width=8cm] {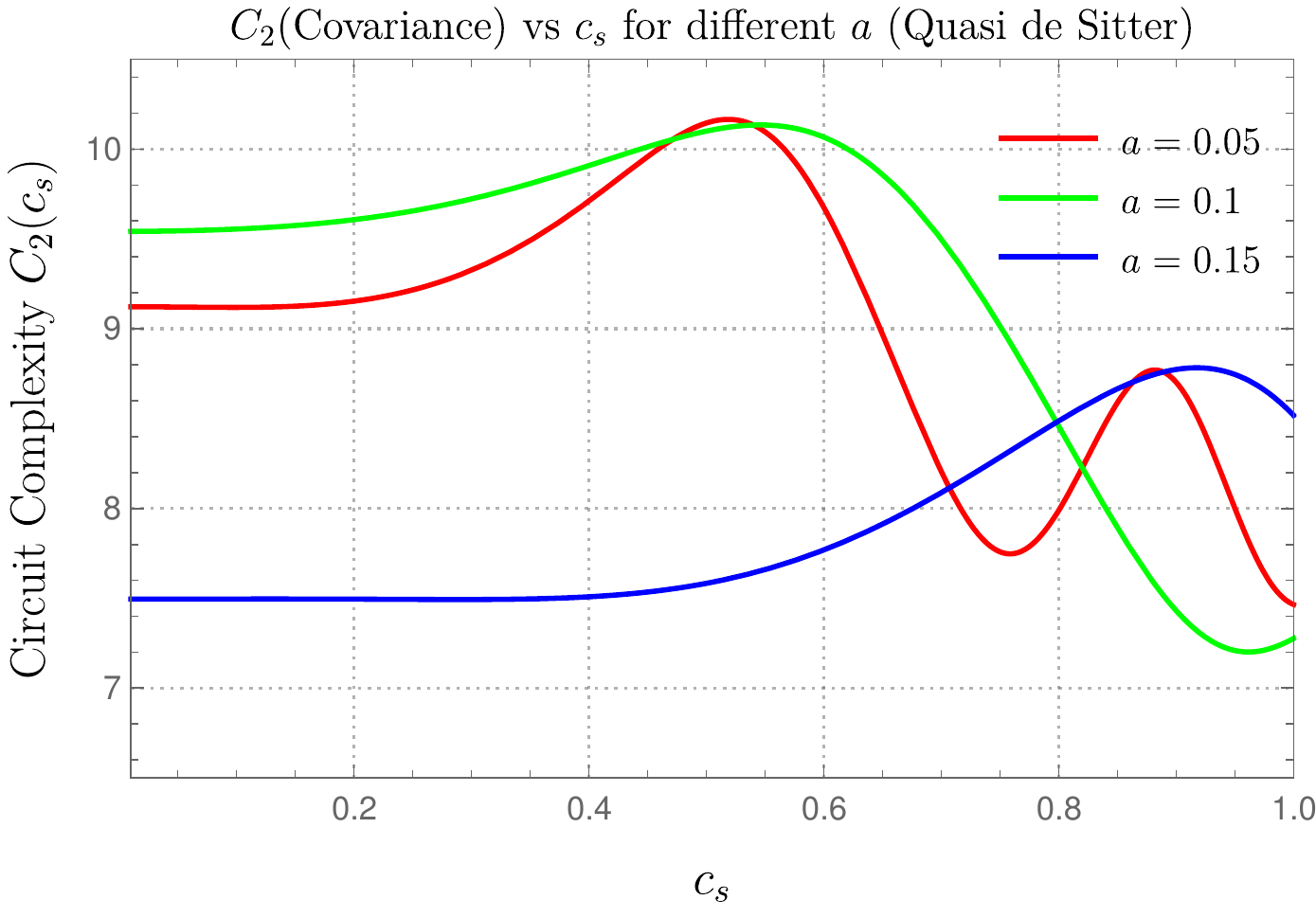}\label{coCvscs2}
	}
	\caption{Behaviour of the circuit complexity from linear cost function ($C_{1}$) and quadratic cost function ($C_{2}$) with respect to the effective sound speed parameter $c_s$ for the different values of the cosmological scale factor $a$ fixed at an early time scale of our universe.  Here we have used the covariance matrix method. }
	\label{coCvscs}
\end{figure*}

\begin{figure*}[htb]
	\centering
	\subfigure[\rm $C_{1}$ ~vs~$a$]{
		\includegraphics[width=16.5cm,height=7cm] {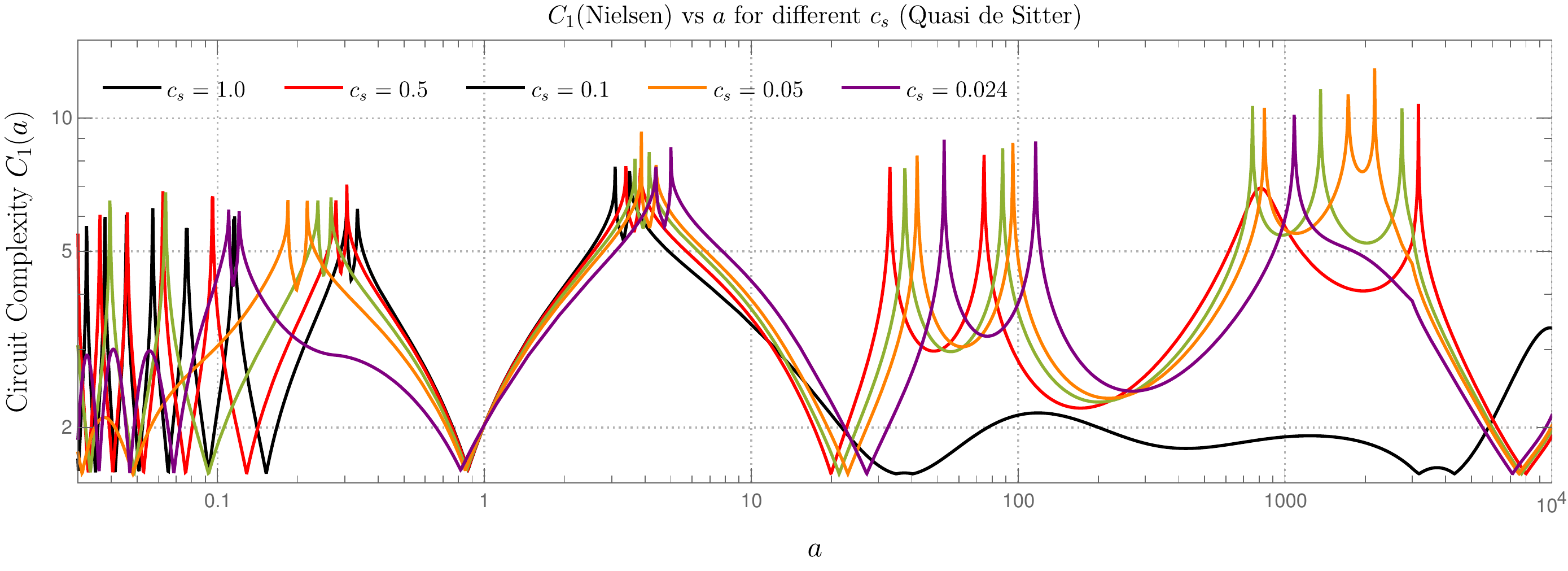}\label{NielCvsa1}
	}
	\subfigure[\rm $C_{2}$ ~vs~$a$]{
		\includegraphics[width=16.5cm,height=7cm] {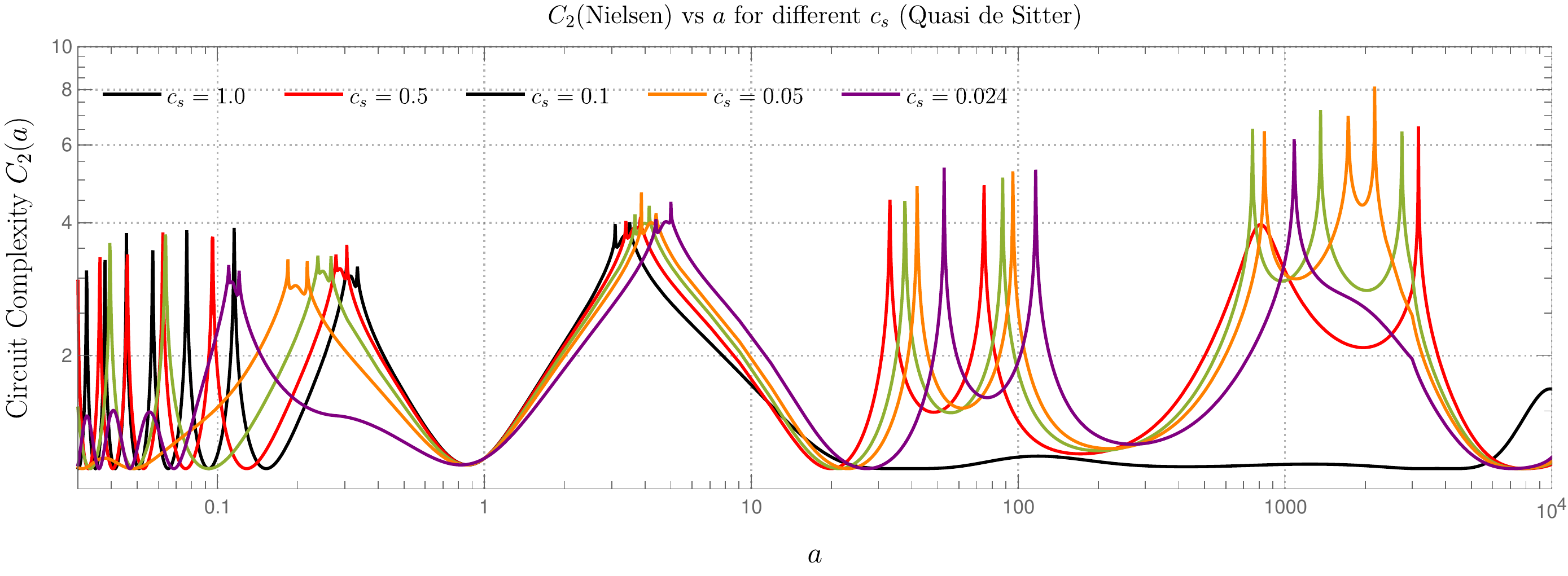}\label{NielCvsa2}
	}
	\caption{Behaviour of the circuit complexity from linear cost function ($C_{1}$) and quadratic cost function ($C_{2}$) with respect to cosmological scale factor $a$ for the different values of the effective sound speed parameter $c_s$ lying within the window $0.024\leq c_s\leq 1$.  Here we have used Nielsen's wave-function method. }
	\label{NielCvsa}
\end{figure*}

\begin{figure*}[htb!]
	\centering
	\subfigure[\rm $C_{1}$ ~vs~$c_s$]{
		\includegraphics[width=13cm]{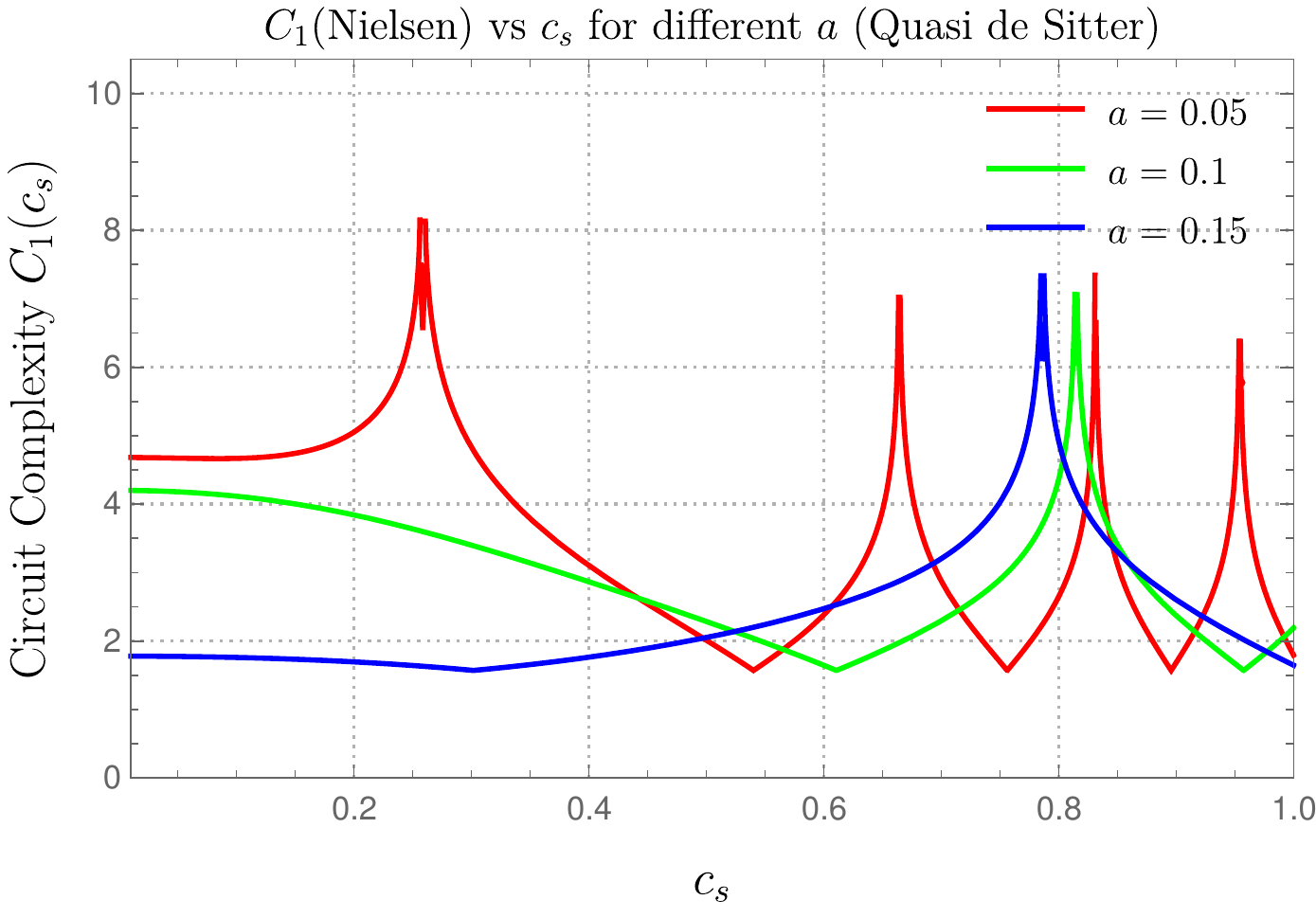}\label{NielCvscs1}
	}
	\subfigure[\rm $C_{2}$ ~vs~$c_s$]{
		\includegraphics[width=13cm]{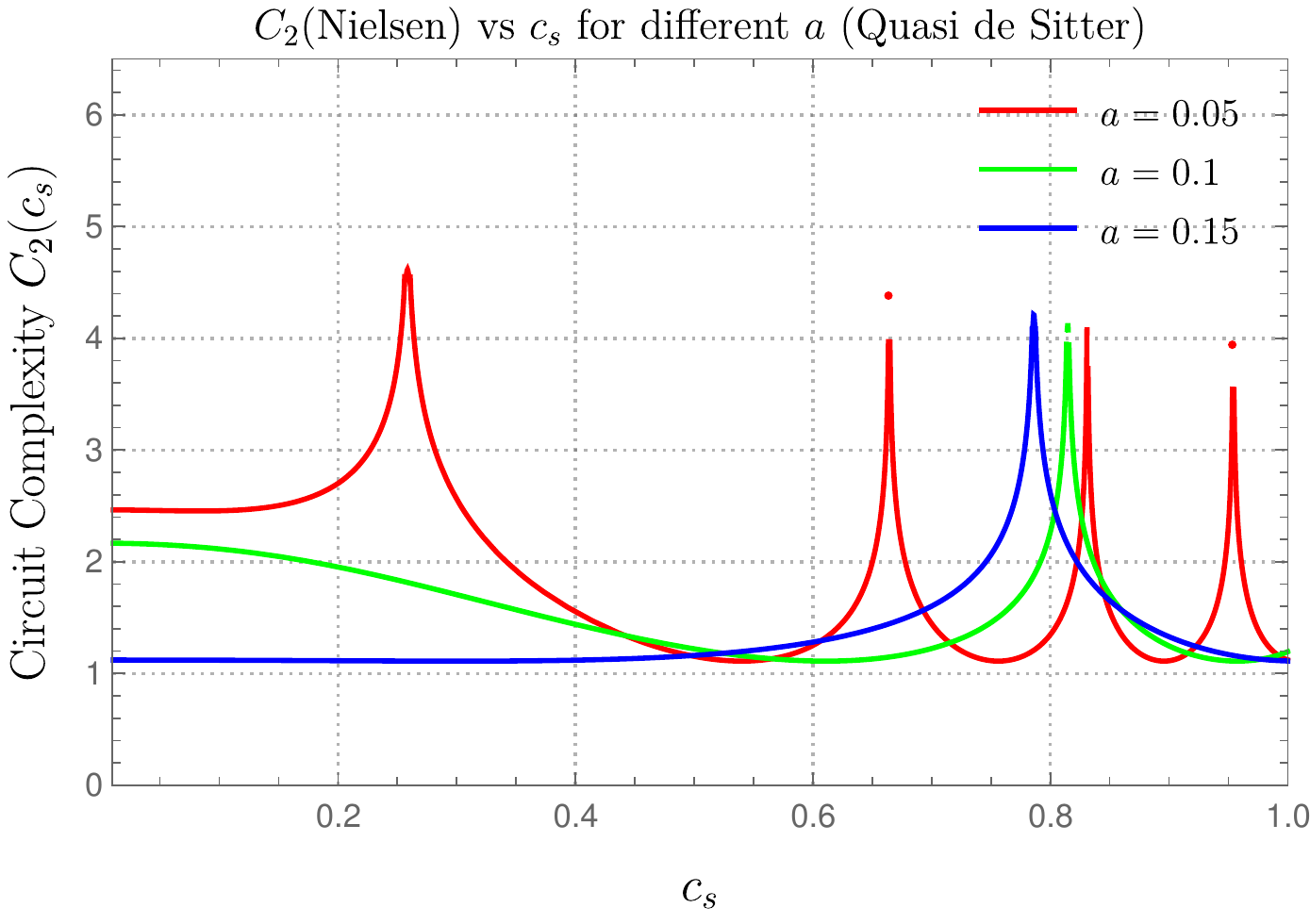}\label{NielCvscs2}
	}
	\caption{Behaviour of the circuit complexity from linear cost function ($C_{1}$) and quadratic cost function ($C_{2}$) with respect to the effective sound speed parameter $c_s$ for the different values of the cosmological scale factor $a$ fixed at an early time scale of our universe.  Here we have used Nielsen's wave-function method.}
	\label{NielCvscs}
\end{figure*}

\begin{figure*}[htb!]
	\centering
	\subfigure{
		\includegraphics[width=8cm] {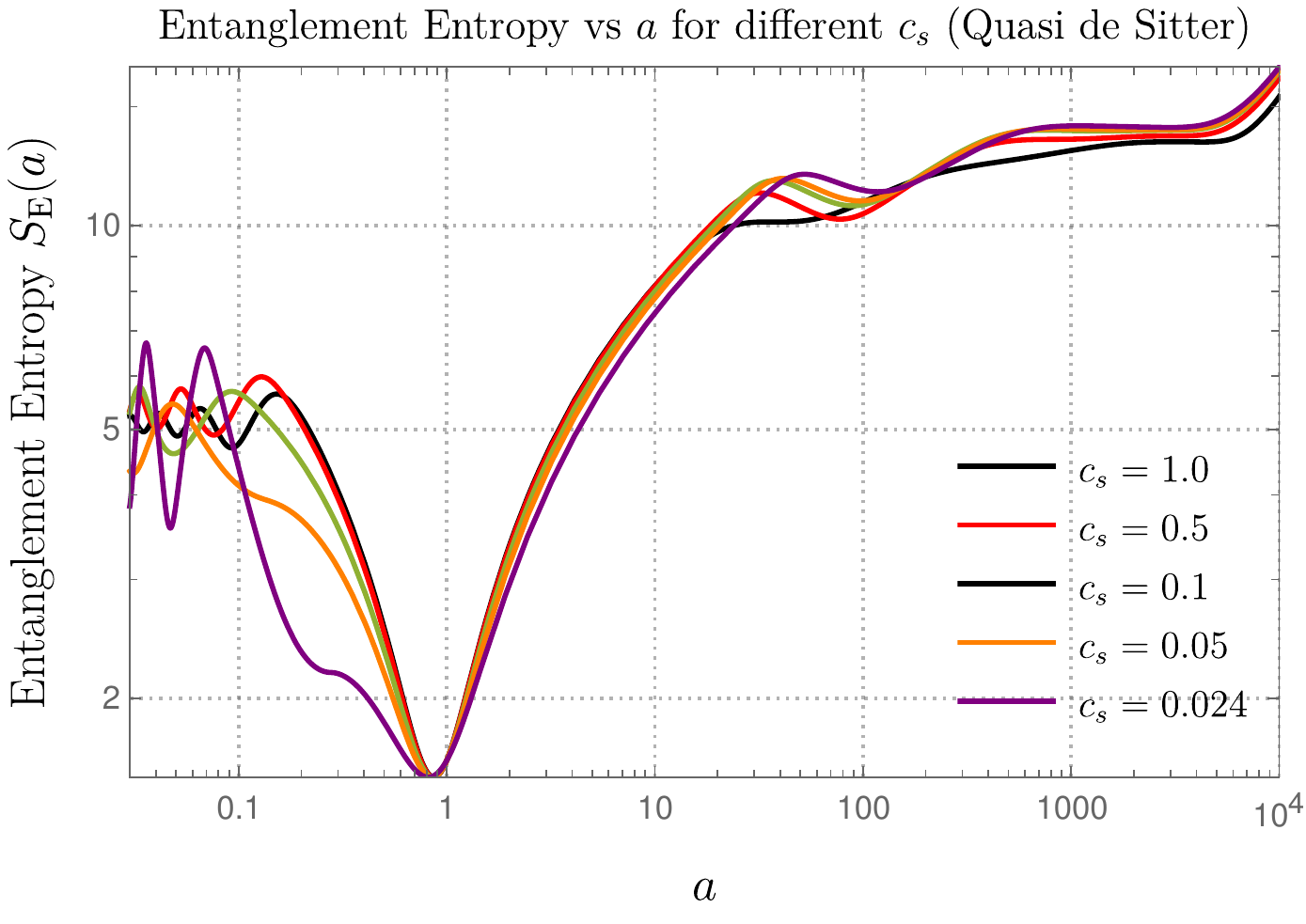}\label{entvsacs1}
	}
	\subfigure{
		\includegraphics[width=8.2cm] {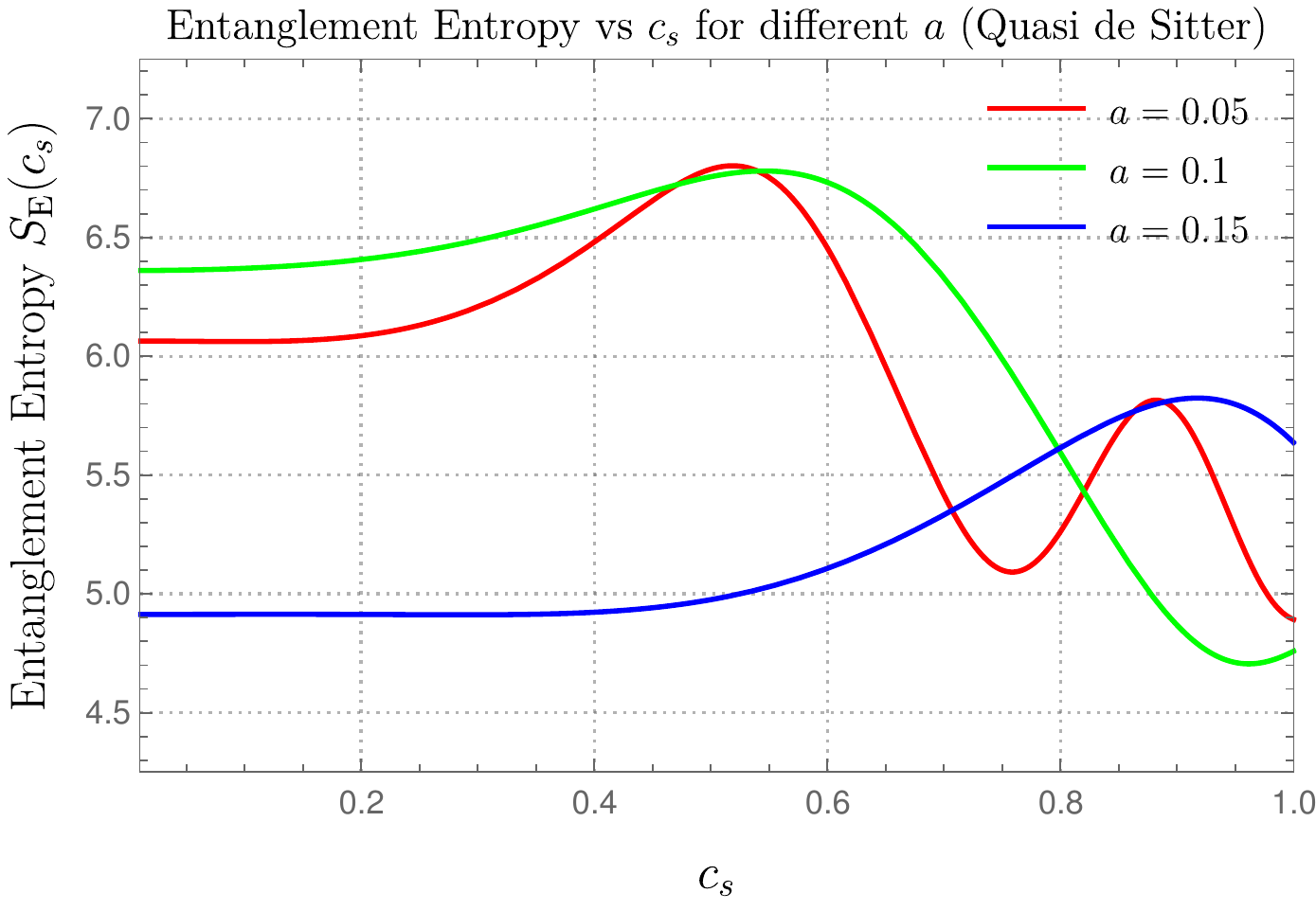}\label{entvsacs2}
	}
		\subfigure{
		\includegraphics[width=8.2cm] {ent_vs_cs.pdf}\label{entvsacs2}
	}
	\caption{Behaviour of the entanglement entropy ($S_{\rm E}$) with respect to cosmological scale factor $a$ and effective sound speed $c_s$ respectively. }
	\label{entvsacs}
\end{figure*}

\begin{figure*}[htb]
	\centering
	\subfigure[\rm $dC_1/da$ ~vs~$a$]{
		\includegraphics[width=8cm] {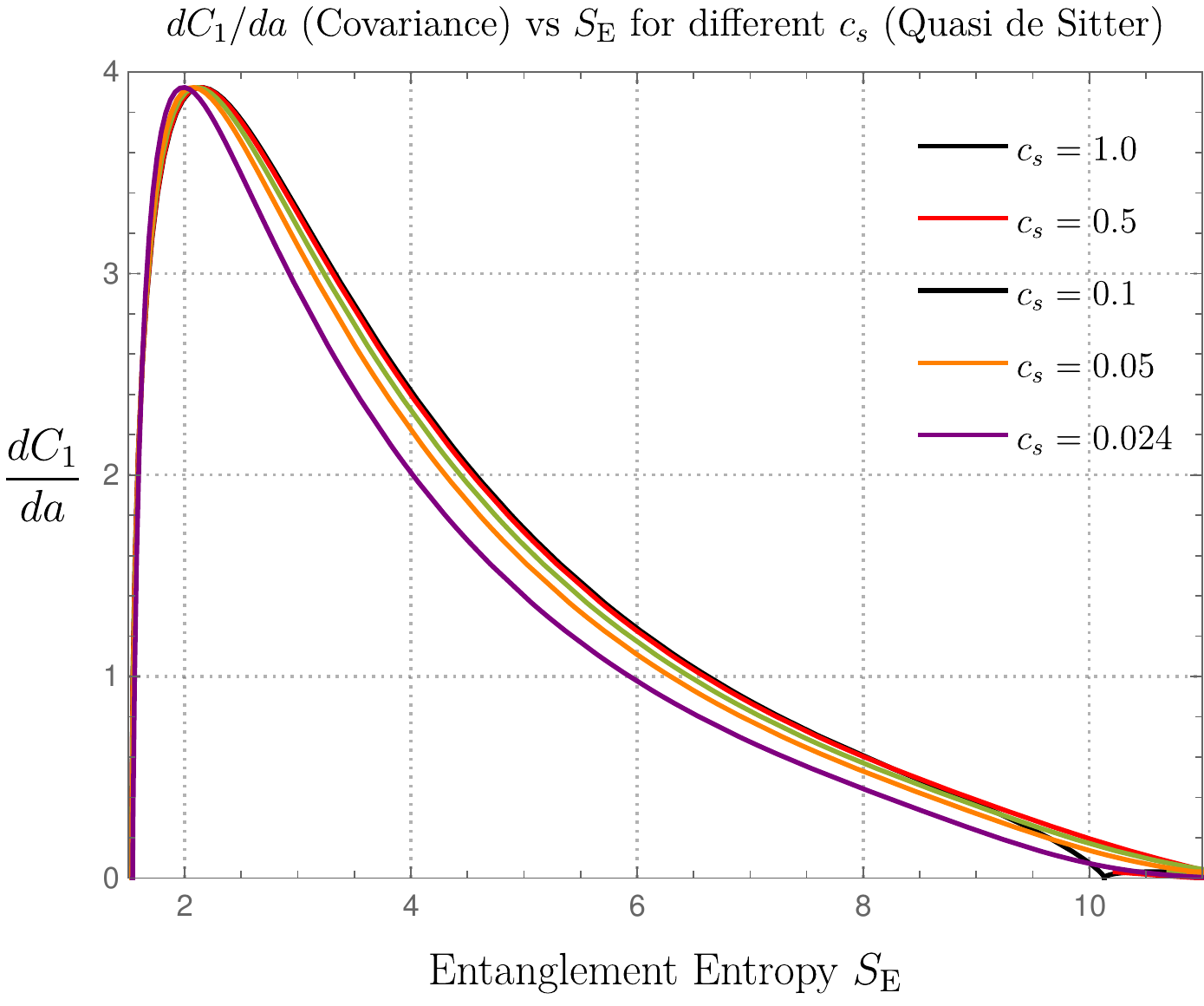}\label{dcdavss1}
	}
	\subfigure[\rm $dC_1/da$ ~vs~$c_s$]{
		\includegraphics[width=8.35cm] {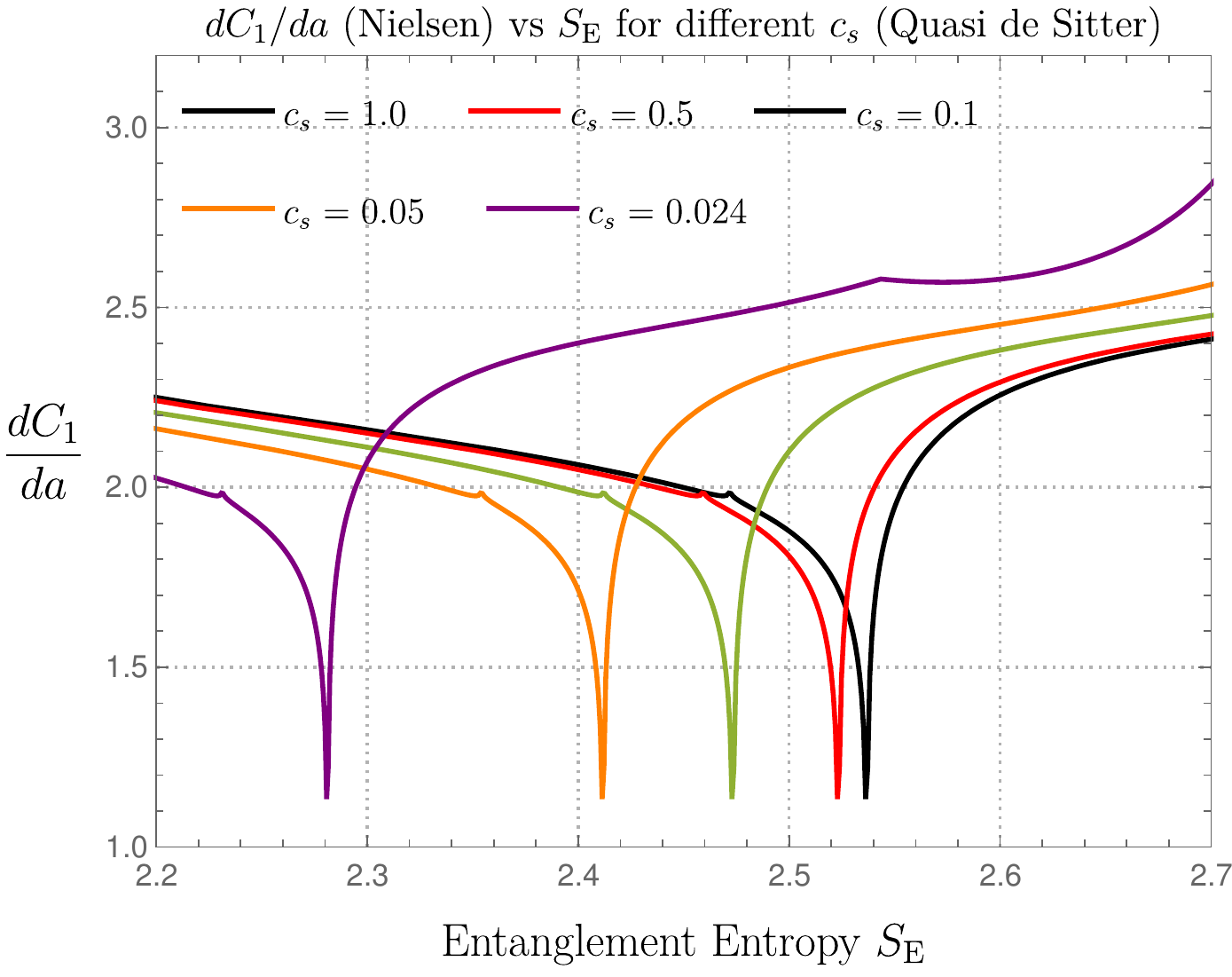}\label{dcdavss2}
	}
	\caption{Behaviour of the rate of change of complexity from linear cost function measure with respect to entanglement entropy ($S_{\rm E}$) computed from the covariance matrix method and Nielsen's wave-function method for the different effective sound speed $c_s$ respectively.  }
	\label{dcdavss}
\end{figure*}

In this section, we perform the numerical analysis of the quantum circuit complexity measure and the entanglement entropy measure computed from the present COSMOEFT setup using the two-mode squeezed state formalism. As  mentioned before, instead of considering  the evolution with respect to the conformal time scale,  here we perform the numerical analysis using the cosmological scale factor for the quasi de Sitter space time in $3+1$ dimensions. This is because numerically solving the evolution of the squeezing amplitude $r_{\bf k}$ and the squeezing angle $\phi_{\bf k}$ from the previously mentioned coupled differential equations (\ref{eq:evolution1a}) and (\ref{eq:evolution1b}) for a given boundary condition is difficult due to the presence of numerical instabilities.  Here we fix the boundary condition at the late time scale $\tau=\tau_0$ which fix the squeezing amplitude and the squeezing angle at $r_{\bf k}(\tau_0)=1$ and $\phi_{\bf k}(\tau_0)=1$.  On the other hand,  numerically solving the evolution of the squeezing amplitude $r_{\bf k}$ and the squeezing angle $\phi_{\bf k}$ with respect to the scale factor as stated in 
Eq  (\ref{eq:evolution2a}) and (\ref{eq:evolution2b}) comparatively simpler for a given boundary condition.  Here the boundary condition is fixed here at the late time scale $\tau=\tau_0$ where $a(\tau_0)=a_0=1$ which fix the squeezing amplitude and the squeezing angle at $r_{\bf k}(a_0=1)=1$ and $\phi_{\bf k}(a_0=1)=1$.  This boundary condition helps us to know the cosmological evolution of squeezing amplitude $r_{{\bf k}}(a)$ and squeezing angle $\phi_{{\bf k}}(a)$ at any value of the scale factor.  Solving these coupled differential equations with respect to the cosmological scale factor is comparatively simpler than solving with respect to the conformal time scale as the numerical instabilities are less and capture more information in the cosmological scale.  To solve the problem with respect to the cosmological scale factor we have used the change of variable from the conformal time scale to the scale factor by using Eq (\ref{schange}).

In figure (\ref{rkphikvsa1}) and figure (\ref{rkphikvsa2}),  we have plotted the behaviour of the squeezing amplitude parameter $r_{\bf k}$ and the squeezing angle parameter $\phi_{\bf k}$ with respect to the dynamical cosmological scale factor $a$ of the quasi de Sitter space from our present COSMOEFT setup.  In these plots, we have considered different values of the speed of sound lying within the window $0.024\leq c_s\leq 1$.  Here $c_s=1$ corresponds to the single scalar field canonical models and the rest of the values describe a wide class of non-canonical scalar field models in the present framework.  From figure (\ref{rkphikvsa1}), it is possible to get the information about the behaviour of the squeezing amplitude parameter $r_{\bf k}$ at different cosmological scales ranging from very small to large values of $a$,  which describe past, present and future of our quasi de Sitter universe.  For very early universe (extremely small values of $a$) squeezing amplitude parameter $r_{\bf k}$ shows irregular oscillatory behaviour and the amplitude of the oscillations becomes large for smaller values of effective sound speed $c_s$ whereas, for $c_s=1$ the amplitude becomes smaller compared to all other amplitudes obtained considering different sound speeds $c_s$. Further suppression in the  amplitude is not allowed due to the restriction from the causality.  For the value of the scale factor lying within the range $0.1<a<1$ squeezing amplitude parameter $r_{\bf k}$ decrease considerably for different values of $c_s$.  
However,  it is visible from this plot that in this region, the behaviour of $r_{\bf k}$ for the different values of $c_s$ is distinguishable.  Particularly, one can observe sufficient distinguishability for smaller values of the effective sound speed,  such as $c_s=0.024$ and $c_s=0.05$.  Then at the present time scale where the corresponding value of the scale factor $a=1$, the squeezing amplitude parameter $r_{\bf k}$ shown for different values of $c_s$ reach  a coincident minimum  which is small but non-zero.  As we have pointed out before, we fix the boundary condition at this point. Later, within the range $1<a<10^4$ which corresponds to the future universe, the squeezing amplitude parameter $r_{\bf k}$  increases sufficiently by following almost similar behaviour for different values of $c_s$,  which are less distinguishable.  Though in between $30<a<10^4$ some small distinguishable features can be observed from the figure (\ref{rkphikvsa1}). Similarly, in  figure (\ref{rkphikvsa2}), the behaviour of the squeezing angle parameter $\phi_{\bf k}$ is shown at different cosmological scales ranging from very small to large values of $a$.   For very early universe (\textit{i.e} with small values of $a$), the squeezing  angle $\phi_{\bf k}$ shows a sharp growth  and the change is quite distinguishable for different values of the sound speed $c_s$ lying within the previously mentioned window. However, the distinguishability is more prominent for  smaller values of $c_s$.  This feature is shown within the range  $0.1<a<1$. At exactly $a=1$, the observed behaviour  for different values of $c_s$  becomes indistinguishable which is expected due to the boundary condition.  In between $1<a<30$,  again a growing behaviour is observed but as we have mentioned, in this region, the effects of the  speed of sound become sub-dominant. Later,  within the range $30<a<10^4$, marginal distinguishable features  can be observed. 
This analysis of the squeezing  parameters $r_{\bf k}$ and $\phi_{\bf k}$ as a function of the dynamical scale $a$ is extremely useful as it tells us, in which region the effect of the  speed of sound  parameter $c_s$ is  prominent. Thus, using this analysis, one can make a distinction between various types of theories which is described by a single COSMOEFT setup which is essential to physically constraint some of them.   
 
 In figure (\ref{rkphikvscs1}) and figure (\ref{rkphikvscs2}),  we have plotted the behaviour of the squeezing amplitude parameter $r_{\bf k}$ and the squeezing angle parameter $\phi_{\bf k}$ with respect to the effective  speed of sound parameter $c_s$ lying within the range $0.024\leq c_s\leq 1$,  which is restricted by cosmological observation from Planck and causality constraint.  Here we have plotted the behaviour for three different values of the scale factor $a$ which is fixed at the early time  in the cosmological evolutionary scale.  From figure (\ref{rkphikvscs1}), the dynamical behaviour of the squeezing amplitude parameter $r_{\bf k}$ is depicted for smaller and larger values of the sound speed parameter $c_s$.   For  small values of $c_s$, we can observe that the squeezing amplitude parameter $r_{\bf k}$ is almost constant. Whereas, for the large values of $c_s$, it shows a significant feature,  initially it  increases to a maximum value and then it decreases.  Due to the causality constraints, the plots are shown up to  $c_s=1$.   Another  interesting feature that can be noticed here is  that, depending on the value of the scale factor, multiple peaks can appear within the allowed range of $c_s$.  For example, with  the scale factor fixed at $a=0.05$, we can observe two peaks with different  amplitudes. Such features are  interesting   as they carry important signatures pertaining to   the system under consideration.   
 In figure (\ref{rkphikvscs2}), the dynamical behaviour of the squeezing angle parameter $\phi_{\bf k}$ is shown as a function of the  speed of sound  parameter $c_s$.   For  smaller values of $c_s$, one can observe that the squeezing angle parameter $\phi_{\bf k}$ is almost constant and later, it shows a slow increasing behaviour  as  $c_s$ is increased. Within the range  $0.024\leq c_s\leq 1$, the squeezing angle parameter $\phi_{\bf k}$  always remains negative. Also we note here that the different $\phi_{\bf k}$s obtained considering the fixed values of the scale factor remain quite distinguishable throughout the  allowed window of $c_s$.

 In figures (\ref{coCvsa1}) and (\ref{coCvsa2}),  we have plotted the behaviour of the circuit complexity from linear $C_1$ and the quadratic cost function $C_2$ computed using the covariance matrix method with respect to the dynamical cosmological scale factor $a$ of the quasi de Sitter space from our present COSMOEFT setup. Similar to the previous cases,  we have considered different values of the speed of sound parameter lying within the window $0.024\leq c_s\leq 1$.   The behaviour of both of the measures of the circuit complexity with respect to the variation of the scale factor $a$ is  similar to the squeezing amplitude parameter $r_{\bf k}$ as depicted in figure (\ref{rkphikvsa1}).  This is quite expected because  of the fact that $C_1=\sqrt{2}~C_2=4~r_{\bf k}$. Hence, the difference can be observed only in the overall amplitude of the measure of quantum circuit complexity.  Also, one should note that  both of the measures give similar features in the present context. Similar reasoning is applicable for the figures 
(\ref{coCvscs1}) and  (\ref{coCvscs2}), where  the behaviour of the circuit complexity from the linear $C_1$ and the quadratic cost function $C_2$ for the covariance matrix method is shown as a function of  the effective  speed of sound parameter $c_s$ lying within the range $0.024\leq c_s\leq 1$ and one obtains only a trivial difference  in the overall amplitude compared to the squeezing amplitude parameter $r_{\bf k}$.

 Now, in figures (\ref{NielCvsa1}) and  (\ref{NielCvsa2}),  we have plotted the linear $C_1$ and the quadratic cost function $C_2$ computed using the Nielsen's wave-function method as a function of the dynamical cosmological scale factor $a$ of the quasi de Sitter space with fixed values of $c_s$ from the window $0.024\leq c_s\leq 1$.  For smaller values of the scale factor (for $a<1$) the behaviour of both the measure $C_1$ and $C_2$  shows similar behaviour.  Only the differences are found in the smoothness at the transition points where the behaviour changes from an increasing trend to a decreasing trend or vice versa.  For the smaller values of $a$, the frequency of the increasing and the decreasing trend is very high for both of the measures and it is hard to distinguish the individual effects appearing from the different values of the  speed of sound $c_s$. At the point $a=1$, curves obtained  from different values of $c_s$ coincide with each other which is expected due to the  fixed  boundary condition.  As we approach  larger values of scale factor $a$,  the modifications arising from the different values of the  speed of sound parameter $c_s$ become more and more distinguishable. For example, within the range $20<a<10^4$, the trends for  $c_s=0.024$ is markedly different compared to the trends with $c_s=1$.  Also, the rising and falling behaviour in both of these plots appear in an aperiodic fashion. As mentioned before, these distinguishable features are extremely helpful to differentiate the models having $c_s=1$ and $c_s<1$.
 
 In figure (\ref{NielCvscs1}) and (\ref{NielCvscs2}), the cost functions are shown for fixed scale parameter as a function of the effective  speed of sound  $c_s$  within the range $0.024\leq c_s\leq 1$.  For the smaller values of the sound speed, the behaviour of both the measure $C_1$ and $C_2$ computed using Nielsen's wave-function method shows similar behaviour and only  difference that  can be observed is in terms of the smoothness of the plots at the transition points.  As we change the value of the  speed of sound, we observe multiple peaks in irregular intervals having different smoothness for $C_1$ and $C_2$ and for  larger values of $c_s$, these multiple peaks appear more frequently. 
 
  In figure (\ref{entvsacs1}) we have plotted the behaviour of entanglement entropy with respect to the dynamical cosmological scale factor for the different values of the effective sound speed $c_s$.  On the other hand,  in figure (\ref{entvsacs2}) we have plotted the behaviour of entanglement entropy with respect to the effective sound speed $c_s$ for fixed values of the cosmological scale factor which we fixed at the early time scale.  Here the entanglement entropy is evaluated from the von-Neumann measure.  The behaviour of entanglement entropy is exactly similar to that of the squeezing amplitude parameter $r_{\bf k}$ and the circuit complexity measure ($C_1,C_2$) computed from the covariance matrix method in the present context. The only noticeable difference that we found is that for both of these plots, the amplitude of the entanglement entropy computed from the von-Neumann measure is larger than $r_{\bf k}$ but  smaller than both circuit complexities,  $C_1$ and $C_2$ computed from covariance matrix method. From the analytical expression for the entanglement entropy from the von-Neumann measure, it is  difficult to understand why one should expect such  similarities with the behaviour of the two complexity measures computed from the covariance matrix method.  The possible explanation lies within the dependency of both of them on the squeezing amplitude parameter $r_{\bf k}$.  Though from the analytical expressions it seems like the entanglement entropy from the von-Neumann measure and two measures of circuit complexity computed from the covariance matrix method behave in a completely different way, the numerical study from the present setup tells us that the overall features of both of them are  similar apart from having different amplitudes.

  Finally,  in figure (\ref{dcdavss1}) and figure (\ref{dcdavss2}), we have shown the behaviour of the rate of change of complexity function from linear cost function with respect to the entanglement entropy from the COSMOEFT setup.  Here the circuit complexity and the entanglement entropy are computed using (1) the covariance matrix method and the von-Neumann measure for the figure (\ref{dcdavss1}) and    (2) the Nielsen's wave-function method and the  von-Neumann measure for the figure (\ref{dcdavss2}). These parametric plots are obtained treating the effective  speed of sound $c_s$ as a parameter  lying within the window $0.024\leq c_s\leq 1$.

  In the present context, both of these plots are important because it will help us to know, for our system, the underlying connection between the quantum circuit complexity and the entanglement entropy. It should be mentioned here that computing the rate of change of the circuit complexity function with respect to the scale factor is  difficult as we have seen from figure (\ref{coCvsa1}) and  (\ref{NielCvsa1}) that the behaviour of the circuit complexity itself with respect the scale factor is not at all smooth.  Due to having multiple peak like features in both of these plots, computing the derivatives with respect to the scale factor at the transition point is numerically challenging.  Instead of considering  a large range of values for the scale factor, in this computation we have concentrated only on a specific range  which can be  efficiently handled in the numerics.  For the  figure (\ref{dcdavss1}), we have considered the range $0.3<a<100$ and for the figure (\ref{dcdavss2}) we have taken  $0.3<a<2$. Clearly,  for the plot using the covariance matrix method, one can cover a larger interval of $a$ compared to the plot obtained using Nielsen's wave-function method.  Particularly in this context, it seems that the covariance matrix method gives a better result compared to Nielsen's wave-function method as numerical instabilities can be handled easily for it. However, we would rather argue, the scenario is exactly the opposite and this can be justified by a  crucial observation from the analytically derived expressions for the quantum circuit complexity from the linear and the quadratic cost functions using both methods.  The result obtained from the covariance matrix method captures the contribution of the squeezing amplitude parameter $r_{\bf k}$.  On the other hand,   the result obtained from Nielsen's wave-function method capture both the information from the squeezing amplitude parameter $r_{\bf k}$ and the squeezing angle/phase parameter $\phi_{\bf k}$. However, due to  this contribution from the phase factor in the later, one observes the aperiodic oscillations which makes the numerics challenging at the transition points.  But if we concentrate only in the region where the circuit complexity function only grows with the scale factor and does not encounter any transition,  then it is possible to easily extract the growth rate of the circuit complexity using the Nielsen's wave-function method.  Now from the figure (\ref{dcdavss1}), it can be observed that initially within a very small value of the entanglement entropy, the rate of change of the circuit complexity with respect to the scale factor sharply increases to a maximum value where the  effects from different values of the effective  speed of sound $c_s$ are not at all distinguishable. Later, it shows a decaying trend for a wide range of values of the entanglement entropy where one can explicitly identify the distinguishable effects from different values of  $c_s$.

  On the other hand, in  the figure (\ref{dcdavss2}), one can distinguish  the  effects arising  from different values of the  speed of sound $c_s$ throughout  the considered range of the entanglement entropy. 
  Also in each plot, in this case, one can observe  a dip like feature which eventually gets converted to an increasing trend.  This behaviour is observed for all values of $c_s$.  Due to the very different natures observed in the figures (\ref{dcdavss1}) and  (\ref{dcdavss2}), a non-linear relationship is expected between the circuit complexity and the entanglement entropy computed from both the covariance matrix method and Nielsen's wave-function method.  The prime motivation for finding a relationship between the circuit complexity and the entanglement entropy lies in the conjecture proposed by Susskind in the context of black hole physics.  In the introduction of this article, we have already pointed out that  the growth rate of complexity is equal to the product of the entropy of the black hole and the corresponding Hawking temperature.   In black hole physics, the growth rate of complexity is computed with respect to the underlying time scale.  However,  in the present context, instead of computing this rate with respect to the cosmological time scale, we have used the cosmological scale factor in terms of which we have studied the evolution of the complexity function.  Considering the covariance matrix method in the limit of vanishing  squeezing amplitude  $r_{\bf k}\rightarrow 0$, one can write the following non-linear relations among the equilibrium temperature, the entanglement entropy and the circuit complexity measure for our EFT setup:
  \bea 
  && \underline{\textcolor{Sepia}{ \rm Results~for~ small ~r_{\bf k}:}}\nonumber\\
  C_1&=&\sqrt{2}~C_2=4~\exp\bigg(-\frac{\Omega_{\bf k}}{4T}\bigg),\\
  S&=&\bigg(1+\frac{\Omega_{\bf k}}{2T}\bigg)~\exp\bigg(-\frac{\Omega_{\bf k}}{2T}\bigg)\nonumber\\
  &=&\frac{C^2_1}{16}\bigg[1-2\ln\bigg(\frac{C_1}{4}\bigg)\bigg]\nonumber\\
&=&\frac{C^2_2}{8}\bigg[1-2\ln\bigg(\frac{C_2}{2\sqrt{2}}\bigg)\bigg].\eea

\section{\textcolor{Sepia}{\textbf{\large Covariance Vs Nielsen's method in COSMOEFT\label{method}}}}    
In this section, our objective is to compare the results obtained from the covariance matrix method and Nielsen's wave function method in the context of the present COSMOEFT setup. In the following, we compare the two methods  point wise :
\begin{itemize}
\item  The results obtained in the covariance method do not depend on the squeezing angle/phase $\phi_{\bf k}$ whereas, for Nielsen's method, it will explicitly depend on this parameter.

\item In the case of the covariance method circuit complexity measures linearly depend on the squeezing amplitude parameter $r_{\bf k}$.  On the other hand,  in Nielsen's method, a non-linear dependence on the squeezing amplitude parameter $r_{\bf k}$ is observed.  

\item For the covariance method, the explicit details of the wave function of the squeezed state does not appear in the final expression for the circuit complexity measure.  On the other hand,   for Nielsen's method, circuit complexity measures explicitly depend on the details of the wave function.   It turns out that the circuit complexity in Nielsen's method is parametrized by both squeezing parameters $r_{\bf k}$ and $\phi_{\bf k}$.  

\item  In the covariance method circuit complexities computed from the linear and the quadratic measure is approximately the same as $C_1=\sqrt{2}~C_2$, which is very simple.  On the other hand in Nielsen's method circuit complexities are related via the relation,  $C^2_1-C^2_2=f(r_{\bf k},\phi_{\bf k})$,  where in general $f(r_{\bf k},\phi_{\bf k})$ are the non-linear functions of both squeezing parameters $r_{\bf k}$ and $\phi_{\bf k}$.

\item  The numerical analysis shows that, in the case of the covariance method, the circuit complexity exactly behaves like the squeezing amplitude $r_{\bf k}$ apart from having a small change in amplitude.  On the other hand,  just by looking at the behaviour of the squeezed parameters $r_{\bf k}$ and $\phi_{\bf k}$, it is difficult to guess the behaviour of the circuit complexity computed from Nielsen's method.  The frequency of oscillation of the aperiodic behaviour in the circuit complexity computed from both the methods is large initially and the individual effects for the different sound speeds $c_s$ are not distinguishable.  On the other hand,  it turns out that this frequency will decrease a lot and one can be able to distinguish the effects of the different sound speed contributions explicitly from the behaviour at a larger value of the scale factor.   However,  in the case of the covariance method, we obtain a growing behaviour and also the aperiodic feature is  lost.  

\item The causality constraint on the circuit complexity computed from both of these methods results in different behaviour in terms of the smoothness of each of the functions at the transition point.   Results obtained from the covariance method are smooth.  On the other hand, sharp changes are observed at the transition points when the Nielsen's method is considered.   Apart from that,  the outcomes of both methods show phase shifts as can be seen from the plots obtained for different values of the scale factor.  The causality constraint restricts us to  consider the region $c_s\leq 1$ and the cosmological observation from Planck, sets the lower limit at $c_s\geq 0.024$.  

\item  The Entanglement entropy is solely determined by the squeezing amplitude $r_{\bf k}$ in the present context.   Since the computed results of the circuit complexity measure from the covariance method are also solely determined in terms of the squeezing amplitude $r_{\bf k}$,  it is easy to obtain  a  relationship between the circuit complexity measure and the entanglement entropy measure.  On the other hand,  the computed results of the circuit complexity measure from Nielsen's method parametrized by both the squeezing parameters $r_{\bf k}$ and $\phi_{\bf k}$,  it seems impossible to obtain a simple relationship between the circuit complexity measure and the entanglement entropy measure in this case.  

\item  The rate of change of the circuit complexity measure with respect to the scale factor varies smoothly with respect to the entanglement entropy for different values of the sound speed $c_s$ in the case of the covariance method.  The individual effects of the different sound speeds are not visible for the lower values of the entanglement entropy. However, such effects become prominent for the large values of the entanglement entropy.   On the other hand,   the rate of change of the circuit complexity measure with respect to the scale factor varies sharply at the transition point with respect to the entanglement entropy for different values of the sound speed $c_s$ in the case of Nielsen's method.   The individual effects of the different sound speeds are visible at all values of the entanglement entropy considered.

\item  For the small values of the squeezed amplitude parameter $r_{\bf k}$, one can  determine analytically the  relationship among the equilibrium temperature, the entanglement entropy measure and the quantum circuit complexity measure using the covariance matrix method.  On the other hand, establishing such a simple connection  seems to be  impossible  using Nielsen's wave-function method  due to the non-linear dependencies on both the squeezing parameters $r_{\bf k}$ and $\phi_{\bf k}$ in the expressions.  Otherwise,   the entanglement entropy and the equilibrium temperature are solely parametrized by the squeezing amplitude parameter $r_{\bf k}$.  
\end{itemize}

  \section{\textcolor{Sepia}{\textbf{ \large Conclusion\label{sec:Conclusions}}}}

In the following, we list the summary and the concluding remarks  point-wise:
\begin{itemize}
\item  First we have developed an effective field theory framework of cosmological perturbation,  which we identified as COSMOEFT using St$\ddot{\text{u}}$ckelberg trick which generates the scalar modes from Goldstone modes in this framework.  In this formalism, we have restricted up to the two derivative terms in the metric.  Here the Goldstone modes are generated from the broken time diffeomorphism symmetry.  This 
EFT framework at the level of cosmological perturbation is described by an effective fluid whose sound speed is $c_s$.  Also, the causality constraint demands that for such an effective fluid the sound speed has to satisfy $c_s\leq 1$.  On the other hand,  from the cosmological observation from Planck, one can fix the lower bound of the effective sound speed at $c_s\geq 0.024$.   For this reason,  we have considered only the range $0.024\leq c_s\leq 1$ for the analysis.  In this analysis,  $c_s=1$ corresponds to the single scalar field canonical models and $c_s<1$ signifies a class of non-canonical scalar field models.  
	
	\item Next we have computed the quantized Hamiltonian of this EFT setup and found that the interacting part of the Hamiltonian can be conveniently parametrized by two parameters using  the framework of the squeezed state formalism, namely, the amplitude $r_{\bf k}$ and the phase $\phi_{\bf k}$.  We have used this framework as the underlying  physical principle in this work.
	
	\item We have computed the expressions for the circuit complexity from two different measures- using linear and quadratic cost function from the covariance matrix method and the Nielsen's wave function method using the  two-mode squeezed state formalism.  Also, we have computed the entanglement entropy from this setup.  
	
	\item From our analysis we  find that the results obtained by 
	 Nielsen's wave function method provides a better understanding of the quantum circuit complexity compared to the results obtained from the covariance matrix method.   Nielsen's wave function method captures more information than the covariance matrix method.   This is because the circuit complexity in Nielsen's wave function method is parametrized in terms of both the squeezing parameters $r_{\bf k}$ and phase $\phi_{\bf k}$.  On the other hand,  in the case of the covariance matrix method, the result is parametrized only in terms of the squeezing amplitude $r_{\bf k}$.
	 
\item For the small values of squeezing amplitude $r_{\bf k}$, we have obtained an  analytical relationship among different measures of the quantum circuit complexity, the  entanglement entropy and the equilibrium temperature using the results obtained from the covariance matrix method. 
We have found a non-trivial non-linear temperature dependence in this case.  On the other hand, using the results obtained from Nielsen's wave function method, we are unable to obtain such simple relationship due to the non-linear dependencies on both the squeezing parameters in the computed  linear and quadratic cost function measures.  

\item However,  to establish some connection among different measures of the quantum circuit complexity,  the entanglement entropy and the equilibrium temperature using Nielsen's wave function method, we have computed the rate of change of the circuit complexity function with respect to the scale factor numerically and plotted with respect to entanglement entropy considering  a wide range of values.

	\item The present analysis helps us to know the underlying features of the present EFT setup which describes a wide class of canonical and non-canonical scalar field models.  Using the usual cosmological analysis,  such as studying the cosmological correlation functions,  amplitudes of the spectrum of the correlators, and its scale-dependent or invariant features, sometimes it becomes extremely  difficult to distinguish various models available in the literature.  The main reason is, all of these cosmological models  provide degenerate predictions for the cosmological observations.  To break this degeneracy, the higher point correlators have to be constrained from observation,  which is not yet quite efficiently possible due to lack of required statistical accuracy in these measurements and sometimes, also due to lack of  possible resources to do the job.  On the other hand,  one can try to look for  additional information from the models that can be probed  with the available tools and in that way, it will be possible to rule out several available models. In this work, our motivation has been to provide such theoretical tools,  particularly the quantum information theoretic tools which are useful to distinguish the effects of the variation in  the  speed of sound parameter $c_s$.
	 
\end{itemize}

\textbf{Future Prospects:}
\begin{itemize}
    
\item In the present work, we have analyzed the causality constraints on circuit complexity in the COSMOEFT framework and we have restricted the initial quantum vacuum state as the well-known Bunch Davies or Euclidean state but one can also study the Causality constraint by considering other choices of initial quantum vacuum states such as $\alpha-$vacuum and $\alpha-\gamma$ vacuum (also called Motta-Allen vacuum).

\item Also in the present context, our study is mainly focused on the analysis of causality constraint on the circuit complexity with background space as de sitter space but it would be interesting to study other kinds of model having different scale factors such as Radiation, Matter, Inflation, Bouncing, Cyclic and Blackhole gas model etc because each of the scale factors carries significant information of the physics of the early universe. 

\item In this article,  a model-independent analysis has been considered. But using the explicit details from the models like DBI \cite{Alishahiha:2004eh,Choudhury:2012yh,Mazumdar:2001mm,Choudhury:2002xu,Choudhury:2015hvr}, 
Tachyon \cite{Mazumdar:2001mm,Choudhury:2002xu,Choudhury:2015hvr}, Galileon \cite{Chow:2009fm,DeFelice:2010nf} and two interacting quantum systems in de Sitter space \cite{Choudhury:2022btc}, one can explore the similar type of analysis. This type of analysis will help us to know about the role of the UV completion in the context of circuit complexity.

\item Circuit Complexity has been studied as deformations in the euclidean path integrals for CFTs. Similar deformations also appear in the context of cosmological perturbations. One could also study these path integral optimization in de Sitter space.

\item Recently in \cite{Bhattacharyya:2020iic,Haque:2021kdm,Haque:2021hyw} Circuit Complexity, Krylov complexity \cite{Adhikari:2022oxr,Adhikari:2022whf} and entanglement have been explored in the continuous variable system, the open quantum systems, in interacting QFTs and also in the quenched quantum systems \cite{Choudhury:2022dox}. It is interesting to pursue similar  analysis for the quantum complexity and the entanglement in the context of cosmological perturbations.

\end{itemize}

\textbf{Acknowledgement:}
SC would like to thank the Institute of Physics,  Bhubaneswar for supporting with Visiting Scientist position.  SC and AM would also like to thank the School of Physical Sciences, National Institute for Science Education and Research (NISER),  Bhubaneswar where part of the work was done.  SC also thank all the members of our newly formed virtual international non-profit consortium Quantum Aspects of the Space-Time \& Matter (QASTM) for elaborative discussions. NP and AR would like to thank the members of the QASTM Forum for useful discussions.
Last but not least,  we would like to acknowledge our debt to the people belonging to the various part of the world for their generous and steady support for research in natural sciences.  
\\

\section{\textcolor{black}{Appendix.1: Brief review of EFT}}\label{appen1}
In this section we will discuss two approaches towards the construction of the effective field theories, which is useful for understanding the COSMOEFT framework. 
	\begin{enumerate}
		\item \underline{\textcolor{Sepia}{\bf Top-down~method:}}\\
		 
		 In this case,  the key idea is to start with a UV complete framework which has many degrees of freedom.   Using this setup, one can derive an EFT below the high energy cut-off of the EFT by removing the unwanted fields using the path integration method.   For more details on this see refs.~\cite{Baumann:2014nda,Choudhury:2016wlj,Choudhury:2017glj}. 
	\\	 \\
		 To implement this idea, let us consider a  simple toy model of interacting scalar fields which is minimally coupled to the classical gravitational background via the space-time metric and this setup is in principle UV complete.  In this  toy model, let us consider a light scalar field $\phi$ which has a mass $m_{\phi}<M_p$ and many heavy scalar fields $\Phi_{i}\forall i=1,2,\cdots, N$ with mass $M_{\Phi_{i}}>M_p$.  In principle, one can formulate this setup in arbitrary $d+1$ dimensional gravitational space-time.  But for simplicity and particularly since we are interested in 4-dimensional COSMOEFT setup, let us consider only the $3+1$ dimensional gravitational space-time.  In this context,  the light field can be easily probed by the EFT framework as its mass is below the Planckian UV cut-off and for this reason, this field belongs to the visible sector of the theory.  On the other hand,  the heavy scalar fields cannot be probed by using the EFT framework which belongs to the hidden sector of the theory.  So to construct the EFT below the UV Planckian cut-off scale, one needs to path integrate over all the hidden sector of heavy fields.  In principle, this path integration can be performed in Lorentzian signature.  But it turns out that technically it is very complicated to perform such path integral.  This is because finding out the proper saddle points is very  difficult in the Lorentzian signature.  So the next best possible way to do this path integration over hidden sector of heavy scalar fields is in Euclidean signature which turns out to be  useful in terms of finding proper saddles for the path integration.  In this case,   the starting point of the interacting toy model is described by the following representative action~\cite{Baumann:2014nda,Choudhury:2016wlj,Choudhury:2017glj}:
		 	\begin{widetext}
		\bea \label{toy1} \textcolor{Sepia}{\bf Toy~model:}~~~S[\phi,\Phi_{i},g_{\mu\nu}]=\int d^4x \sqrt{-g}\left[\frac{M^2_p}{2}R+{\cal L}_{\rm vis}[\phi]+\sum^{N}_{i=1}{\cal L}^{(i)}_{\rm hid}[\Phi_{i}]+\sum^{N}_{j=1}{\cal L}^{(j)}_{\rm int}[\phi,\Phi_{j}]\right]~,\eea
			\end{widetext}
		where $g_{\mu\nu}$ represents the background gravitational metric in $3+1$ dimensional space-time,  ${\cal L}_{\rm vis}[\phi]$ represents the Lagrangian density which describes the visible sector of the toy model having only light scalar field $\phi$,  $ {\cal L}^{(i)}_{\rm hid}[\Phi_{i}]~\forall i=1,2,\cdots, N$ represents the Lagrangian density which describes the hidden sector of the toy model having only heavy scalar fields $\Phi_i~\forall i=1,2,\cdots, N$, and ${\cal L}^{(j)}_{\rm int}[\phi,\Phi_{j}]~\forall j=1,2,\cdots, N$ is the Lagrangian density representing the interaction between the light and the heavy scalar fields. In principle, any type of interaction can appear which is allowed by the underlying symmetry in the present context.  
	\\ \\	
		Now the next job is to perform the path integral in Euclidean signature using appropriate saddles on the heavy fields appearing in the UV complete toy model described in Eqn~(\ref{toy1}).   Consequently,  we are left with only an EFT theory of the light fields whose corresponding partition function is given by the following expression:
		\begin{widetext}
		\bea \textcolor{Sepia}{\bf Euclidean~partition~function:}~~~Z_{\rm EFT}\left[\phi,g_{\mu\nu}\right]&=&\prod^{N}_{j=1}\int\left[{\cal D}\Phi_{j}\right]e^{-S_{\rm E}[\phi,\Phi_{j},g_{\mu\nu}]}=\exp\left(-S^{E}_{\rm EFT}\left[\phi,g_{\mu\nu}\right]\right),~~~~~\eea
		\end{widetext}
		where in the above expression, $S_{\rm E}[\phi,\Phi_{j},g_{\mu\nu}]$ is the Euclidean version of the Lorentzian action $S[\phi,\Phi_{j},g_{\mu\nu}]$ which can be obtained by replacing $S[\phi,\Phi_{j},g_{\mu\nu}]\rightarrow i S_{\rm E}[\phi,\Phi_{j},g_{\mu\nu}]$.  After performing the path integration in the Euclidean signature over the heavy hidden sector fields, one can further compute the effective action or commonly known as the EFT action $S^{E}_{EFT}\left[\phi,g_{\mu\nu}\right]$ which is directly related to the partition function. Here it is important to note that this is a semi-classical treatment where we treat the background gravitational space-time in classical footing and consider the scalar fields belonging to both hidden and visible sectors in a quantum mechanical footing.  Also, it is important to mention here that,  after performing the path integration in the Euclidean signature and obtaining the expression for the Euclidean effective action, one needs to go back to the Lorentzian signature,  which is useful for further computation.  For this, we need to transform $S^{E}_{EFT}\left[\phi,g_{\mu\nu}\right]\rightarrow -i S_{\rm EFT}[\phi,g_{\mu\nu}]$.  After performing this, in the end,  the representative EFT action for the visible sector with light fields is described by the following equation~\cite{Baumann:2014nda,Choudhury:2016wlj,Choudhury:2017glj}:
			\begin{widetext}
		\bea \label{EFT1} \textcolor{Sepia}{\bf EFT~action:}~~~S_{EFT}\left[\phi,g_{\mu\nu}\right]=\int d^4x \sqrt{-g}\left[\frac{M^2_p}{2}R+{\cal L}_{\rm vis}[\phi]+\sum^{N}_{p=1}{\cal C}^{(p)}(g)\frac{{\cal O}^{(p)}[\phi]}{M^{\Delta-4}}\right]~.\eea
			\end{widetext}
Here, the first two terms remain the same and only the effect can be observed in the last series of terms which are written in terms of the EFT operators originated as an outcome of the previously mentioned path integration procedure.  The coupling strength of these operators is identified as ${\cal C}^{(p)}(g)~\forall p=1,2,\cdots, N$ which is, in principle, dimensionless and depends on the scale $g$ of the UV  complete toy theory.   In technical language, these coefficients are identified as the Wilson coefficients and the associated operators are Wilson operators.  Here the EFT operators are given by ${\cal O}^{(p)}[\phi]~\forall p=1,2,\cdots, N$ where $\Delta$ represents the mass dimensional local EFT operators which are suppressed by $M^{\Delta-4}$ scale.  It is important to mention here that,  the mathematical structure of the Wilson operators completely depend on the starting point where we have provided the toy model of the UV complete theory.  With a different toy model, the final mathematical form of these EFT operators will be different.  However,  the overall form of the representative EFT action in eqn~\ref{EFT1} will not change.  

		\item\underline{\textcolor{Sepia}{\bf Bottom-up~method:}}~~\\ \\
		In this case,  the key idea is to start with a  model-independent effective action which is valid below the UV Planckian cut-off scale and the structure of the EFT action is restricted by symmetry.  Using this setup, the next goal is to find out a UV complete theory allowed by the symmetries~\cite{Baumann:2014nda,Choudhury:2016wlj}.  Consequently, it is possible to determine the Wilson coefficients in terms of the UV complete theory.  In this work, we follow this procedure.  Though we have not found the exact form of the UV complete theory from the Wilson coefficients but this approach helps us to provide the causality constraint in terms of the effective  speed of sound parameter $c_s=\sqrt{\dot{p}/\dot{\rho}}=\sqrt{\partial p/\partial \rho}$ for COSMOEFT particularly in the parameter region where $c_s=1$ or $c_s<1$.   Here the $\dot{}$ corresponds to the derivative with respect to the physical time which  explicitly appears in the spatially flat FLRW space-time metric having the quasi de Sitter solution,  which is given by:
		\bea ds^2=-dt^2+a^2(t)d{\bf x}^2,\eea
		where we have followed the $(-,+,+,+)$ metric signature along with the velocity of light $c=1$ in natural units.   Here $a(t)=\exp(Ht)$ is the solution of the scale factor for the quasi-de Sitter background and  the Hubble parameter $H$ is not  a constant. In the COSMOEFT setup $p$ and $\rho$ correspond to the effective pressure and the density for an effective fluid,  which can be  easily computed from the $(ij)$ and $(00)$ components of the energy-momentum  tensor,  provided, the explicit form of the effective action of the matter sector or the EFT operators which mimic the same role are known.  
		
		Now, it is to be noted that in this context, causality breaks when  $c_s>1$,  and  we will not consider that in this work.  In literature, it is usually identified as the a-causal theory.  Additionally, it is important to note that though we  look into the region where the effective  speed of sound $c_s<1$, we  exclude the value $c_s=0$ in this work because this corresponds to ghost type of scalar field theories within the framework of EFT which is studied in the refs.~\cite{Gorji:2021isn,Arkani-Hamed:2003juy} in other contexts.  Also, this value is redundant from cosmological observation using Planck data,  which fixes the lower bound at $c_s=0.024$.  This means that, in terms of the EFT theories, the effective  speed of sound  lies within the window,   $0.024\leq c_s\leq 1$,  as supported by both cosmological observation and causality. When $c_s=1$ is considered, it describes a single scalar field slow roll model in the COSMOEFT setup.  On the other hand,  for $c_s<1$, we have  a wide class of non-canonical scalar field models within the framework of COSMOEFT.  Thus, both of these possibilities can be probed by using the constraint $c_s\leq 1$ within the present EFT setup.
	\end{enumerate}

\section{\textcolor{black}{\textbf{\large Appendix.2: Construction of COSMOEFT}}} 
\label{sec:E2}
In this section,  our prime objective is to give a review of constructing
the most generic COSMOEFT action written in the background of the quasi-de Sitter space-time.  The  objective to use this specific framework in the present context is to generate the scalar modes of the cosmological perturbations which will be useful for our  study. In our EFT setup, there are matter degrees of freedom whose fluctuation will generate such perturbations.  The breakdown of the  time diffeomorphism symmetry will generate the Goldstone modes from the various EFT operators which are nothing but the gravitational fluctuations.  Fortunately, these Goldstone modes mimic the role of the spontaneous symmetry breaking mechanism   in the context of $SU(N)$ gauge theory.   In our discussion, such Goldstone modes serve the purpose of the scalar modes that appear in the  cosmological perturbation theory.  In this setup,  a scalar field transform as a scalar under the following space-time diffeomorphism symmetry within the framework of the General Theory of Relativity (GTR):
\begin{widetext}
	\be\label{xxxc1} {\textcolor{Sepia}{\textbf{Space-time~diffeomorphism~ symmetry:}}~~~~ x^{\nu}\Longrightarrow  \tilde{x}^{\nu}=x^{\nu}+\xi^{\nu}(t,{\bf x})~~~\forall~ \nu=0,1,2,3}~.\ee   
	\end{widetext}
	Here $\xi^{\nu}(t,{\bf x})$ represents the space-time diffeomorphism parameter.  Now if we think of a parallel situation where we are interested in the scalar fields,  then the perturbation on the field $\delta \phi$ transforms (1) as a scalar  only under the spatial part of the space-time diffeomorphism symmetry and (2) non-linearly only under the temporal part of the space-time diffeomorphism symmetry.
These spatial and temporal parts of the diffeomorphic transformations can be separately written as:
\begin{widetext}
\bea
\textcolor{Sepia}{\bf Spatial~diffeomorphism~ symmetry:}~~	t&&\Longrightarrow \tilde{t}=t,~x^{i}\Longrightarrow   \tilde{x}^{i}=x^{i}+\xi^{i}(t,{\bf x})~~~\forall~ i=1,2,3\\
&&\longrightarrow 
\textcolor{Sepia}{\bf Outcome:}~~~\delta\phi\Longrightarrow \widetilde{\delta\phi}=\delta\phi,\\	
\label{tdiff}\textcolor{Sepia}{\bf Time~diffeomorphism~ symmetry:}~~	t&&\Longrightarrow \tilde{t}=t+ \xi^{0}(t,{\bf x}),~x^{i}\Longrightarrow \tilde{x}^{i}=x^{i}~~~\forall~ i=1,2,3\\
&&\longrightarrow
	\textcolor{Sepia}{\bf Outcome:}~~~\delta\phi\Longrightarrow \widetilde{\delta\phi}=\delta\phi +\dot{\phi}_{0}(t)\xi^{0}(t,{\bf x}).
\eea
\end{widetext}
In the above  expressions, $\xi^{0}(t,{\bf x})$ and $\xi^{i}(t,{\bf x})\forall i=1,2,3$ represent the spatial and the temporal diffeomorphism parameters.  Here we choose the gravitational gauge such that,  $\phi(t,{\bf x})=\phi_{0}(t),$ where $\phi_{0}(t)$ is the background time dependent scalar field in homogeneous isotropic FLRW cosmological space-time.  This further demands that in this gauge choice: \be {\textcolor{Sepia}{\bf Unitary~gauge~fixing~condition:}~~~~~\delta \phi(t,{\bf x})=0}~,\ee 

In the framework of cosmology, this is the unitary gravitational gauge fixing criteria which implies that all information on the field content is preserved in the background metric.  A similar  fact also appears in the context of $SU(N)$ gauge theory where the Goldstone modes transform non-linearly in the spontaneous symmetry breaking mechanism.  In the present context, similarly, the temporal
diffeomorphism symmetry transformation  captures this non-linear effect.  Now, to construct the COSMOEFT, we need to follow a few steps:
\begin{enumerate}
	\item First of all the EFT operators has to be a function of the gravitational space-time metric $g_{\mu\nu}$.   
	\item  Now using the derivatives of the space-time metric $g_{\mu\nu}$, one can compute the Riemann tensor $R_{\mu\nu\alpha\beta}$, the Ricci tensor $R_{\mu\nu}$ and the Ricci scalar $R$ which are also the components of the EFT action.
	\item EFT operators have to be invariant under the spatial diffeomorphism symmetry transformation.  For this reason, terms containing $g^{00}$ and 
	$\delta g^{00}=\left(g^{00}+1\right)$ are allowed in EFT action.

	\item Additionally, one can consider the extrinsic curvature $K_{\mu\nu}$ and its fluctuation $\delta K_{\mu\nu}$,  which is explicitly defined before.
\end{enumerate}

   Consequently,  the
COSMOEFT action can be expressed as:
\begin{widetext}
\begin{equation}\label{EFTmodel1}
{
	\begin{array}{rl}
	S&=\displaystyle\int d^{4}x \sqrt{-g}\left[\underbrace{\frac{M^2_p}{2}R}_{\textcolor{Sepia}{\bf Einstein~Hilbert~term}}+\underbrace{M^2_p \dot{H} g^{00}-M^2_p \left(3H^2+\dot{H}\right)}_{\textcolor{Sepia}{\bf Mimics~the~ role~of~kinetic~and~ potential~term~for~a~scalar~field}}~~~~~~~~\right.\\&
	\left.\displaystyle~~~~~~~~~~~~~~~~~~+\underbrace{\sum^{\infty}_{n=2}\frac{M^4_n(t)}{n!}
	(\delta g^{00})^n-\sum^{\infty}_{q=0}\frac{\bar{M}^{3-q}_1(t)}{(q+2)!}\delta g^{00}
	\left(\delta K_{\mu}^{\mu}\right)^{q+1}-\sum^{\infty}_{m=0}\frac{\bar{M}^{2-m}_2(t)}{(m+2)!}
	\left(\delta K_{\mu}^{\mu}\right)^{m+2}+\cdots}_{\textcolor{Sepia}{\bf Fluctuation~ part~which~will~directly~contribute~in~finding~scalar~perturbation}} \right].
	\end{array}}
\end{equation}
\end{widetext}
where the dots correspond to the higher-order fluctuations in which we are not interested at present as we have restricted our analysis to operators which contain only  two derivatives in the space-time metric.  Here $\delta K_{\mu\nu}$ and $\delta g^{00}$ is already defined before.
  
In this context, the extrinsic curvature $K_{\mu\nu}$, the unit normal $n_{\mu}$ and the
induced metric $h_{\mu\nu}$ is defined as:
\begin{widetext}
\bea \textcolor{Sepia}{\bf Extrinsic~curvature:}~~~K_{\mu \nu}&=&h^{\sigma}_{\mu}\nabla_{\sigma} n_{\nu}\nonumber\\
&=&\left[\frac{\delta^{0}_{\mu}\partial_{\nu}g^{00}+\delta^{0}_{\nu}\partial_{\mu}g^{00}}{2(-g^{00})^{3/2}}
+\frac{\delta^{0}_{\mu}\delta^{0}_{\nu}g^{0\sigma}\partial_{\sigma}g^{00}}{2(-g^{00})^{5/2}}-\frac{g^{0\rho}\left(\partial_{\mu}g_{\rho\nu}+\partial_{\nu}g_{\rho\mu}-\partial_{\rho}g_{\mu\nu}\right)}{2(-g^{00})^{1/2}}\right],\\
 \textcolor{Sepia}{\bf Induced~metric:}~~~h_{\mu \nu}&=&g_{\mu \nu}+n_{\mu} n_{\nu},\\
 \textcolor{Sepia}{\bf Unit~normal:}~~~n_{\mu}&=&\frac{\partial_{\mu}t}{\sqrt{-g^{\mu \nu}\partial_{\mu}t \partial_{\nu}t}}
=\frac{\delta_{\mu}^0}{\sqrt{-g^{00}}}.\eea	
\end{widetext}

 Now, as we are interested to compute the two- and three-point correlation function, we have restricted it to the following truncated EFT action~:
\begin{widetext}
\begin{equation}\label{EFTmodel2}
{
	\begin{array}{rl}
 \textcolor{Sepia}{\bf Required~COSMOEFT~action:}~~~ S&=\displaystyle\int d^{4}x \sqrt{-g}\left[\frac{M^2_p}{2}R+M^2_p \dot{H} g^{00}-M^2_p \left(3H^2+\dot{H}\right)+\frac{M^{4}_{2}(t)}{2!}\left(g^{00}+1\right)^2+\cdots\right].
	\end{array}}
\end{equation}
\end{widetext}
where the dotted terms are not interesting  for the present purpose as they will not appear in the expression for the effective  speed of sound $c_s$,  which will become  more clear in the next section.  But for other purposes, the other higher order fluctuating contributions are very useful.  For example,  to find the three-point and the four-point correlations from scalar perturbations, these higher order terms are necessarily needed. { \textcolor{black}{ For more details on these aspects see refs.~\cite{Senatore:2010jy}.}}  Last but not the least, the corresponding gravitational metric components  describing spatially flat FLRW space-time having the quasi de Sitter solution is explicitly written in the introduction,  which is used here as a background gravitational space-time.  In this case, we have $\sqrt{-g}=a^3(t)=\exp(3Ht)$  in the required COSMOEFT action.  Here it is important to note that,  the Hubble parameter $H$ is slowly varying with time and is not a constant for the quasi-de Sitter space-time.  For this reason, in the later part of the discussion in the article, we have introduced a quantity,  $\epsilon=-\dot{H}/H^2$,  which measures the deviation from the constant behaviour of the Hubble parameter which appears in the context of the de Sitter space.  This parameter is known as the slow roll parameter and our analysis is valid in the region $\epsilon\ll 1$ where all the approximations of the cosmological perturbation theory work well within the framework of COSMOEFT.   

\section{\textcolor{black}{\textbf{\large Appendix.3: COSMOEFT and Goldstone~Boson}}}
\label{v2b}
In the section, we elaborate the St$\ddot{\text{u}}$ckelberg Trick for the broken time diffeomorphism which will generate the Goldstone modes.  We also discuss how the Goldstone action from the EFT can be constructed which will mimic the role of the scalar modes for the cosmological perturbation.  For better understanding, we divide the discussion into two different subsections which deal with both of these issues in detail.
\subsection{\textcolor{Sepia}{\textbf{ \large St$\ddot{\text{u}}$ckelberg Trick}}}
\label{v2b2}
Let us start with the time diffeomorphism symmetry as stated in Eq (\ref{tdiff}) under which the Goldstone mode ($\pi(t, {\bf x})$) transforms as:

\bea
\pi(t, {\bf x})\rightarrow\tilde{\pi}(t, {\bf x})=\pi(t, {\bf x})-\xi^{0}(t,{\bf x}).
\eea
where $\xi^{0}(t,{\bf x})$ represents the local parameter which is appearing in Eq (\ref{tdiff}).  In this discussion, this Goldstone modes mimic the role of the scalar modes in cosmological perturbation.  Then the corresponding unitary gauge fixing condition is given by:
\begin{widetext}
\be {\textcolor{Sepia}{\bf Unitary~gauge~fixing:}~~~~~\pi(t,{\bf x})=0~~~~\Rightarrow~~~~\tilde{\pi}(t,{\bf x})=-\xi^{0}(t,{\bf x})}~,\ee
\end{widetext} 
which will be useful further to construct the EFT action.

Now it is  useful to mention how the components of the space-time metric, the  Ricci tensor,  the Ricci scalar, the  perturbation on the extrinsic curvature, the  time-dependent coefficients and the slowly varying Hubble parameter transform under the application of the broken time diffeomorphism symmetry: 
\begin{enumerate}
	\item \underline{\textcolor{Sepia}{\bf Transformation of space-time metric:}} \\ Under the  broken time diffeomorphism symmetry, the contravariant and the covariant metric transform as:
	\begin{widetext}
\bea
	\textcolor{Sepia}{\bf Contravariant~metric:}~~	{g}^{00}\Longrightarrow&& 
		\tilde{g}^{00}=(1+\dot{\pi}(t,{\bf x}))^2 {g}^{00}+2(1+\dot{\pi}(t,{\bf x})){g}^{0 i}\partial_{i}\pi(t,{\bf x})+{g}^{ij}\partial_{i}\pi(t,{\bf x})\partial_{j}\pi(t,{\bf x}),\\ 
		{g}^{0i}\Longrightarrow&&\tilde{g}^{0i}=
		(1+\dot{\pi}(t,{\bf x})){g}^{0i}+{g}^{ij}\partial_j \pi(t,{\bf x}),\\
		{g}^{ij}\Longrightarrow&&\tilde{g}^{ij}={g}^{ij}.
	\\
		\textcolor{Sepia}{\bf Covariant~metric:}~~	g_{00}\Longrightarrow  &&\tilde{g}_{00}=(1+\dot{\pi}(t,{\bf x}))^2 g_{00},\\
		g_{0i}\Longrightarrow&&  \tilde{g}_{0i}=(1+\dot{\pi}(t,{\bf x})){g}_{0i}+{g}_{00}\dot{\pi}(t,{\bf x})\partial_{i}\pi(t,{\bf x}),\\ g_{ij}\Longrightarrow&&  \tilde{g}_{ij}={g}_{ij}+{g}_{0j}\partial_{i}\pi(t,{\bf x})+{g}_{i0}\partial_{j}\pi(t,{\bf x}).
	\eea
	\end{widetext}
	\item \underline{\textcolor{Sepia}{\bf Transformation of Ricci scalar and Ricci tensor:}}  \\ Under the  broken time diffeomorphism symmetry, the  Ricci scalar and the spatial component of the Ricci tensor on 3-hypersurface transform as:
	\begin{widetext}
\bea
			\textcolor{Sepia}{\bf Ricci~ scalar:}~~ {}^{(3)}R&&\Longrightarrow\displaystyle {}^{(3)}\tilde{R}={}^{(3)}R+\frac{4}{a^2}H(\partial^2\pi(t,{\bf x})),\\
			\textcolor{Sepia}{\bf Spatial~ Ricci~ tensor:}~~ {}^{(3)}R_{ij}&&\Longrightarrow{}^{(3)}\tilde{R}_{ij}={}^{(3)}R_{ij}+H(\partial_{i}\partial_{j}\pi(t,{\bf x})+\delta_{ij}\partial^2\pi(t,{\bf x})).
			\eea
	\end{widetext}
	\item \underline{\textcolor{Sepia}{\bf Transformation of extrinsic curvature:}} \\ Under the  broken time diffeomorphism symmetry the trace and the spatial, time and a mixed component of the extrinsic curvature transform as:
	\begin{widetext}
\bea
		\textcolor{Sepia}{\bf Trace~of~ extrinsic~ curvature:}~~ \delta K&&\Longrightarrow \widetilde{\delta K}=\displaystyle \delta K-3\pi\dot{H}-\frac{1}{a^2}(\partial^2\pi(t,{\bf x})),\\
		\textcolor{Sepia}{\bf Spatial~ extrinsic~ curvature:}~~\delta K_{ij}&&\Longrightarrow\widetilde{\delta K}_{ij}=\delta K_{ij}-\pi(t,{\bf x})\dot{H}h_{ij}-\partial_{i}\partial_{j}\pi(t,{\bf x}),\\
		\textcolor{Sepia}{\bf Temporal~ extrinsic~ curvature:}~~\delta K^{0}_{0}&&\Longrightarrow\widetilde{\delta K}^{0}_{0}=\delta K^{0}_{0},\\
		\textcolor{Sepia}{\bf Mixed~ extrinsic~ curvature:}~~\delta K^{0}_{i}&&\Longrightarrow\widetilde{\delta K}^{0}_{i}=\delta K^{0}_{i},\\
		\textcolor{Sepia}{\bf Mixed~ extrinsic~ curvature:}~~\delta K^{i}_{0}&&\Longrightarrow\widetilde{\delta K}^{i}_{0}=\delta K^{i}_{0}+2Hg^{ij}\partial_j\pi(t,{\bf x}).
	\eea
	\end{widetext}
	\item \underline{\textcolor{Sepia}{\bf Transformation of time-dependent EFT coefficients:}}  \\ Under the  broken time diffeomorphism symmetry, the time-dependent EFT coefficients, after the canonical normalization $\pi_c(t,{\bf x})=Q^2(t)\pi(t,{\bf x})$, transform as:
	\begin{widetext}
\bea
		&&\textcolor{Sepia}{\bf Time-dependent~coefficients}:\nonumber\\
		&&~~~~~~~~~~~~~~~~Q(t)\Longrightarrow Q(t+\pi(t,{\bf x}))=\displaystyle \sum^{\infty}_{n=0}\frac{\pi^{n}(t,{\bf x})}{n!}\frac{d^{n}Q(t)}{dt^n}=\displaystyle \sum^{\infty}_{n=0}\underbrace{\frac{\pi^{n}_c(t,{\bf x})}{n!Q^{2n}(t)}}_{\textcolor{Sepia}{\bf Suppression}}\frac{d^{n}Q(t)}{dt^n}\approx Q(t)~.\eea
\end{widetext}
Here $Q(t)$ represents the time-dependent coefficients in the EFT action.
\item \underline{\textcolor{Sepia}{\bf Transformation of Hubble parameter:}} \\ Under the  broken time diffeomorphism symmetry, the Hubble parameter transforms as:
\begin{widetext}
\bea
	&&\textcolor{Sepia}{\bf Hubble~parameter}:\nonumber\\
	&&~~~~~~~H(t)\Longrightarrow H(t+\pi(t,{\bf x}))=\displaystyle \sum^{\infty}_{n=0}\frac{\pi^{n}}{n!}\frac{d^{n}H(t)}{dt^n}=\displaystyle\left[1-\underbrace{\pi(t,{\bf x}) H(t) \epsilon-\frac{\pi^2(t,{\bf x})H(t)}{2}\left(\dot{\epsilon}-2\epsilon^2\right)+\cdots}_{\textcolor{Sepia}{\bf Correction~terms}}\right]H(t)~.\eea
\end{widetext}
Here $\epsilon$ is the slow-roll parameter which we have defined before.
\end{enumerate}

Now, to construct the COSMOEFT action, we need to understand the decoupling limit in a more detailed fashion.  In this limit, the mixing contributions between the gravity and the Goldstone modes can easily be neglected.  To justify the validity of this statement, let us start with the contribution from the operator $-\dot{H}M_{p}^2g^{00}$  in the EFT action
which is essential  for further computation.

Under the  broken time diffeomorphism symmetry,  this operator  transforms in the following fashion:
\begin{widetext}
\be\label{ddd}  -\dot{H}M_{p}^2g^{00}\Longrightarrow -\tilde{\dot{H}}M_{p}^2\tilde{g}^{00}\approx-\dot{H}M_{p}^2\left[ (1+\dot{\pi}(t,{\bf x}))^2g^{00}+\left(2(1+\dot{\pi}(t,{\bf x}))\partial_i \pi(t,{\bf x}) g^{0i}+g^{ij}\partial_i\pi(t,{\bf x}) \partial_j \pi(t,{\bf x})\right)\right]~.\ee
\end{widetext}
Now after perturbation, the temporal component of the metric can be written as:
\be g^{00}=\bar{g}^{00}+\delta g^{00},\ee
 where the time component of the background quasi de Sitter metric is given by,  $\bar{g}^{00}=-1$ and the perturbation is characterized by $\delta g^{00}$.  Further using this in Equation~(\ref{ddd}), we get a kinetic contribution,  $M_{p}^2\dot{H}\dot{\pi}^2\bar{g^{00}}$ and a mixing contribution,  $M_{p}^2\dot{H}\dot{\pi}\delta g^{00}$. Furthermore, we use a canonical normalized metric perturbation, 
$\delta g^{00}_c=M_{p}\delta g^{00}$,
and one can write the mixing contribution as,  $M_{p}^2\dot{H}\dot{\pi}\delta g^{00}= \sqrt{\dot{H}}\dot{\pi}_c\delta g^{00}_c$.  Here,  above the energy scale $E_{mix}=\sqrt{\dot{H}}$,  one can easily neglect this mixing term in the decoupling limit.  One can also consider mixing contributions $M_{p}^2\dot{H}\dot{\pi}^2\delta{g^{00}}$ and $\pi M_{p}^2\ddot{H}\dot{\pi}\bar{g}^{00}$, which can be rewritten after the canonical normalization as, 
$M_{p}^2\dot{H}\dot{\pi}^2\delta{g^{00}}=\dot{\pi}_c^2\delta{g^{00}_c}/M_{p}$ and $\pi M_{p}^2\ddot{H}\dot{\pi}\bar{g}^{00}=\ddot{H}\pi_c\dot{\pi}_c\bar{g}^{00}/\dot{H}$.  For $E>E_{mix}$, one can neglect the contribution from $M_{p}^2\dot{H}\dot{\pi}\delta{g^{00}}$ term. In  the decoupling limit,  Equation~(\ref{ddd}) can be further simplified as:
\be\label{ddd1} -\dot{H}M_{p}^2g^{00}\Longrightarrow-\tilde{\dot{H}}M_{p}^2\tilde{g}^{00}\approx-\dot{H}M_{p}^2g^{00}\left[\dot{\pi}^2-\frac{1}{a^2}(\partial_i\pi)^2\right].\ee
The above-mentioned simplified expression is  useful for further calculation as it appears explicitly  in the EFT action.
   
   \subsection{\textcolor{Sepia}{\textbf{\large The Goldstone Action}}}
\label{v2b4}
In the decoupling limit, the Goldstone action ($S_{\pi}$) is given by:
   \bea 
   S_{\pi}&=&\displaystyle \int d^{4}x ~a^3\left[-M^2_{p}\dot{H}\left(\dot{\pi}^2-\frac{1}{a^2}(\partial_{i}\pi)^2\right)
   	+2M^4_2 \dot{\pi}^2\right]\nonumber\\
   	&=&\int d^{4}x ~a^3~\left(-\frac{M^2_p\dot{H}}{c^2_s}\right)\left[\dot{\pi}^2
   -c^2_s\frac{1}{a^2}(\partial_{i}\pi)^2
   \right],~~~~~\eea
 where we define the  effective  speed of sound parameter $c_{s}$  as:
   \bea\label{cs}\textcolor{Sepia}{\bf Effective~sound~ speed}: c_{s}\equiv \frac{1}{\displaystyle \sqrt{1-\frac{2M^4_2}{\dot{H}M^2_p}}}.\eea 
   Here if we set $M_{2}=0$ then we have $c_{s}=1$, which is true for the canonical single-scalar field  slow-roll models.  But if $M_{2}\neq 0$ then we get $c_{s}\neq1$.  But since we are only interested within the window,  $0.024\leq c_s\leq 1$,  using this the following constraint on the parameter $M_{2}$ can be derived as:
   \be -867.556\leq \frac{M^4_2}{\dot{H}M^2_p}\leq 0.\ee

Here the linear spatial part of the metric perturbation can be written as:
   \bea\label{W1} g_{ij}=a^{2}(t)\left[\left(1+2\zeta(t,{\bf x})\right)\delta_{ij}+\gamma_{ij}\right]~~\forall~~~i=1,2,3,~~\eea
  where $a(t)$ is the scale factor which we have defined before in the quasi de Sitter background.  Here $\zeta(t,{\bf x})$ represents the curvature perturbation,
   $\gamma_{ij}$ is the transverse and trace-less tensor perturbation. Under the  broken time diffeomorphism symmetry, the  scale factor $a(t)$ transforms as:
   \bea\label{W2} a(t)\Longrightarrow a(t-\pi(t,{\bf x}))
   &\approx & a(t)\left(1-H\pi(t,{\bf x})\right)~. \eea
  
  This further implies that the curvature perturbation $\zeta(t,{\bf x})$ can be expressed in terms of the Goldstone modes $\pi(t,{\bf x})$ as:
  \be \label{W3} \zeta(t,{\bf x})=-H\pi(t,{\bf x}).\ee
 Consequently,  the Goldstone action can be expressed in terms of the curvature perturbation variable $\zeta(t,{\bf x})$  as:
  \bea \label{GM}S_{\zeta}=\int d^{4}x ~a^3~\left(\frac{M^2_p\epsilon}{c^2_s}\right)\left[\dot{\zeta}^2
   -c^2_s \frac{1}{a^2}(\partial_{i}\zeta)^2
   \right].~~~~~~~\eea
This action can be further used to study the squeezed state formalism in the present context.

\bibliography{referencesnew}
\bibliographystyle{utphys}

\end{document}